\documentclass{aastex62}[manuscript]
\newcommand\msun{M$_\sun$}

\newcommand\rearth{R$_\earth$}

\submitjournal{ApJ}

\shorttitle{}
\shortauthors{Anderson et al.}

\begin{document}

\title{Higher Compact Multiple Occurrence Around Metal-Poor M-Dwarfs and Late K-Dwarfs}

\correspondingauthor{Sophie G. Anderson}
\email{sophiega@mit.edu}

%1st Author
\author{Sophie G. Anderson}
\affiliation{Massachusetts Institute of Technology, 77 Massachusetts Avenue, Cambridge, MA 02139, USA}

%Other authors
\author[0000-0001-7730-2240]{Jason A. Dittmann}
\affiliation{51 Pegasi b Fellow, Earth and Planetary Sciences, Massachusetts Institute of Technology, 77 Massachusetts Avenue, Cambridge, MA 02139, USA}
\author{Sarah Ballard}
\affiliation{University of Florida, 211 Bryant Space Science Center, Gainesville, FL 32611, USA}
\author{Megan Bedell}
\affiliation{Center for Computational Astrophysics, Flatiron Institute, 162 5th Avenue, New~York, NY 10010, USA}

\begin{abstract}
%========= What is an abstract?=========
%This is the most important part of the paper.
%No one reads a paper without first reading the abstract.
%Often only the abstract is read...

%It must be short (less than 250 words typically)
%But it must also tell the reader:
%    What was Done,
%    What Hypothesis was being investigated
%    What was in fact measured
%    What Hypothesis (or model) was supported by your measurements
%    Why this is important, and broader impact
%This is the hardest part of a paper to write, and should be written last since it summarizes your entire paper.

The planet-metallicity correlation serves as a potential link between exoplanet systems as we observe them today and the effects of bulk composition on the planet formation process. Many observers have noted a tendency for Jovian planets to form around stars with higher metallicities; however, there is no consensus on a trend for smaller planets. Here, we investigate the planet-metallicity correlation for rocky planets in single and multi-planet systems around Kepler M-dwarf and late K-dwarf stars. Due to molecular blanketing and the dim nature of these low mass stars, it is difficult to make direct elemental abundance measurements via spectroscopy. We instead use a combination of accurate and uniformly measured parallaxes and photometry to obtain relative metallicities and validate this method with a subsample of spectroscopically determined metallicities. We use the Kolmogorov-Smirnov (KS) test, Mann-Whitney U test, and Anderson-Darling test to compare the compact multiple planetary systems with single transiting planet systems and systems with no detected transiting planets. We find that the compact multiple planetary systems are derived from a statistically more metal-poor population, with a p-value of 0.015 in the KS test, a p-value of 0.005 in the Mann-Whitney U test, and a value of 2.574 in the Anderson-Darling test statistic, which exceeds the derived threshold for significance by a factor of 25. We conclude that metallicity plays a significant role in determining the architecture of rocky planet systems. Compact multiples either form more readily, or are more likely to survive on Gyr timescales, around metal-poor stars. 

%We find that compact multiple systems prefer metal-poor M and late K-dwarfs, with 66\% of compact multiple systems (compared to 50\% of single-planet systems) in orbit around a star more metal poor than the average star in the \'$M_K$ vs. \'$G_{BP}$ - $G_{RP}$ plane. The \textbf{Kolmogorov-Smirnov (KS)} test values comparing the compact multiple systems with the field population of stars is 0.028 in our derived metallicity diagnostic from \'$M_K$ and \'$G_{BP}$ - $G_{RP}$. 

\end{abstract}
\section{Introduction}
Planets are ubiquitous in the Milky Way (\cite{Howard2012}, \cite{Dressing2015}). Understanding planetary systems in detail, and how their planet types and system architecture vary as a function of stellar type, bulk composition, and other external parameters, is the next step in understanding the planet formation and migration process. % and whether systems like the Solar System are intrinsically common. 
\cite{Fisher05} conducted a metallicity study of 1040 FGK-type stars with data from Keck Observatory, Lick Observatory, and the Anglo-Australian Telescope. They found that the probability that a solar-type star hosts a massive planet depends on metallicity. Among stars with higher than solar metallicity, there is a consistent increasing trend in massive gas giant planet occurrence. The correlation is tied to total planet mass, and it also applies to multi-planet systems. However, there is no correlation between host star metallicity and orbital period \citep{Fisher05}. 

The Kepler space telescope launched in 2009 as a statistical mission to uncover the transiting planet population in the Milky Way \citep{Borucki_Kepler}. Kepler surveyed a single patch of sky continuously from 2009-2013, discovering thousands of exoplanets.\footnote{\url{https://exoplanetarchive.ipac.caltech.edu/docs/counts_detail.html}} \citet{Howard2012} looked at solar-type stars with temperatures between 4100 and 6100 K and found occurrence rates of 0.130, 0.023, and 0.013 planets per star for planets with orbital periods of less than 50 days and radii of 2-4, 4-8, and 8-32 $R_\earth$, respectively. For longer period (237-500 days) and Earth-sized (0.75-1.5 \rearth{}) planets, \citet{Hsu2019} identified an occurrence rate of $<$0.27. Smaller planets are more common than larger planets, and total planet occurrence rate also increases for lower mass stars. Overall, 30$\%$ of FGK-type stars host planets with $R_\texttt{planet}$ $>$\rearth{} and orbital period $<$ 400 days \citep{Zhu18}. \citet{Mulders2015} found that the number of small planets per star is a factor of 3.5 times larger for M-dwarfs than for FGK type stars. Planets of radius 0.5-4 \rearth{} with an orbital period of less than 50 days occur at a rate of 0.9 planets per star around the smallest stars \citep{Dressing2013}, and a rate of 2.5 planets per M-dwarf with radii of 1-4 \rearth{} and periods $<$ 200 days \citep{Dressing2015}. \citet{Hardegree19} found planet occurrence rates of 0.86, 1.36, and 3.07 for M3 V, M4 V, and M5 V stars in the Kepler field respectively.

Armed with the larger Kepler statistical sample, it is possible to re-examine the planet-metallicity correlation in finer detail. \cite{Schlaufman2011} proposed that the planet-metallicity correlation may extend to smaller planets. They found that Kepler stars hosting giant planets and K-dwarfs with small exoplanet candidates are significantly redder in $g$-$r$ color than the field population of stars. % in a color-color plane, they found evidence that giant exoplanet candidate host stars and K-dwarfs with small exoplanet candidates are significantly redder in g-r color.
This suggests that exoplanet host stars, including small planet hosts, are preferentially more metal rich. However, \cite{Mann2012} later found that this study was systematically contaminated by giant stars, which questions the validity of that result. 

While several additional studies have confirmed that there exists a correlation between host star metallicity and likelihood of hosting a short period giant planet \citep{Everett2013, Buchhave2014}, a clear trend for smaller planets has been difficult to establish \citep{Buchhave2012}. Both \citet{Wang2014} and \citet{Buchhave2014} found evidence of a universal planet-metallicity correlation using metalliticies of Kepler planet hosts. \citet{Buchhave2014} used newly gathered spectra to compare planet hosts to non-planet hosts, while \citet{Wang2014} used metallicites derived from the Kepler Input Catalog (KIC) as a proxy for their target's true metallicities after showing rough agreement with the measurements from \citet{Buchhave2014}. \citet{Wang2014} found that terrestrial planets ($R_\texttt{planet}$ $\leq$ 1.7 \rearth) are 1.72 times more likely to form around metal rich stars. However, a different statistical method used by \cite{Schlaufman2015} found no correlation between stellar metallicity and the likelihood of hosting a small planet. 

There is some evidence that bulk metallicity is also correlated with other planetary system properties. \cite{Mulders16} found a correlation between host star metallicity and exoplanet orbital period using the Kepler sample and spectroscopic metallicities from the LAMOST survey \citep{LAMOST}. Hot planets with periods of $<$10 days exist with a much higher frequency around metal rich stars, whereas cooler exoplanets have host star metallicities closer to that of the sun. \cite{Mulders16} found no evidence of a direct planet-metallicity correlation, but indicated that they would need a larger sample to investigate this further. 

%Paragraph 2:
%What sort of reasons might there be for a planet-metallicity correlation? What theory papers are there describing how such an effect could work and what did they say?
%\cite{Dawson16}
%\cite{Ballard16}
%\cite{Munoz18}
%\cite{Morton14b}
%Paragraph 3:
%What were the first planet-metallicity relation studies and what did they find? From where did their sample come from? What size of planets are these results applicable for? 
%Paragraph 4:
%How has the study of planet-metallicity correlations evolved over time? Summarize our understanding of planet occurrence as a function of stellar metallicity for giant planets as well as the numerous (and sometimes contradictory) attempts to do this for smaller planets. When comparing these studies, note the differences in the sample or methods as well. I rearranged the order of these sections a little. 
%\cite{Brewer18}
%\cite{MunozRomero18}
%Paragraph 5: 
%The two most recent planet-metallicity correlation papers are the Weiss et al. (2018) and the Brewer et al. (2018) papers. They are contradictory. Summarize these papers here, how they differ in their methods and samples, and what their shortcomings might be.

Recently, \cite{Brewer18} and \cite{Weiss18} investigated whether metallicity is correlated with the architecture of planetary systems. Their findings are contradictory. \cite{Weiss18} used data from the California-Kepler Survey (CKS) and Gaia Data Release 2 (DR2) \citep{Gaia2018} for their filtered, high-purity sample of 892 planets around 349 stars. Their sample consisted of solar type stars with effective temperatures ranging from 4500-6300 K. They looked at system architecture, comparing the metallicities of single and multi-planet systems. Using the Anderson-Darling test to compare the distributions of single and multi-planet systems, they found a p-value of 0.29 indicating no significant planet-metallicity correlation.

\cite{Brewer18} spectroscopically derived metallicities for close to 3000 FGK-type stars, filtering for log(g) $>$ 4.0 to combat error due to increasing metallicity with stellar evolution (\cite{Dotter2017}, \cite{Souto2018}). They separated systems with planets into three categories: hot Jupiters, cool Jupiters, and compact multiples (which they defined as 3 or more planets within 1 AU). They found that hot Jupiter occurrence increases with metallicity, consistent with previous studies. The cool Jupiter systems, which contain a Jupiter-sized planet on a wider orbit, were included for comparison purposes and displayed a similar trend. The compact multiple systems, however, showed a significant increase in probability around stars with a metallicity [Fe/H] below -0.3 dex.  

%Paragraph 6:
% Why might looking at smaller mass stars help shed light on the planet metallicity correlation? What sort of planetary systems do M dwarfs form (explain the Kepler dichotomy from Ballard and Johnson's paper).

The \cite{Brewer18} and \citet{Weiss18} results were based on FGK-type stars from the Kepler sample, but M-dwarfs serve as a natural independent sample for testing these results. \cite{Ballard2016} noted the existence of a “Kepler dichotomy” for planetary systems around M-dwarf stars. These systems come in two varieties: systems with a single close-in transiting planet and those with an average of 5 tightly packed planets. \cite{Romero2018} used the Kepler sample along with metallicities from the California-Kepler Survey to look at a possible metallicity-related cause of the Kepler dichotomy: external giant planets. They found that single and multi-planet systems had similar metallicities, indicating that giant planets are likely not the cause of the dichotomy (giant planet systems would have higher metallicities, as suggested by results from \citealt{Fisher05, Buchhave2014}, and others). The results from \cite{Brewer18} imply that the Kepler dichotomy may partially be explained by metallicity differences in the host stars. We seek to test this claim. 

%Paragraph 7:
%Kepler has given us a statistical sample of M dwarf planetary systems, but how do we get the other piece - metallicity? Historically, how have we measured M dwarf metallicity? How has this developed in recent years via Rojas-Ayala et al. (~2012), Mann et al., and Newton et al.? Since we don't have spectral metallicities for all Kepler M dwarfs, what work on the photometric side has been done (Dittmann et al. 2016), and how can we apply that here?

M-dwarfs make up roughly 70$\%$ of the stars in the Galaxy \citep{Henry2006} and are also represented in the Kepler planet sample, although as only about 2.5\% of the targets observed \citep{Hardegree19}. M-dwarfs and late K-dwarfs are quite dim and the optical region of their spectra is blanketed with molecular lines, which makes obtaining a detailed spectrum and determining their metallicity difficult. \citet{Lepine2007} and \citet{Woolf2009} were some of the first to suggest the use of optical spectroscopic indices as a proxy for metallicity, the former using a calibration of the TiO/CaH ratio, and the latter with the infrared Ca II triplet and K I line. Since the publication of the KIC, numerous studies have increased our ability to measure the metallicities of low mass stars. \citet{Ayala2012} estimated [Fe/H] and [M/H] metallicities for 133 nearby M-dwarfs using the spectral lines of Na I and Ca I, and the \(H_2O–K2\) index. They suggested that the CaH1 index could be a promising optical tool for determining metallicity. \cite{Mann2014} used 44 binary systems to calibrate a metallicity diagnostic with features found in $K_s$ band spectra. \cite{Newton2014} introduced a new metallicity calibration using the sodium doublet at 2.2 $\mu$m, with an accuracy of 0.12 dex. These spectroscopic calibrations were created using M-dwarf - G-dwarf binaries with known G-dwarf metallicity, assuming that the two stars formed from the same material and consequently have the same metallicity. Work from \citet{Terrien1} and \citet{Terrien2} investigated the utility of spectral lines in the infrared as unbiased metallicity indicators and has produced the largest compilation to date of spectroscsopic metallicites of M-dwarfs. \cite{Dittmann2016} used these results to further calibrate a photometric method to determine metallicity using trigonometric distance measurements, spectroscopic metallicity measurements, and 2MASS infrared magnitude measurements. We will use a similar photometric method in this work to qualitatively estimate relative metallicities for our M and K-dwarf sample. 

%Paragraph 8:
%The final missing piece to getting photometric metallicities to the Kepler M dwarfs is having a distance measurement to the stars to get an absolute magnitude. Introduce the Gaia mission here and explain how its parallax measurements will enable this study

%Paragraph 9:
%Lay out the structure of the rest of the paper and what you are going to do throughout it.

In Section 2 we present our data sources: the Kepler Input Catalog \citep{KIC}, Gaia Data Release 2 \citep{Gaia2018}, and the NASA Exoplanet Archive \citep{NASAexo}. We also present our sample selection criteria.  Section 3 describes our methods for investigating the existence of a planet-metallicity relation and comparing the metallicity distributions of compact multiples and single-planet systems, and addresses some possible factors that could complicate our findings. In Section 4 we interpret and discuss these results. Finally, in Section 5 we summarize our process and conclusions.

\section{Data and Observations}

In order to utilize the Kepler M and late K-dwarf sample to investigate the planet-metallicity correlation for smaller planets without possible systematic biases, we require a uniform method to determine the relative metallicities of these stars. In our previous work, \citet{Dittmann2016}, we showed that at a given absolute magnitude, redder stars tend to be more metal rich. Additionally, for a reliable comparison, our sample should be free from M-giant contamination, and ensure uniformly and accurately measured parallaxes and magnitudes. 

We compiled an initial list of M-dwarfs and late K-dwarfs from the Kepler Input Catalog (KIC) \citep{KIC}. The most recent version of the KIC (10) was released in August of 2008 and is available through the MAST data archive. However, there is significant contamination by red giants in this list. To eliminate red giant contamination, we used the work of \citet{Huber2014} and \citet{Berger2018} to clean our sample. \citet{Huber2014} provides surface gravity (log(g)) measurements for these stars, and we eliminated all stars with surface gravity less than 4.5. We chose 4500 K as a temperature cutoff to enable a smooth transition between our results and those from \citet{Brewer18} as their sample used 4500 K as a lower bound. \citet{Berger2018} was the source of the majority of the temperature data, which we used to differentiate dwarf stars from low gravity giants, but we used \citet{Huber2014} temperatures for 682 stars that did not have a temperature in \citet{Berger2018}. Additionally, there were 19 stars near the 4500 K cutoff for which the \citet{Huber2014} and \citet{Berger2018} temperatures were inconsistent, and we used results from \citet{Muirhead2012}, derived from spectra, to verify that 9 of these had temperature under 4500 K and were suitable to add back in to our sample. These methods leave us with a robust list of 7146 M and late K-dwarfs. We used $g$, $r$, $i$, $J$, $H$, and $K_s$ magnitudes as reported by the KIC for our stellar sample. Magnitudes are the only information we pulled from the KIC. For any other stellar parameters, such as mass, temperature, and radius, we used values from \citet{Berger2018}, supplemented by values from \citet{Huber2014} when necessary, all derived from spectra.

While the KIC does include parallax measurements for some of our sample, for this work we uniformly drew all of our parallax measurements from Gaia Data Release 2 \citep{Gaia2018}. Released in April 2018, it includes positions on the sky ($\alpha, \delta$), parallaxes, and proper motions for more than 1.3 billion sources. These data were collected by the European Space Agency's Gaia spacecraft, which launched and began taking data in December 2013. We retrieved the Gaia G magnitude, $G_B$ and $G_R$ magnitudes, and parallax measurements from this database, allowing us to uniformly calculate absolute magnitudes for our M and late K-dwarf sample. The Gaia parallax measurements have typical uncertainties of 0.04 milliarcseconds for sources at G $<$ 15, and around 0.1 mas for sources with G=17. Our sources mostly occur within the range G=13-16. 

We cross-matched Gaia DR2 sources with our sample using co-author MB's publicly available cross-match table,\footnote{\url{http://gaia-kepler.fun}, accessed February 2019.} which includes information from the KIC, Gaia DR2, and the NASA Exoplanet Archive. This cross-match was done by comparing the KIC coordinates to Gaia DR2 coordinates propagated to the KIC reference epoch (J2000) using Gaia proper motions. We utilize the 1 arcsecond matching table.

Finally, some planet systems detected by $Kepler$ are known to be false positives. We eliminated all known false positive systems listed in the NASA Exoplanet Archive in order to obtain our final sample.

In Table 1 we include a summary of the M and late K-dwarf systems used in this work. We have identified 7146 total M and K-dwarfs, 207 of which host transiting planetary systems. There are 153 single planet and 54 compact multiple systems. In Table 2 we present our full sample, which includes Kepler ID, g, r, i, J, H, and K band magnitudes, Gaia B, G, and R magnitudes, parallax, number of Kepler Objects of Interest, number of confirmed planets, Gaia “Goodness of Fit” (see Section \ref{GOF}), surface gravity \citep{Huber2014}, temperature \citep{Berger2018, Huber2014}, and corresponding error measurements for each of our 7146 M and late K-dwarfs.

\begin{table}[h]
\begin{center}
\caption{Stellar Sample}
\label{Sample Summary} 
\begin{tabular}{crr}
\tableline\tableline
No-Planet Systems & 6939 \\
Single-Planet Systems & 153 \\
Compact Multiple Systems & 54 \\
Total & 7146 \\
\tableline 
\tableline
\end{tabular}
\end{center}
\end{table}

\begin{table}
\begin{center}
\caption{Data Table}
\label{Data Table} 
\begin{tabular}{cccccccccccccccccccccccccc}
\tableline\tableline
Kepler ID & g & r & i & J & H & K & Gaia B & Gaia G & Gaia R & Parallax & KOIs & Conf. Planets & Gaia GOF & Surface Gravity & Temperature & g Error & r Error & i Error & J Error & H Error & K Error & Parallax Error & Gaia B Error & Gaia G Error & Gaia R Error\\
4731525 & 15.806 & 14.387 & 13.312 & 11.464 & 10.856 & 10.65 & 15.268 & 14.026 & 12.927 & 13.493 & 0.0 & 0.0 & 0.195 & 0.0 & 3421.0 & 0.025 & 0.02 & 0.02 & 0.021 & 0.021 & 0.018 & 0.024 & 0.003 & 0.0004 & 0.0008\\
8544723 & 17.716 & 16.383 & 15.479 & 13.670 & 13.046 & 12.874 & 17.234 & 16.120 & 15.064 & 2.578 & 0.0 & 0.0 & 0.198 & 0.0 & 3630.0 & 0.025 & 0.02 & 0.02 & 0.022 & 0.024 & 0.03 & 0.042 & 0.006 & 0.0006 & 0.002\\
10848709 & 18.409 & 16.985 & 15.808 & 13.844 & 13.194 & 12.968 & 17.876 & 16.566 & 15.367 & 5.987 & 0.0 & 0.0 & 0.260 & 0.0 & 3309.0 & 0.025 & 0.02 & 0.02 & 0.028 & 0.037 & 0.037 & 0.058 & 0.017 & 0.001 & 0.002\\
7104891 & 16.120 & 14.885 & 14.411 & 12.94 & 12.368 & 12.206 & 15.678 & 14.905 & 14.053 & 3.244 & 0.0 & 0.0 & 0.0 & -0.318 & 4223.0 & 0.025 & 0.02 & 0.02 & 0.021 & 0.021 & 0.016 & 0.025 & 0.004 & 0.0004 & 0.001\\
8607869 & 17.841 & 16.850 & 16.118 & 14.492 & 13.827 & 13.681 & 17.900 & 17.193 & 15.936 & 2.679 & 0.0 & 0.0 & 0.0 & 0.379 & 4223.0 & 0.025 & 0.02 & 0.02 & 0.033 & 0.033 & 0.054 & 0.072 & 0.067 & 0.001 & 0.021\\
...&...&...&...&...&...&...&...&...&...&...&...&...&...&...&...&...&...&...&...&...&...&...&...&...&...\\
\tableline 
\tableline
\end{tabular}
\end{center}
\end{table}

%\subsection{The $Kepler$ M dwarf Sample} sure to add it! Be sure to use citations for catalogs as well!
%Paragraph 1: Describe how you selected the Kepler M dwarfs to use for this study
%Paragraph 2: Describe how after you selected the M dwarfs from where and how you got the g, r, i, z, J, H, K photometry for these stars
%Paragraph 3: Describe how you determined which of these stars had planets and if so, how many planets they had. IE, which planet candidate table did you use and summarize how those planets were detected for the table. What would the likely false positive rate for small Kepler planets around M dwarfs be (there are papers and citations for this). 
%\subsection{Gaia DR2 Parallaxes}
%Paragraph 1: Describe how you took the M dwarfs in the previous section and matched them to Gaia sources with parallaxes.
%Paragraph 2: What is the reliability of these parallaxes from Gaia (see Gaia collaboration papers online).
%Paragraph 3: What is the reliability of your matching criteria? For example - what is the typical separation between a Gaia source and your Kepler source? Do the magnitudes/colors make sense for M dwarfs? That sort of thing. 
%After all this, describe your full sample in general terms. How many total stars do you have? How many systems with planets? How many systems with only 1 transiting planet? What is the typical color and mass of a star in your sample? etc.

\section{Analysis and Results}
The unprecedented precision of the Gaia parallaxes allow us to measure intrinsic luminosity with great precision, and use it to place stars on a color-magnitude diagram. Color-magnitude diagrams allow us to principally measure star mass (through the intrinsic brightness) and temperature (through color), but other physical parameters like activity and metallicity can also have smaller effects on a star's location in the diagram.

\cite{Delfosse2000} showed that an M-dwarf's absolute $K_s$ band magnitude can serve as a proxy for stellar mass and that this is relatively insensitive to other external parameters like metallicity and activity. \cite{Benedict2016} further updated the mass-luminosity relation (MLR) for M-dwarfs using precise stellar mass measurements for 47 stars. Their model relates stellar mass to absolute $V$ and $K_s$-band magnitudes, with K-band magnitudes yielding the most precise determination of stellar mass. The change in metallicity due to change in $K_s$ magnitude has been shown to be very small: \cite{Mann2019} measured 0 +/- 2\% change in absolute K magnitude per dex change in metallicity for solar neighborhood M-dwarfs. This makes the $K_s$ bandpass ideal for isolating the effect of metallicity on a star's location in a color-magnitude diagram, as stars will only shift in the color direction. 

Low temperatures in M-dwarf atmospheres permit complex molecules (principally TiO) to form. These molecules blanket the optical region of the spectrum with absorption features. A more metal-rich stellar atmosphere leads to more optical flux suppression, creating a redder star. \citet{Dittmann2016} showed that there exists an absolute $K_s$-band magnitude metallicity relation \textemdash $MEarth$-$K_s$ \textemdash sensitive to an accuracy of 0.11 dex in metallicity, similar to infrared spectroscopic metallicity indicators from \citet{Newton2014}. A further study by \citet{Dittmann2020} broadened this approach to the Sloan filter system, finding that with precise parallaxes a blue - red color index can be an effective relative metallicity indicator for ranking the metallicity of low mass stars on a uniform basis.  

Throughout our analysis, we use distance from the main sequence of a color-magnitude diagram as a proxy for relative metallicity. In the following sub-section, we justify that decision using a linear approximation to the main sequence. We show that distance from the main sequence correlates with spectroscopically determined metallicity, and therefore can be utilized as a metallicity proxy. In the rest of our analysis, we will use an interpolation as an approximate fit to the main sequence.

\subsection{Metallicity Proxy} 

\begin{figure}
\centering
\includegraphics[scale=0.45]{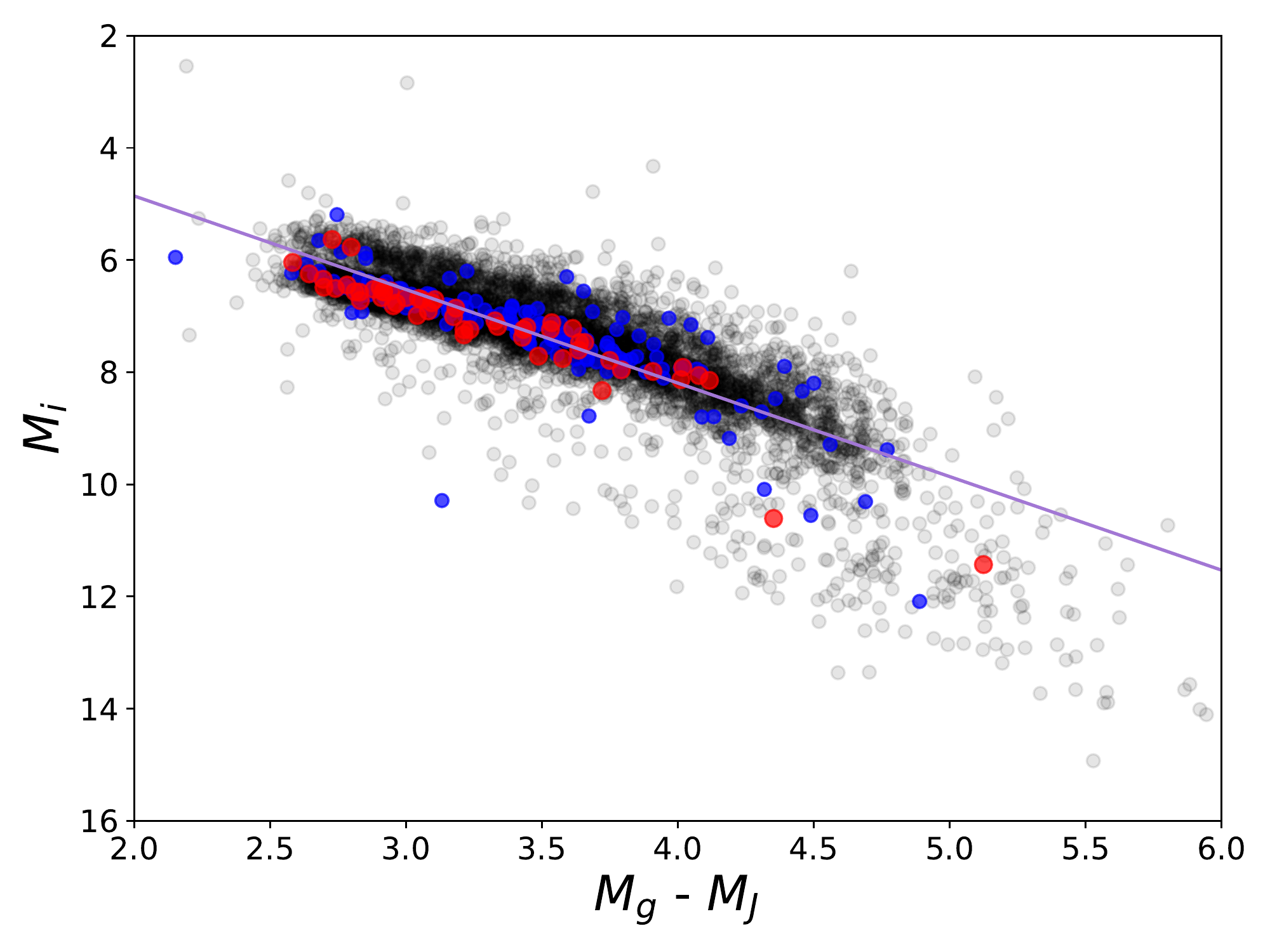}\\
\includegraphics[scale=0.45]{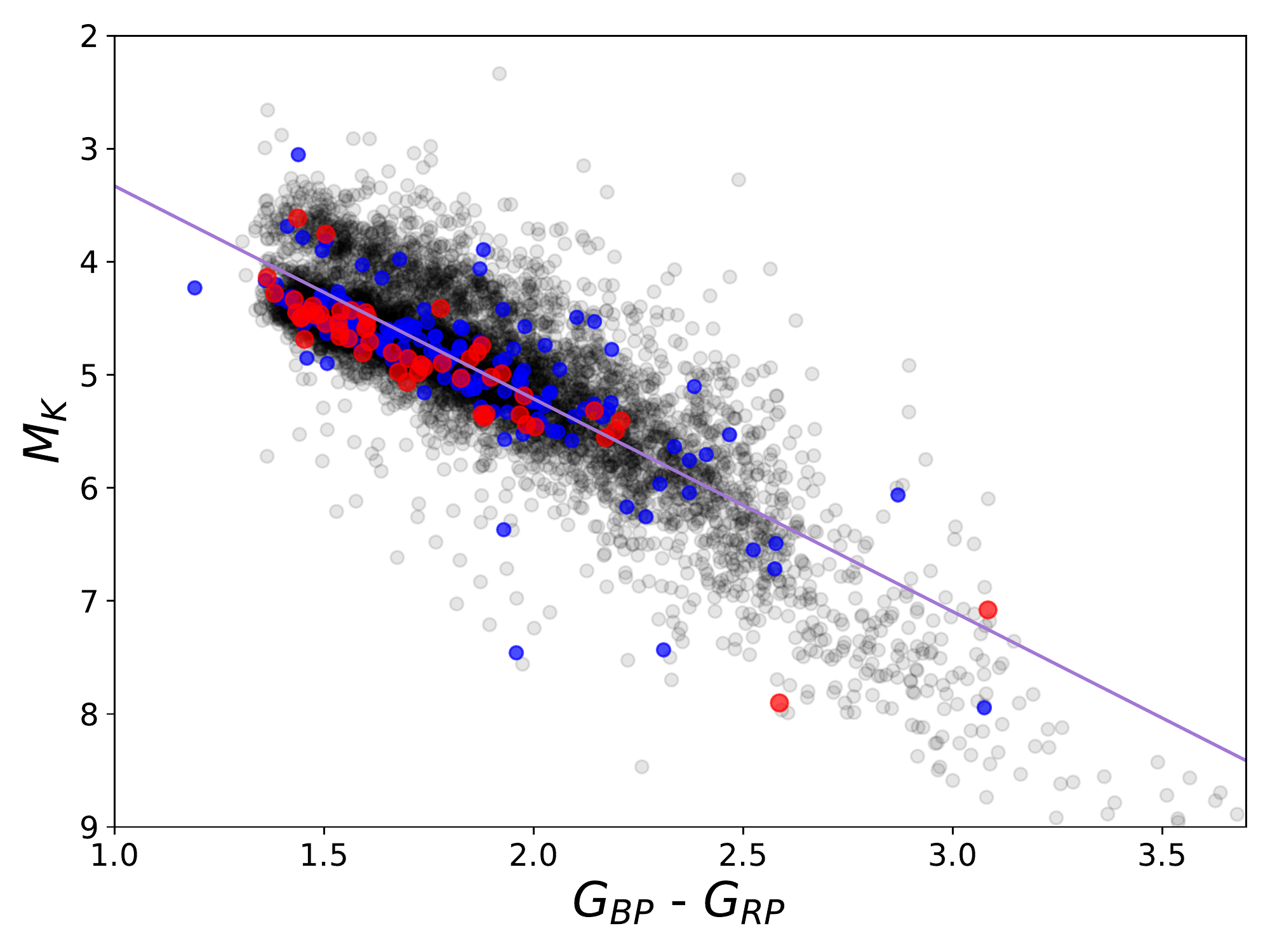}\\
\includegraphics[scale=0.45]{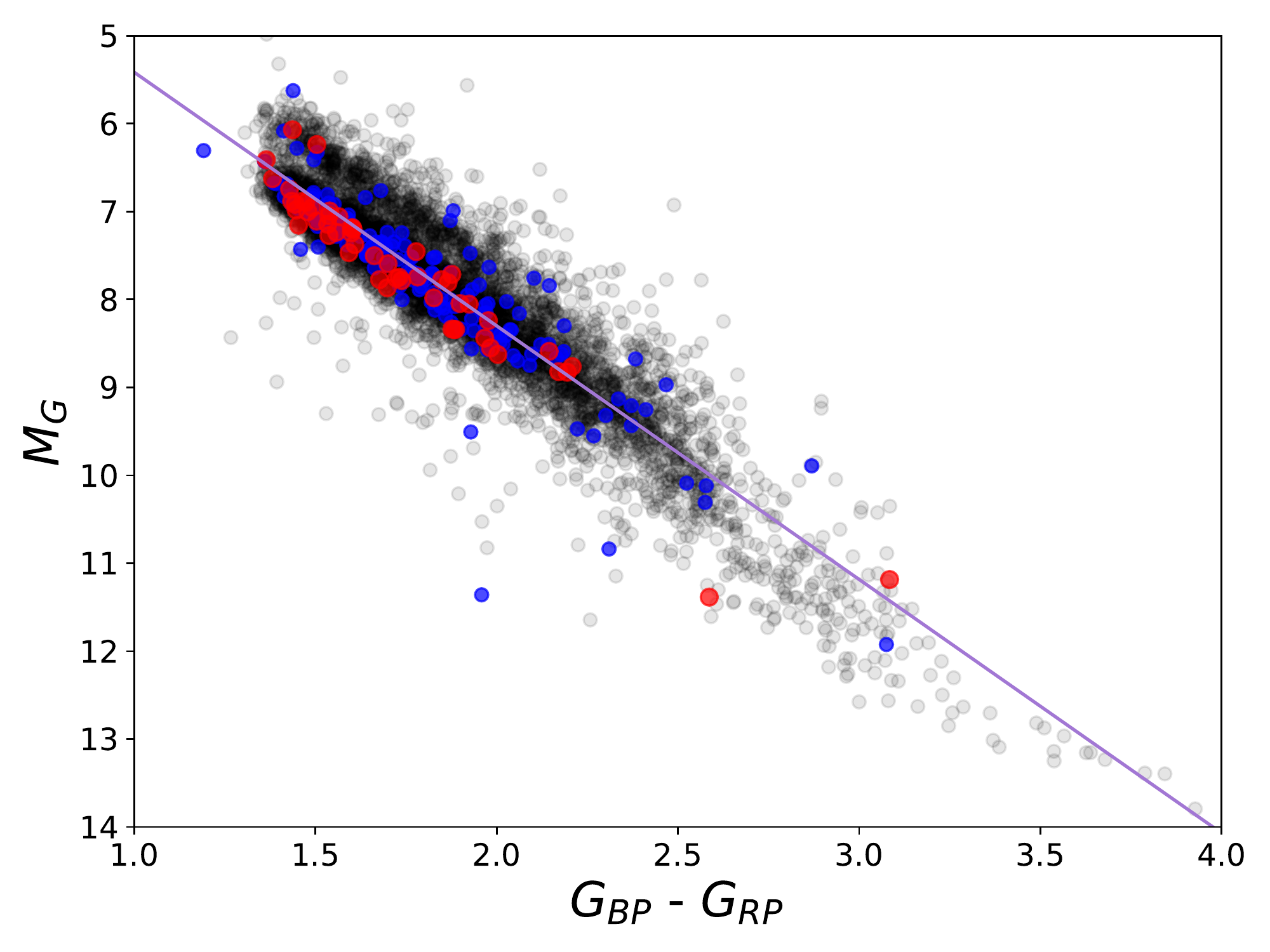}\\
\caption{Color magnitude diagrams in the $M_i$ vs. $M_g - M_J$, $M_K$ vs. Gaia $G_{BP} - G_{RP}$, and Gaia G vs. $G_{BP} - G_{RP}$ bands respectively. The line in each plot is a linear fit to the data, and represents the main sequence at average metallicity. We find that the horizontal distance from this line is correlated with metallicity, and most compact multiple planetary systems occur on the metal-poor side of the line.}
\label{linear_fit_color-mag_plots}
\end{figure}

In Figure \ref{linear_fit_color-mag_plots}, we plot $M_i$ vs. $M_g - M_J$, $M_K$ vs. Gaia $G_{BP} - G_{RP}$, and Gaia $M_G$ vs. $G_{BP} - G_{RP}$ for our full stellar sample. We have 6939 stars with no detected transiting planets, making up the majority of our sample and serving as our control population. We overplot our 153 one-planet systems and 54 compact multiple planetary systems and fit a linear model through the full star sample to represent an approximated main sequence. We use these trends as a base to show that the distance from the main sequence line can be used as a metallicity proxy.

\begin{figure}
\centering
\includegraphics[scale=0.42]{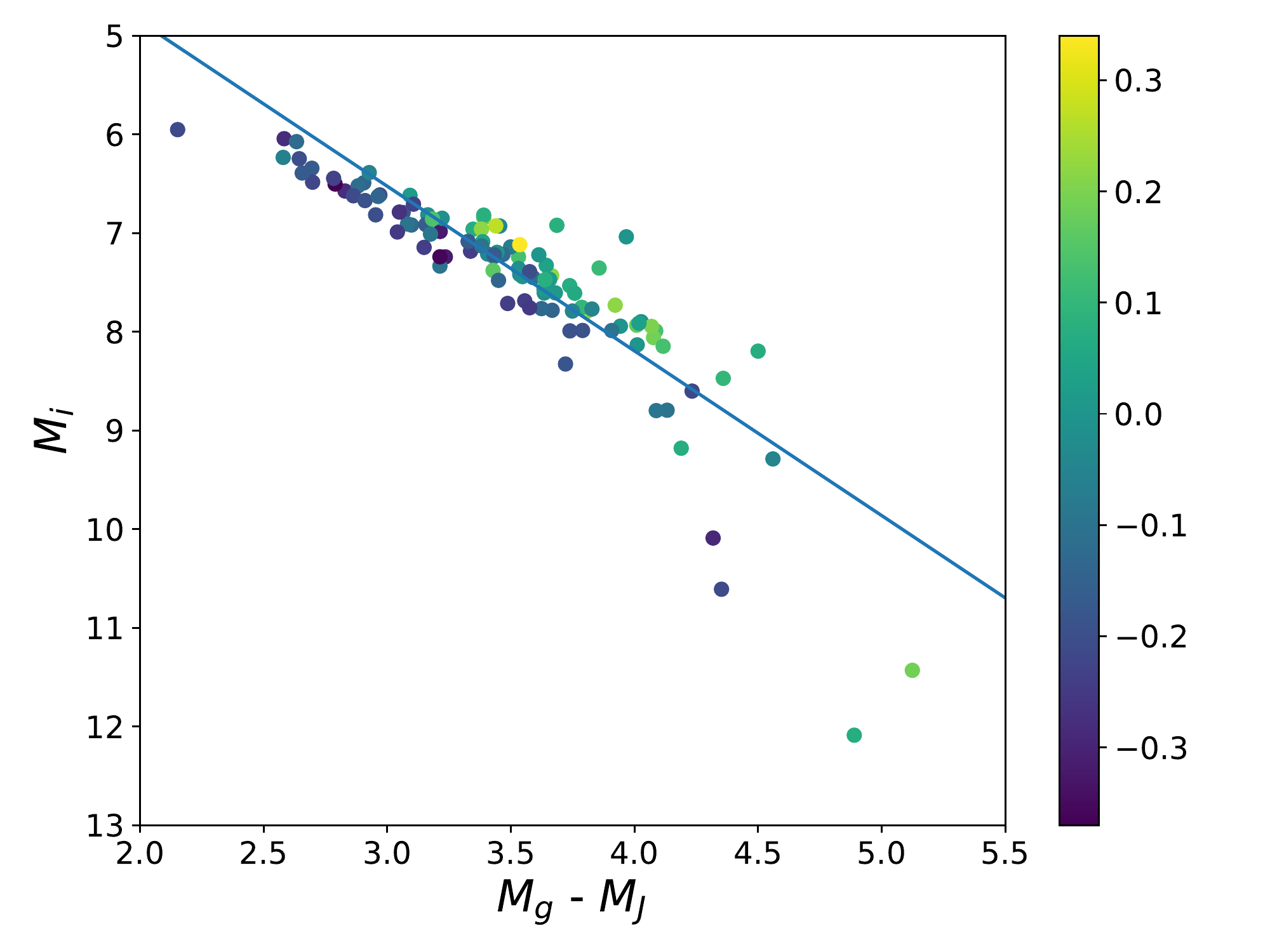}
\includegraphics[scale=0.42]{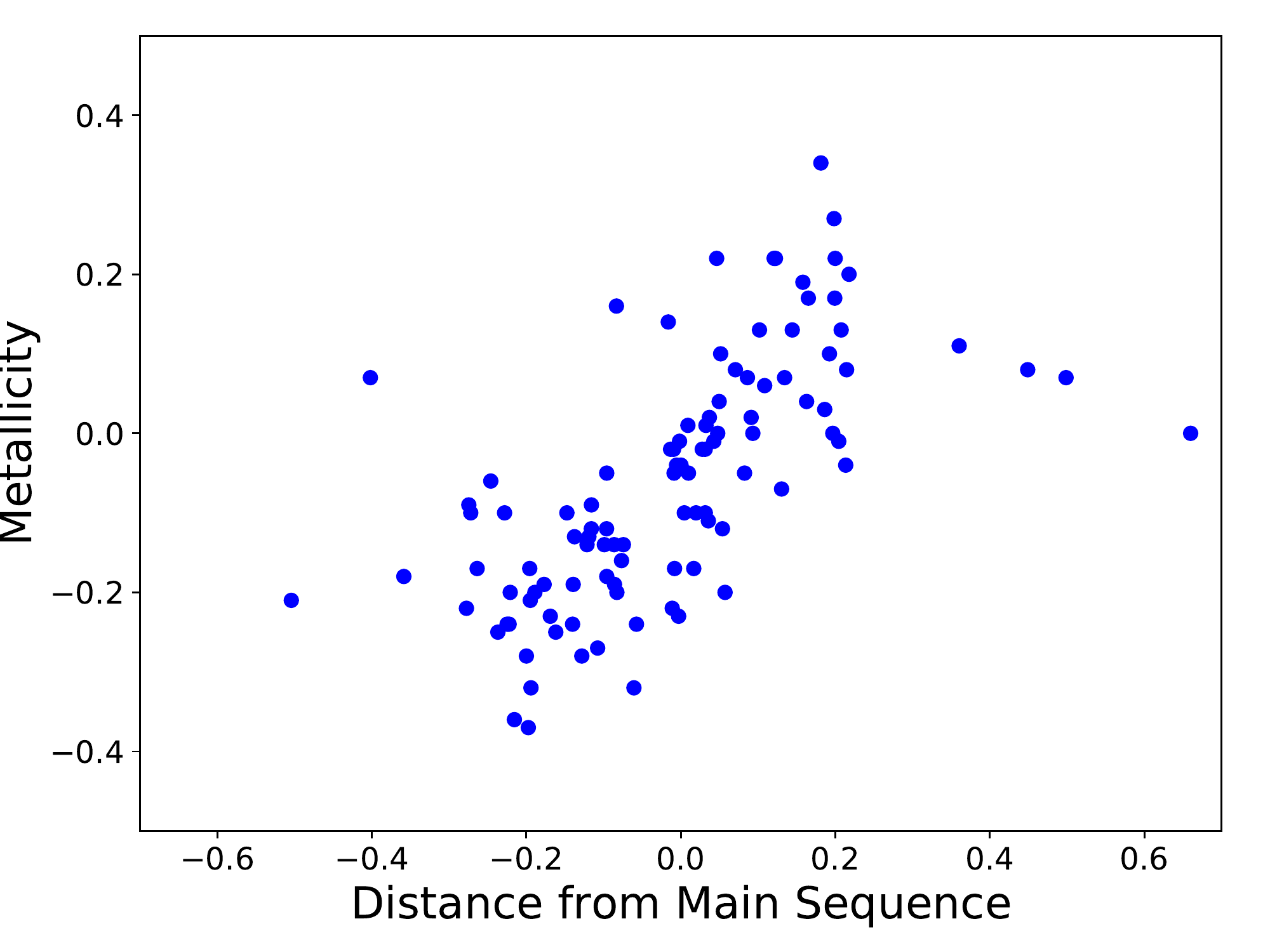}\\
\includegraphics[scale=0.42]{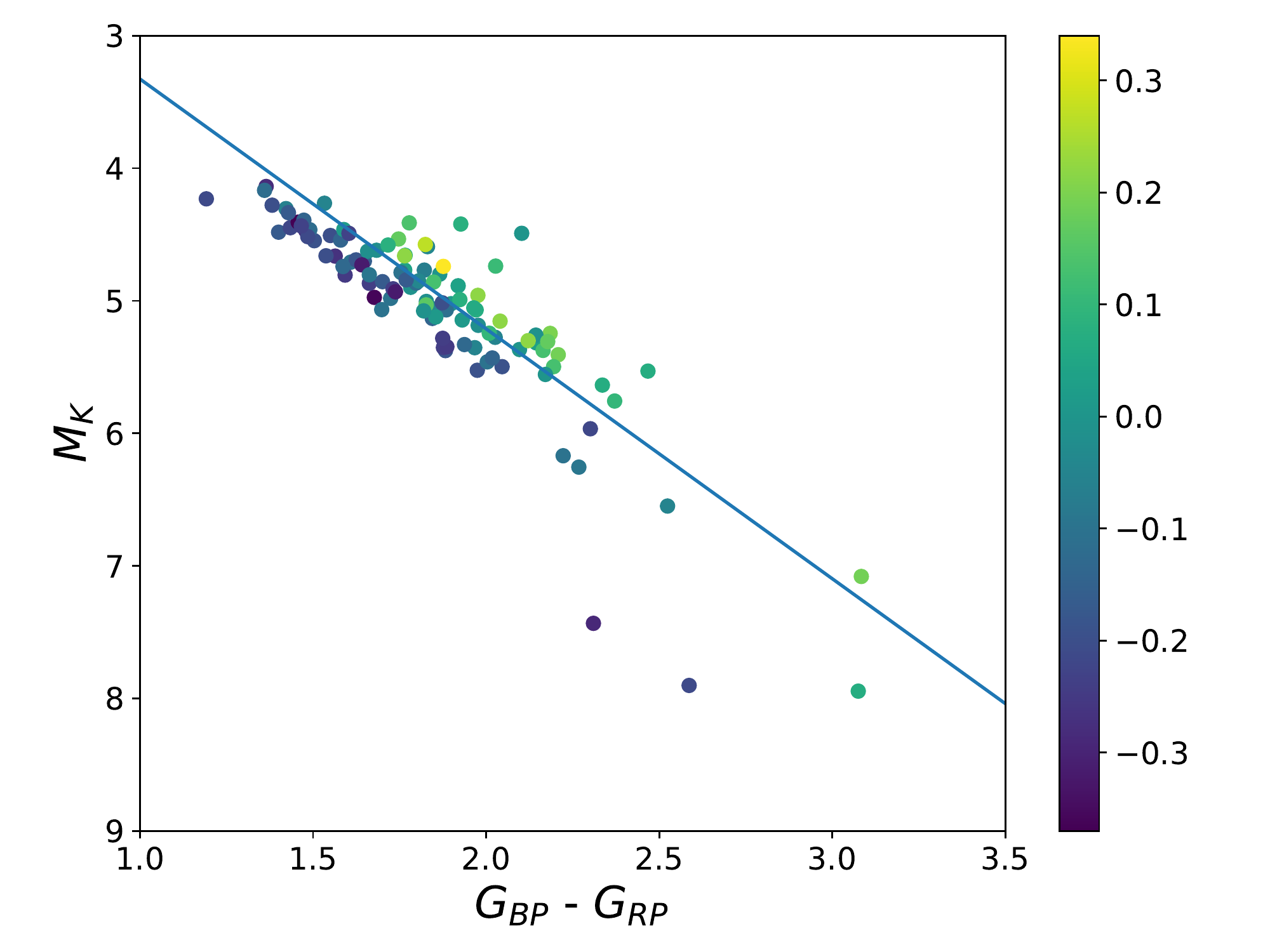}
\includegraphics[scale=0.42]{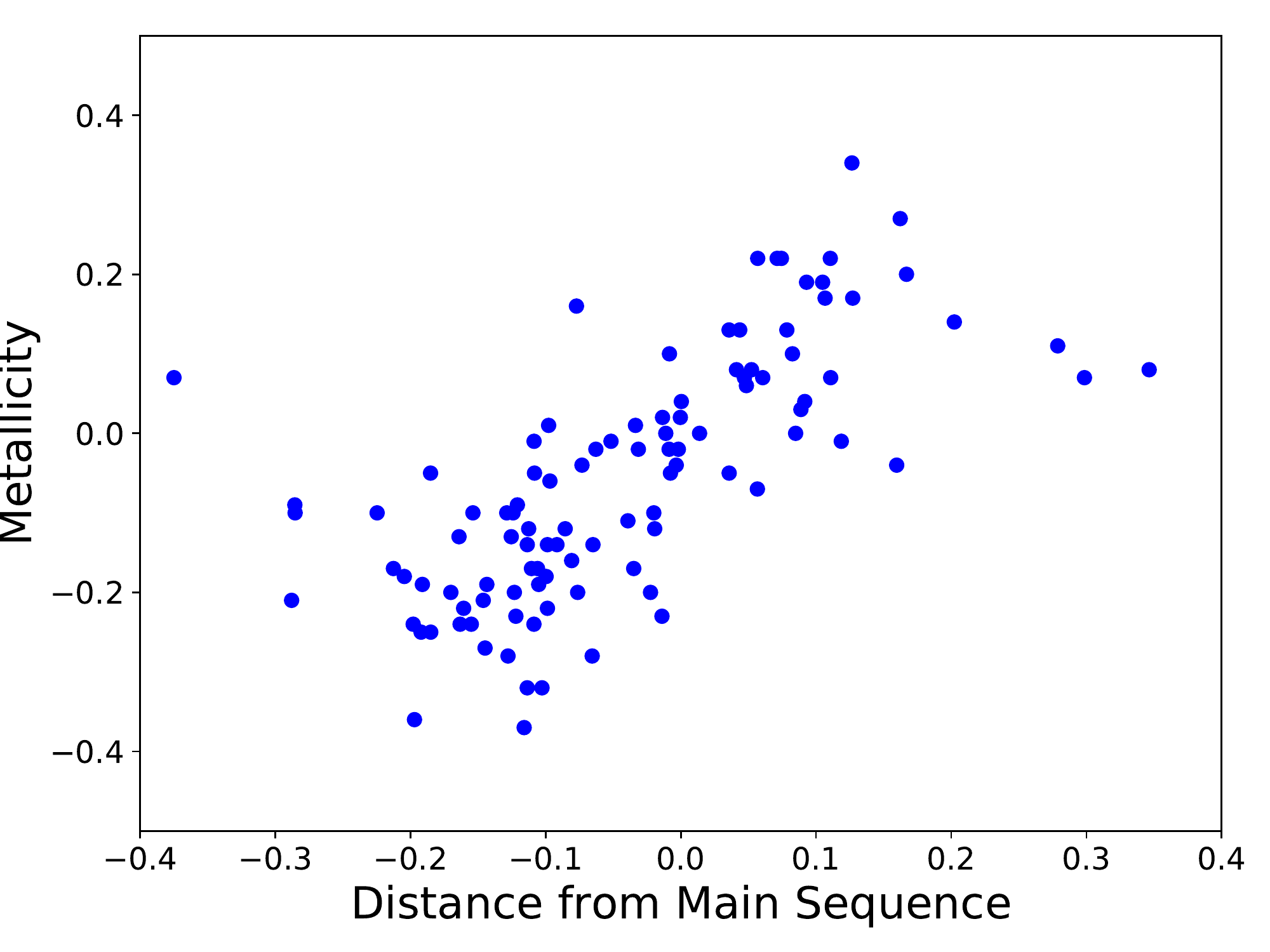}\\
\includegraphics[scale=0.42]{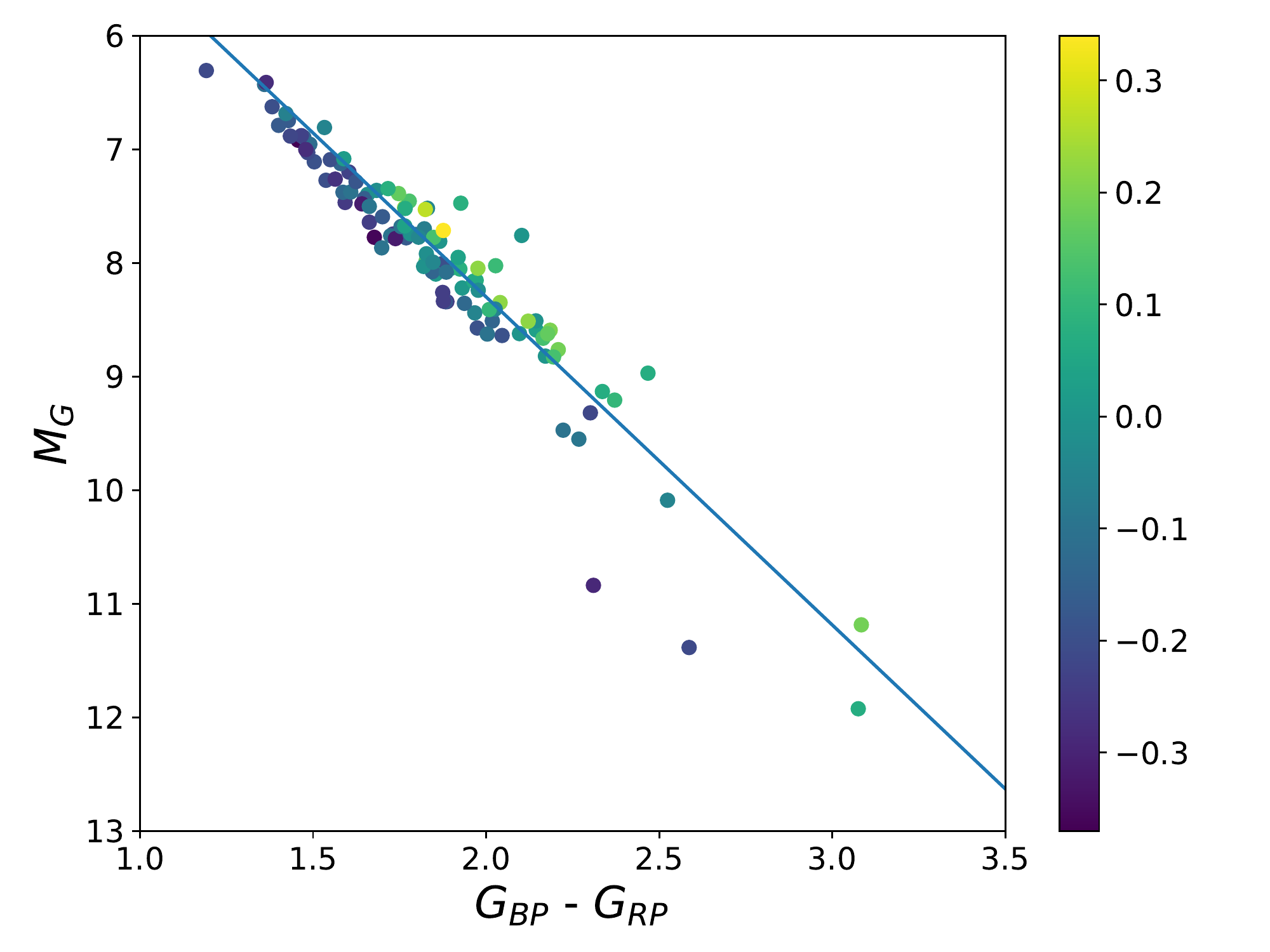}
\includegraphics[scale=0.42]{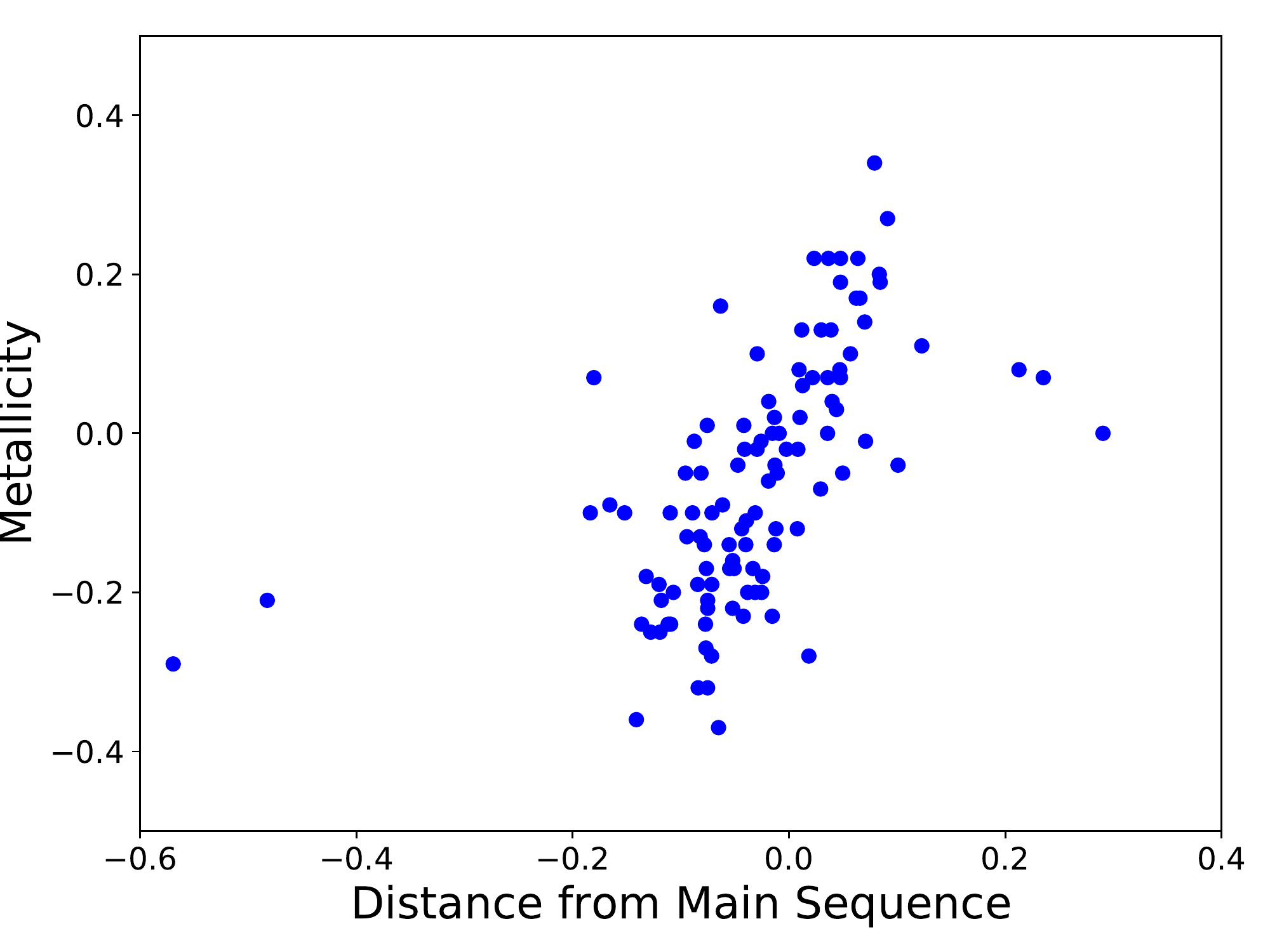}\\
\caption{In the left column, we show color-magnitude diagrams for stars included in both our sample and the \citet{Muirhead2012} or \citet{Muirhead2014} papers, in the $M_i$ vs. $M_g - M_J$, $M_K$ vs. Gaia $G_{BP} - G_{RP}$, and Gaia G vs. $G_{BP} - G_{RP}$ bands. The stars are colored by spectroscopic metallicity \citep{Muirhead2012, Muirhead2014}, and we show an approximate linear fit main sequence generated in each respective band using a color-magnitude plot with only our stellar sample. In the right column, we show the corresponding plot for each band showing metallicity vs. distance from the main sequence line. Both representations show a correlation between metallicity and distance from the main sequence.}
\label{Muirhead_plots}
\end{figure}

We investigate three sets of filters to determine sets that are sensitive to metallicity, including a filter choice that is derived solely from Gaia data products, as a photometric metallicity indicator based on only Gaia may be of wider utility than the present study. Our filter choices are $M_i$ vs. $M_g - M_J$, $M_K$ vs. Gaia $G_{BP} - G_{RP}$, and Gaia $M_G$ vs. $G_{BP} - G_{RP}$ (see Figure \ref{linear_fit_color-mag_plots}). We select the 2MASS $K_s$ band, as the absolute $K_s$ band magnitude has been shown to be relatively unaffected by metallicity \citep{Mann2019}, and is therefore a useful choice to isolate the metallicity effect on the positioning of stars in color-magnitude diagrams. Additionally, the density of spectral lines is fairly limited in the J-band as well \citep{Lindgren2016}. In the future, all sky surveys like TESS \citep{TESS} will provide a larger, more complete sample of M-dwarfs with which to probe the effects of bulk composition on planetary occurrence. We anticipate that Gaia magnitudes will become a more widely used photometric bandpass with which to conduct these and similar studies. Therefore, we find it useful to perform our analysis solely using the optical magnitudes provided by Gaia DR2. Results from \citet{Dittmann2020} suggest that these bandpasses are still sensitive to differences in metallicity at the low-mass end of the main sequence. We further confirm that in the following analysis with spectroscopic metallicities. In Figure \ref{Muirhead_plots}, we show the main sequence diagram and metallicities \citep{Muirhead2012, Muirhead2014} of these stars in all three sets of filter choices. We find that the stars' linear distance from the main sequence is an effective proxy for stellar metallicity.

We note that here and throughout this paper we have not attempted to place these stars on an absolute metallicity calibration scale relative to Solar abundances, but only to use these indices to create a uniformly derived relative metallicity scale with which to perform statistical comparisons.

We used spectroscopic metallicities from \citet{Muirhead2012} and \citet{Muirhead2014} to validate our metallicity proxy method. We took only the stars in the Muirhead papers that overlapped with our star sample and placed them on color-magnitude diagrams in the same $M_i$ vs. $M_g - M_J$, $M_K$ vs. Gaia $G_{BP} - G_{RP}$, and Gaia G vs. $G_{BP} - G_{RP}$ bands (see Figure \ref{Muirhead_plots}). We colored the stars according to spectroscopic metallicity from Muirhead, and overplotted the linear fit trend line from the equivalent color-magnitude diagram made from our full stellar sample as shown in Figure \ref{linear_fit_color-mag_plots}. In this section, we use that line as an approximation for the main sequence. In the left column plots of Figure \ref{Muirhead_plots}, there is a visible pattern showing separation between metal-rich and metal-poor stars.

In the right column of Figure \ref{Muirhead_plots}, we show that distance from the main sequence in the color-magnitude diagram is correlated with the spectroscopic metallicities determined by \citet{Muirhead2012, Muirhead2014}. We find that the redder objects are more metal-rich than their bluer counterparts. These plots confirm the trend seen visually in the color-magnitude diagrams in the left column, and show that there is a relationship between distance from the linear fit main sequence line and metallicity.

\subsubsection{Optimal Distance Measurement Procedure} \label{dist_procedure}
As stated previously,  \citep{Mann2019} found that the K$_s$-band is insensitive to metallicity. As a result, for the $M_K$ vs. Gaia $G_{BP} - G_{RP}$ bands, we used horizontal distance to the line as our proxy for metallicity since this is the only direction that is correlated with metallicity. However, because the main sequence line is consistent regardless of how distance is calculated, perpendicular distance and horizontal distance yield different absolute distance values, but the same overall trend in the distance vs. metallicity plots shown in Figure \ref{Muirhead_plots}. As a result, we choose to use horizontal distance from the line in all bands to maintain consistency. The difference between horizontal and shortest distance to the line is only a scaling factor, and because we use only relative metallicity in our analysis, the two measurements are effectively equivalent.\\\\

Having validated our metallicity proxy using the \citet{Muirhead2012} and \citet{Muirhead2014} spectroscopic metallicities, we expanded our method and conducted statistical analysis on the results shown in Figure \ref{linear_fit_color-mag_plots} to show relative metallicity trends in the larger stellar sample.

\begin{figure}
\centering
\includegraphics[scale=0.45]{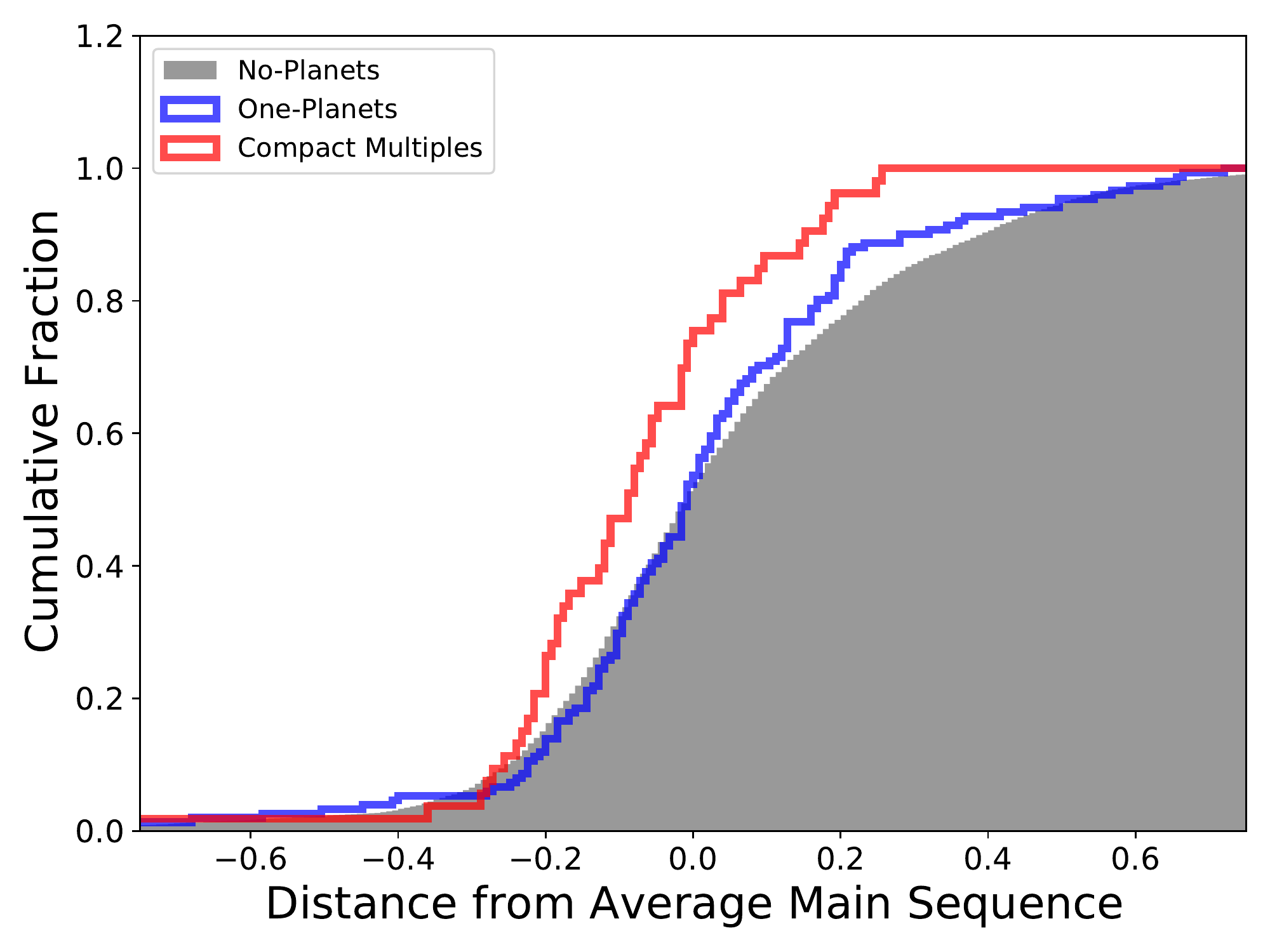}
\includegraphics[scale=0.45]{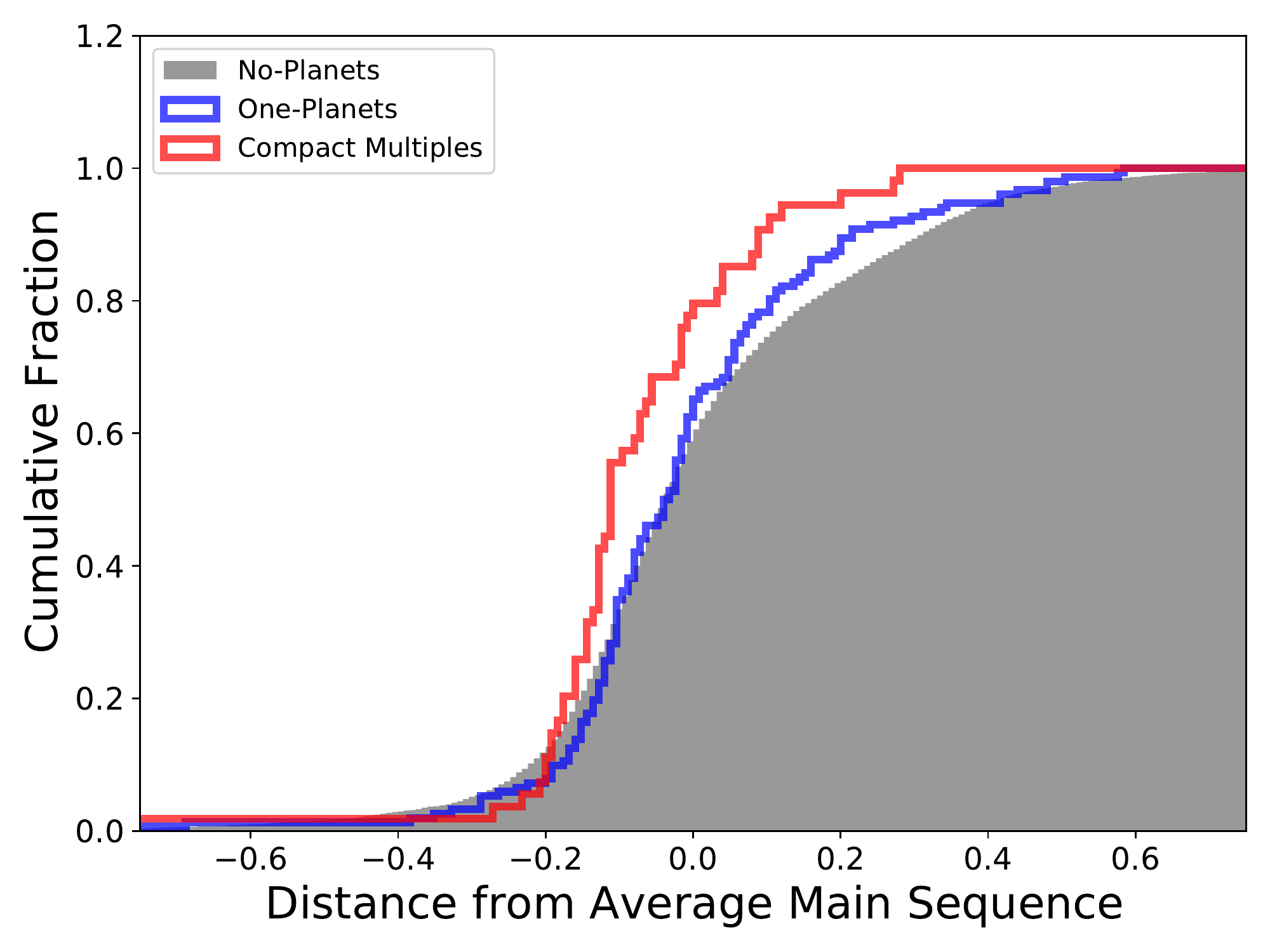}
\includegraphics[scale=0.45]{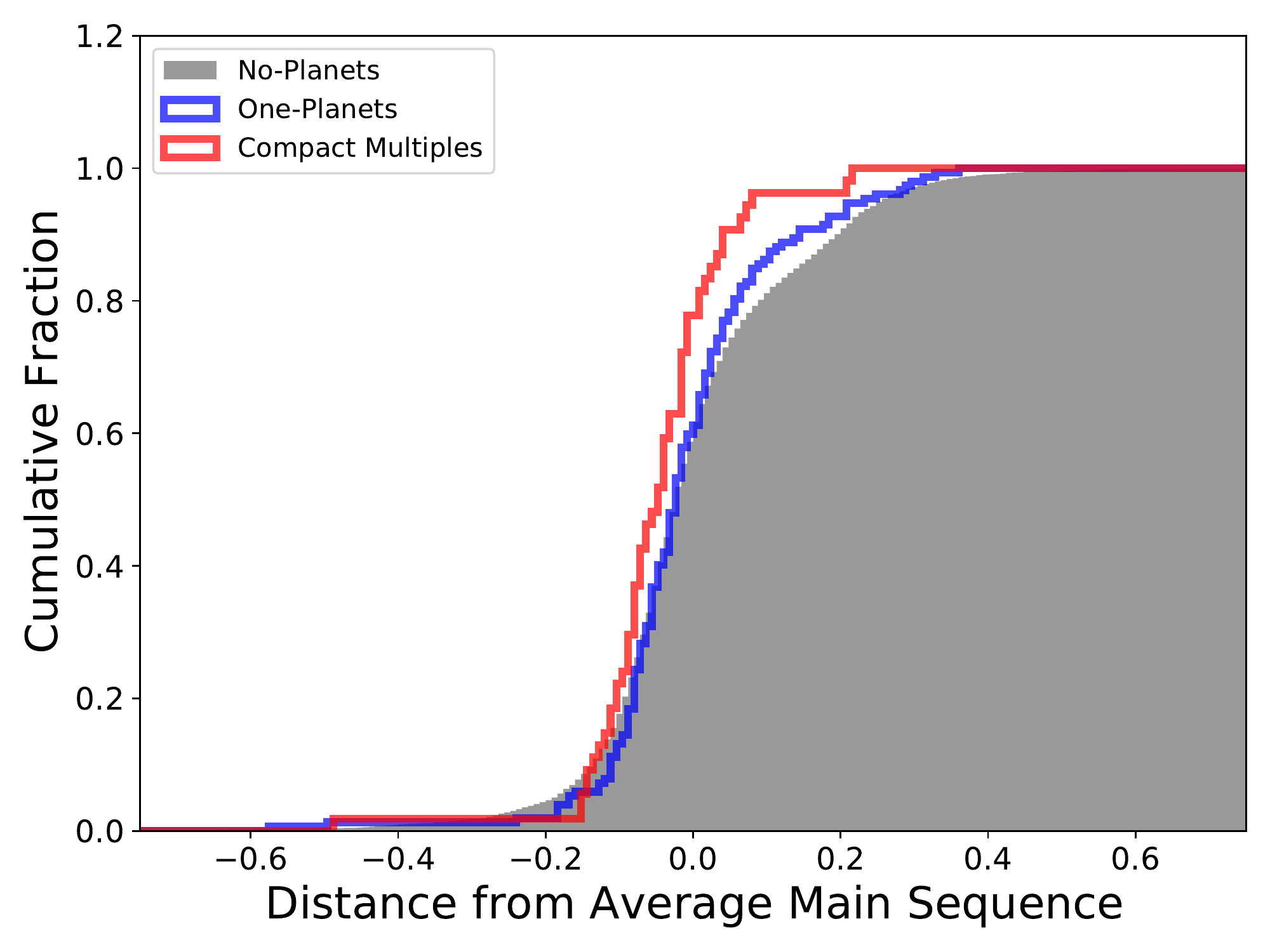}
\caption{Cumulative histograms of the distance from the average metallicity main sequence in the $M_i$ vs. $M_g - M_J$, $M_K$ vs. $G_{BP} - G_{RP}$, and Gaia G vs. $G_{BP} - G_{RP}$ color-magnitude plots respectively. Compact multiples are shown by the red line, single-planet stars in blue, and no transiting planet stars in grey. The compact multiple line deviates from the other two populations because a greater percentage of the total compact multiple sample occurs on the metal-poor, or negative distance, side of the main sequence.}
\label{linear_fit_histograms}
\end{figure}

In Figure \ref{linear_fit_histograms} we construct histograms of the deviation from the main sequence model shown in Figure \ref{linear_fit_color-mag_plots} for each of the three populations (no planets, single planet, and compact multiple systems). By definition, 50\% of the stars with no planets lie to the left of our linear fit main sequence and the other 50\% lie at redder colors. We overplot the single transiting planet systems and compact multiple systems and find that in all bands the compact multiple population, again shown in red, deviates from the single-planet and no-planet populations.

We measure the significance of the difference between these populations with three methods: a Kolmogorov-Smirnov (KS) test, a Mann-Whitney (MW) U-test, and a K-sample Anderson-Darling (AD) test. The KS test is more sensitive to differences in the shapes of the two populations being compared, while the Mann-Whitney U-test is more sensitive to differences in population average, and the Anderson-Darling test is more sensitive to differences in the tails of the distributions. The Anderson-Darling implementation in \textit{SciPy} outputs a statistic and significance level, and the result is significant if the statistic exceeds the significance level. Here, we prefer the the Mann-Whitney U-test because we are measuring differences in the average offset in color between these populations, which correlates with metallicity.

For the $M_K$ vs. Gaia $G_{BP} - G_{RP}$ case, we find that 78\% of compact multiple systems appear on the metal-poor side of the trend line (negative distance) compared with 63\% of single planet-systems, with a KS test p-value for the compact multiple systems compared to no-planet population equal to 0.002.

In the case of the Gaia G vs. $G_{BP} - G_{RP}$ plots, the KS test p-value for the compact multiple population as compared to no-planet systems is 0.026 and 78\% of compact multiples are more metal-poor than average (60\% of single planet systems are the same). Finally, the $M_i$ vs. $M_g - M_J$ plot has a KS test p-value of 0.005 for the compact multiple and no-planet populations, with 74\% on the metal-poor side of the linear fit average metallicity line as compared to 52\% of single planet systems.

We present p-values comparing our three star populations for each of these color-magnitude diagrams in Tables \ref{KS}, \ref{MW}, \ref{AD_stat} and \ref{AD_sig}. Table \ref{KS} displays the values resulting from a KS test. The low p-values for the compact multiple group relative to the 1-planet and no-planet groups indicate that these stars are likely not drawn from the same sample. The slight decrease in significance between the compact multiple-no-planet statistic and the compact multiple-1-planet statistic is primarily due to the diminished sample size when using the single planet systems instead of the zero transiting planet sample. The values for 1-planet and no planet populations are much higher; these stars are likely drawn from the same population. We find that metallicity is not strongly correlated with planet occurrence for single-planet systems. In Table \ref{MW}, we show the statistics for the Mann-Whitney U test, which also tests whether the populations are drawn from the same sample, but is more sensitive to differences in average value than the shape of the distributions. Again, the p-values comparing the compact multiples to the other two populations are much lower, indicating that the compact multiples prefer to form around more blue (metal-poor) stars on average. Finally, in Tables \ref{AD_stat} and \ref{AD_sig}, we show the same results using a K-sample Anderson-Darling test, which is more sensitive than our other tests to the tails of the distributions. We see that the statistic for the compact multiple/no-planet populations exceeds the significance level, but the opposite is true for the no-planet and single-planet populations. This indicates again that compact multiple systems are not drawn from the same distribution, with regard to metallicity as the single-planet and no-planet stars. All three of our statistical tests yield significant results.

% Maybe keep this in? 
%In $M_K$ vs. $G_{BP} - G_{RP}$ space, color is the metallicity sensitive variable, and therefore a horizontal color shift corresponds to a metallicity shift. We can use the Mann-Whitney U statistic in order to determine the offset between the compact multiple population and the field population. We find that the compact multiple systems are offset from the general population by $0.0568 \pm 0.0090$ in $G_{BP}$ - $G_{RP}$, a significance of $6\sigma$. 

\begin{table}%[H]
\begin{center}
\caption{Linear Fit KS Test p-values}
\label{KS} 
\begin{tabular}{crrr}
\tableline\tableline
Comparison & $M_i$ vs. $g - J$ & $M_K$ vs. $G_B - G_R$ & $M_G$ vs. $G_B - G_R$ \\
\tableline
0 and multi-planet & 0.005 & 0.002 & 0.026 \\
single and multi-planet & 0.018 & 0.003 & 0.100 \\
0 and single-planet & 0.191 & 0.519 & 0.623 \\
\tableline 
\tableline
\end{tabular}
\end{center}
\end{table}

\begin{table}%[H]
\begin{center}
\caption{Linear Fit Mann-Whitney U Test p-values}
\label{MW} 
\begin{tabular}{crrr}
\tableline\tableline
Comparison & $M_i$ vs. $g - J$ & $M_K$ vs. $G_B - G_R$ & $M_G$ vs. $G_B - G_R$ \\
\tableline
0 and multi-planet & 0.001 & 0.002 & 0.006 \\
single and multi-planet & 0.002 & 0.003 & 0.012 \\
0 and single-planet & 0.464 & 0.406 & 0.439 \\
\tableline 
\tableline
\end{tabular}
\end{center}
\end{table}

\begin{table}%[H]
\begin{center}
\caption{Linear Fit K-sample Anderson-Darling Test Statistic}
\label{AD_stat} 
\begin{tabular}{crrr}
\tableline\tableline
Comparison & $M_i$ vs. $g - J$ & $M_K$ vs. $G_B - G_R$ & $M_G$ vs. $G_B - G_R$ \\
\tableline
0 and multi-planet & 5.731 & 5.765 & 3.789 \\
single and multi-planet & 4.545 & 4.205 & 2.371 \\
0 and single-planet & 0.070 & 0.240 & -0.030 \\
\tableline 
\tableline
\end{tabular}
\end{center}
\end{table}

\begin{table}%[H]
\begin{center}
\caption{Linear Fit K-sample Anderson-Darling Test Significance Level}
\label{AD_sig} 
\begin{tabular}{crrr}
\tableline\tableline
Comparison & $M_i$ vs. $g - J$ & $M_K$ vs. $G_B - G_R$ & $M_G$ vs. $G_B - G_R$ \\
\tableline
0 and multi-planet & 0.002 & 0.002 & 0.010 \\
single and multi-planet & 0.005 & 0.007 & 0.033 \\
0 and single-planet & 0.250 & 0.250 & 0.250 \\
\tableline 
\tableline
\end{tabular}
\end{center}
\end{table}

While we have shown that a linear fit to the main sequence is anchored to existing spectroscopic metallicity measurements and serves as a useful approximation, this method has significant drawbacks. First, the relatively small number of systems at the reddest colors only minimally influence the placement of this line and are therefore poorly fit. However, there are also very few planetary systems detected by Kepler in this region of the main sequence so this is not significant for our results. Additionally, at the bluest colors of the color-magnitude diagram, the main sequence bifurcates. While it is possible that this is a binary sequence, it disappears at $G_{BP} - G_{RP}$ $>$ 2, which we would not expect. There are confirmed planetary systems around some of these stars, for which the detailed characterization and confirmation process did not yield a positive detection of an additional star. These stars are similarly not well fit by our main sequence line. 

We used the linear fit to validate our metallicity proxy based on spectroscopic metallicities from the two Muirhead papers \citep{Muirhead2012, Muirhead2014}, but for all following analysis we use a more complex interpolated main sequence fit.

\subsection{B-Spline Interpolation}
\label{binary_sequence_section}

We perform the same analysis using distance from the main sequence as a proxy for relative metallicity, but with a more complex interpolated fit to represent the main sequence in the form of a fourth degree B-spline interpolation. We account for the possibility of unresolved binaries with a two-step fitting process to eliminate the visible binary sequence above our primary main sequence.

Binary systems in our stellar sample would be both be location-shifted in a color-magnitude plot and have diluted planet transits, potentially diminishing our ability to detect compact multiple systems. Additionally, the presence of a close stellar companion is known to reduce the planet occurrence rate of these stars and may also have significant influences on the final architecture of the system \citep{Krauss_binaries}. Binarity rate decreases as the stellar mass of the primary star decreases, from approximately 50\% for solar type stars, to 30\% for nearby M-dwarfs, and is lowest for all other stellar types \citep{Winters2019, Duchene2013}. Despite planet occurrence and detectability disfavoring binary systems and the lower binarity rate for these stars, we still find it a potential issue large enough to merit investigation. We examined the distributions of the Gaia Goodness of Fit metric, shown to correlate with binarity, in Section \ref{GOF}, but adjusted our spline fit to account for binaries as well.

Equal mass binary systems lie 0.756 magnitudes brighter than single star systems, with non-equal mass binaries lying at slightly dimmer absolute magnitudes.\footnote{The value of 0.756 magnitudes brighter for an equal magnitude binary is derived from the definition of the magnitude system. An equal mass binary would increase the flux by a factor of 2, which corresponds to a shift of 0.756 magnitudes. This shift occurs in a single photometric band, which we account for by using the vertical shift as a cutoff for stars to exclude as potential unknown binary systems.} In order to remove the possibility of binary influence due to unknown, unresolved binary systems, we performed a B-spline interpolation on the full sample and eliminated all stars (planet hosts and non-planet hosts) more than 0.6 magnitudes brighter than the initial spline fit. We did another B-spline interpolation on the remaining sample and used that line as our average-metallicity main sequence approximation. There is a relatively clear visual distinction between the main sequence and upper main sequence as well. The upper sequence includes 776 stars in the $M_i$ vs. $g - J$ case, 665 stars in the $M_K$ vs. $G_B - G_R$ case, and 622 stars in the $M_G$ vs. $G_B - G_R$ case. We eliminated the systems in the upper branch main sequence from our analysis with the exception of 13 well-characterized systems with no stellar companion \citep{Furlan2017, Ziegler2018}.

\begin{figure}
\centering
\includegraphics[scale=0.43]{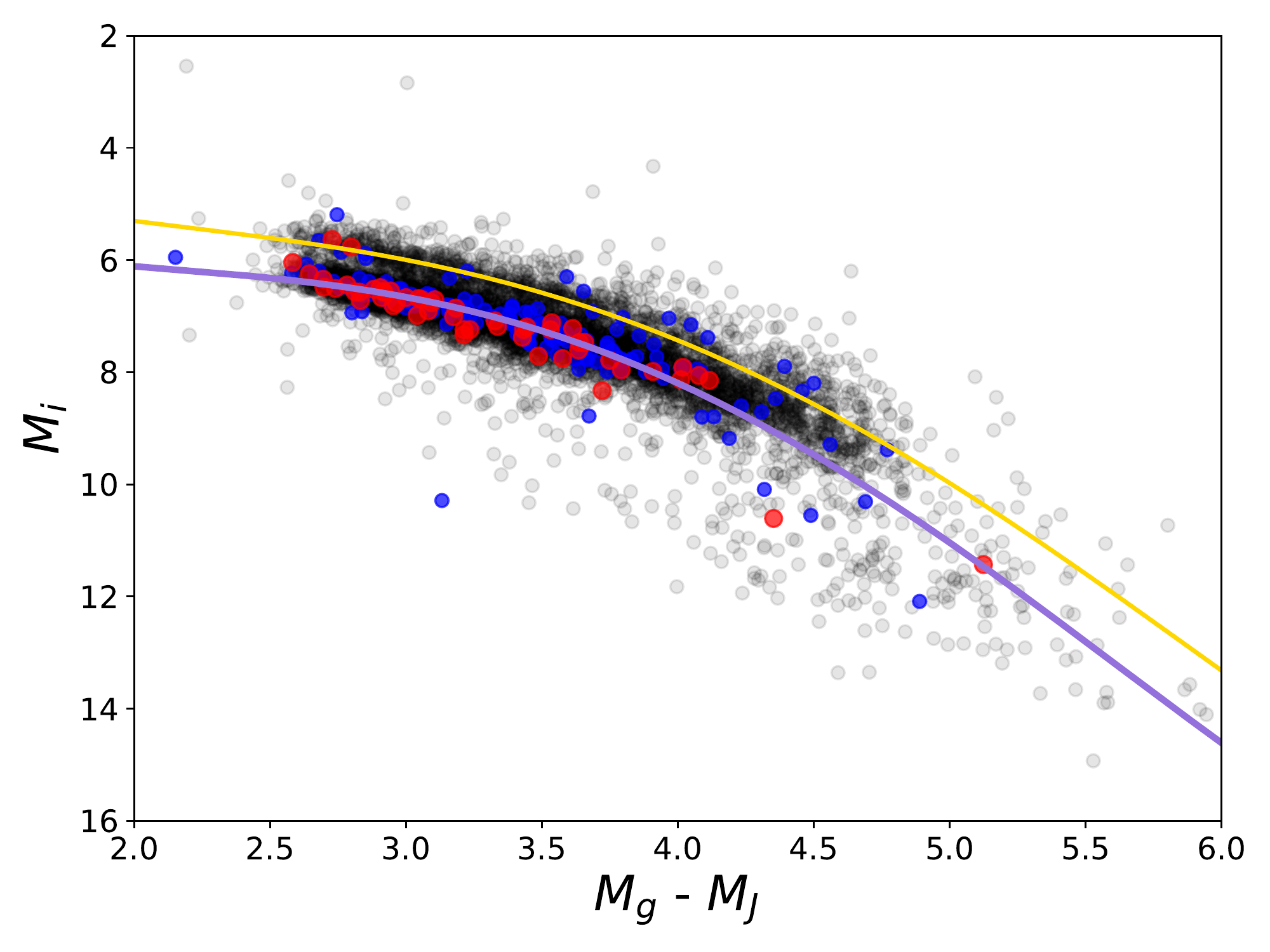}
\includegraphics[scale=0.43]{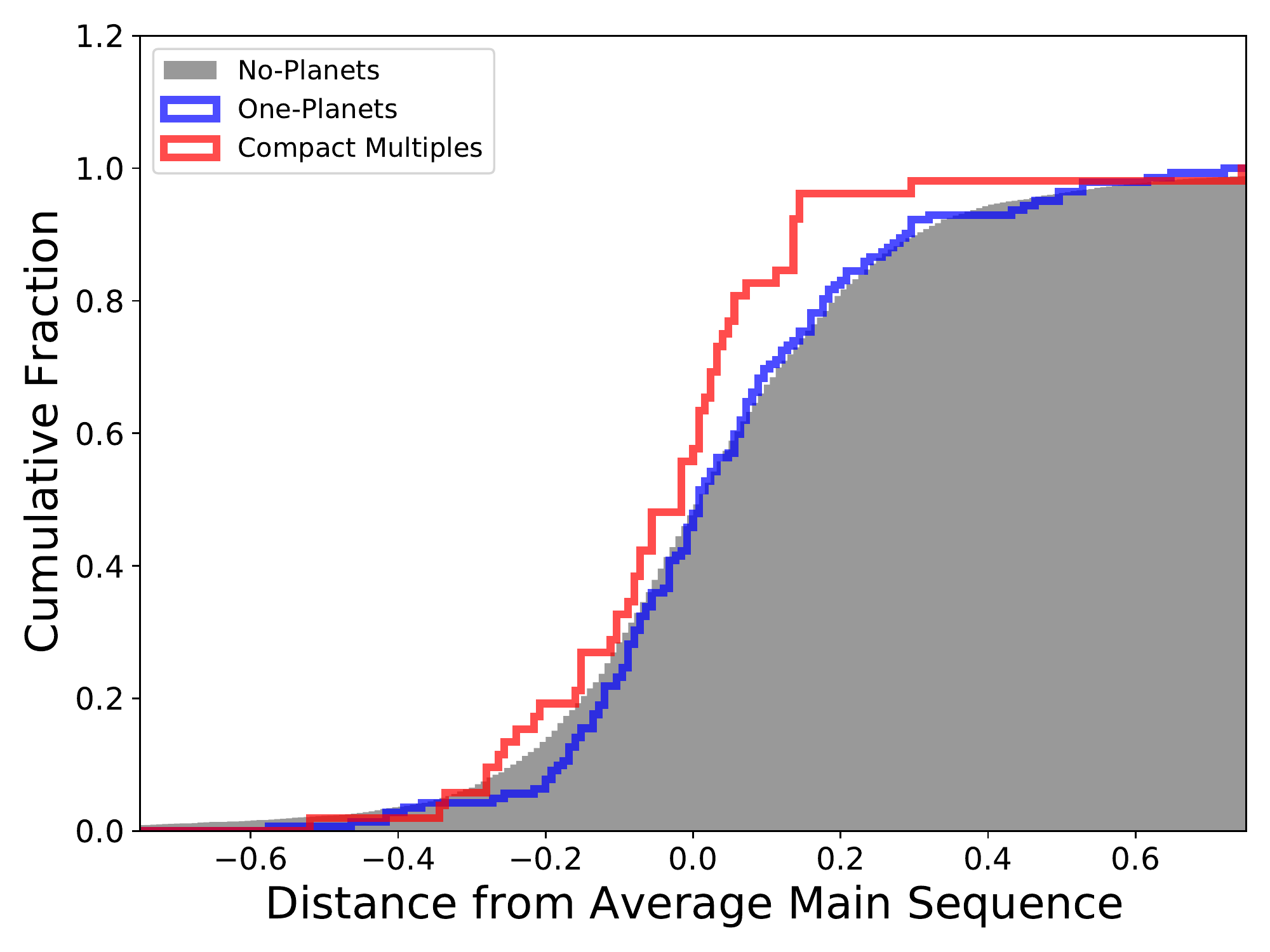}
\caption{Color-magnitude diagram and histogram in the $M_i$ vs. $g - J$ bands. In the color-magnitude diagram on the left, black points represent stars with zero planets, blue points represent stars with a single planet, red points represent compact multiple systems with more than one planet, and the interpolated B-spline fit representing the average metallicity main sequence is shown in purple. The yellow line shows the threshold between the main sequence and the upper sequence, and with the exception of a few stars (citations from \citet{Furlan2017} and \citet{Ziegler2018}), upper sequence stars were eliminated from our analysis. The right panel figure is a cumulative histogram of distance from the main sequence, with red showing compact multiples, blue showing single planet systems, and grey showing the no-planet population. The color-magnitude diagram and histogram both show compact multiples preferentially on the metal-poor, negative distance, side of the main sequence. This is further validated with our three statistical tests, with values listed in Tables \ref{KS_S}, \ref{MW_S}, \ref{AD_stat_S}, and \ref{AD_sig_S}. The KS test p-value for the compact multiple systems as compared to the no-planet systems is 0.015, and the equivalent MW test value is 0.023.}
\label{igJ_Spline}
\end{figure}

\begin{figure}
\centering
\includegraphics[scale=0.43]{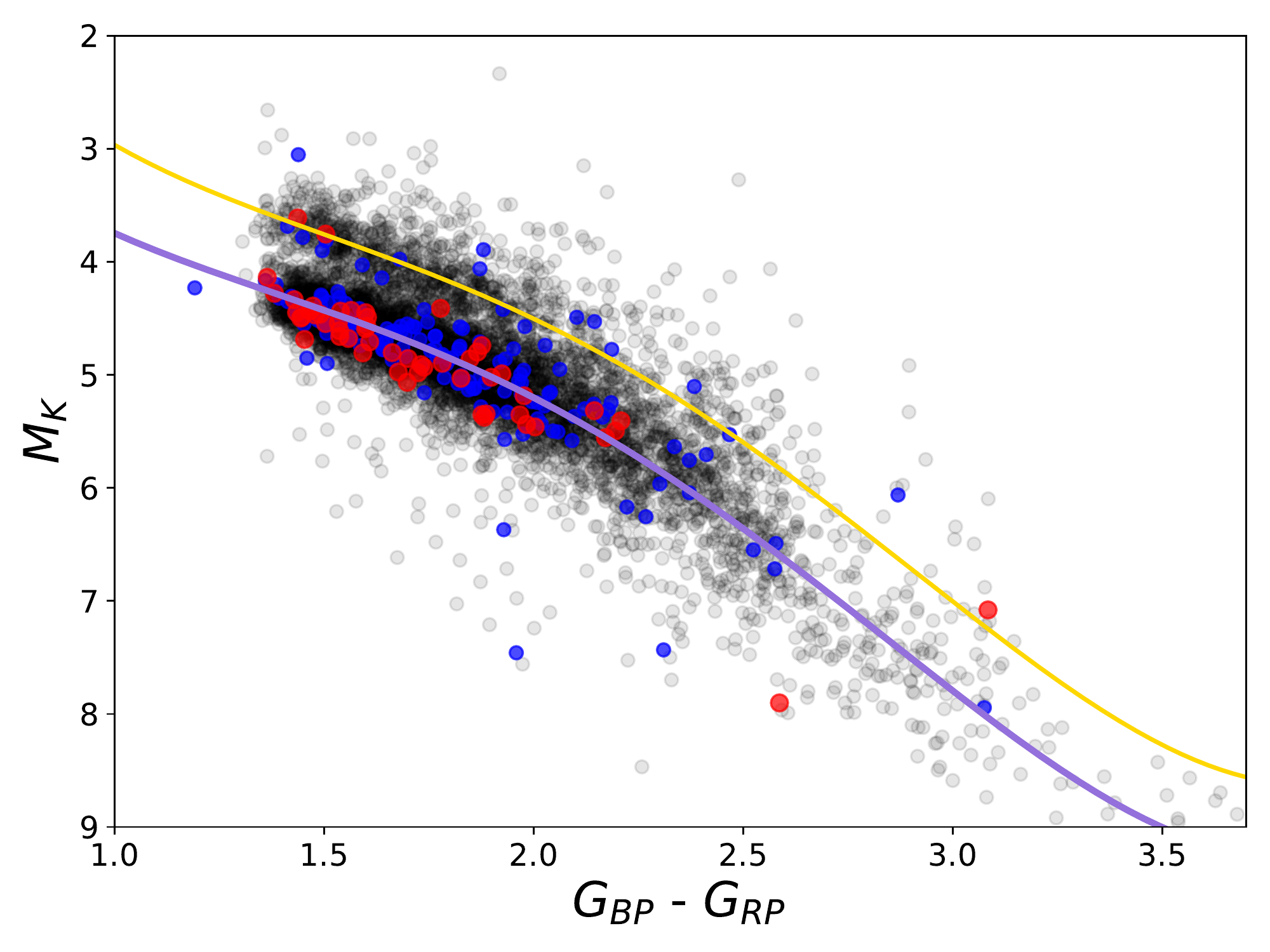}
\includegraphics[scale=0.43]{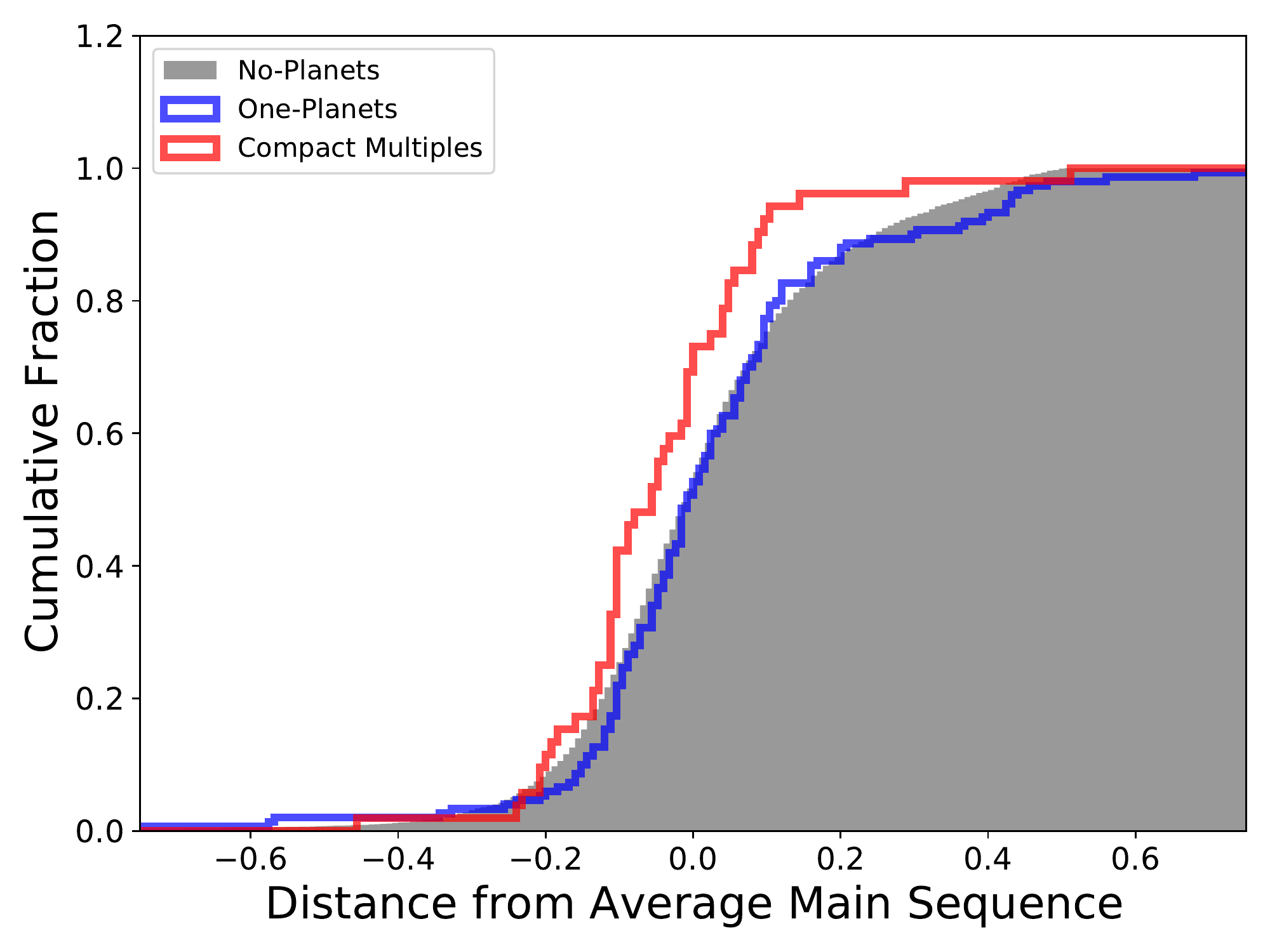}
\caption{Color-magnitude diagram (left) and histogram (right) in the $M_K$ vs. $G_B - G_R$ bands. In both plots, red represents compact multiple systems, blue represents single-planet systems, and black/grey represents no-planet systems. In the color-magnitude diagram, the interpolated B-spline fit is shown in purple and the cutoff between the main sequence and the upper sequence is shown in yellow. The cumulative histogram shows the distribution of distance from the main sequence line for each of the three populations. A larger proportion of compact multiples occur on the metal-poor side of the average metallicity main sequence than either of the other two populations. The statistical values for the KS, MW, and AD tests are listed in Tables \ref{KS_S}, \ref{MW_S}, \ref{AD_stat_S}, and \ref{AD_sig_S}. The KS test p-value for the compact multiple systems as compared to the no-planet systems is 0.016, and the equivalent MW test value is 0.005.}
\label{KBR_Spline}
\end{figure}

\begin{figure}
\centering
\includegraphics[scale=0.43]{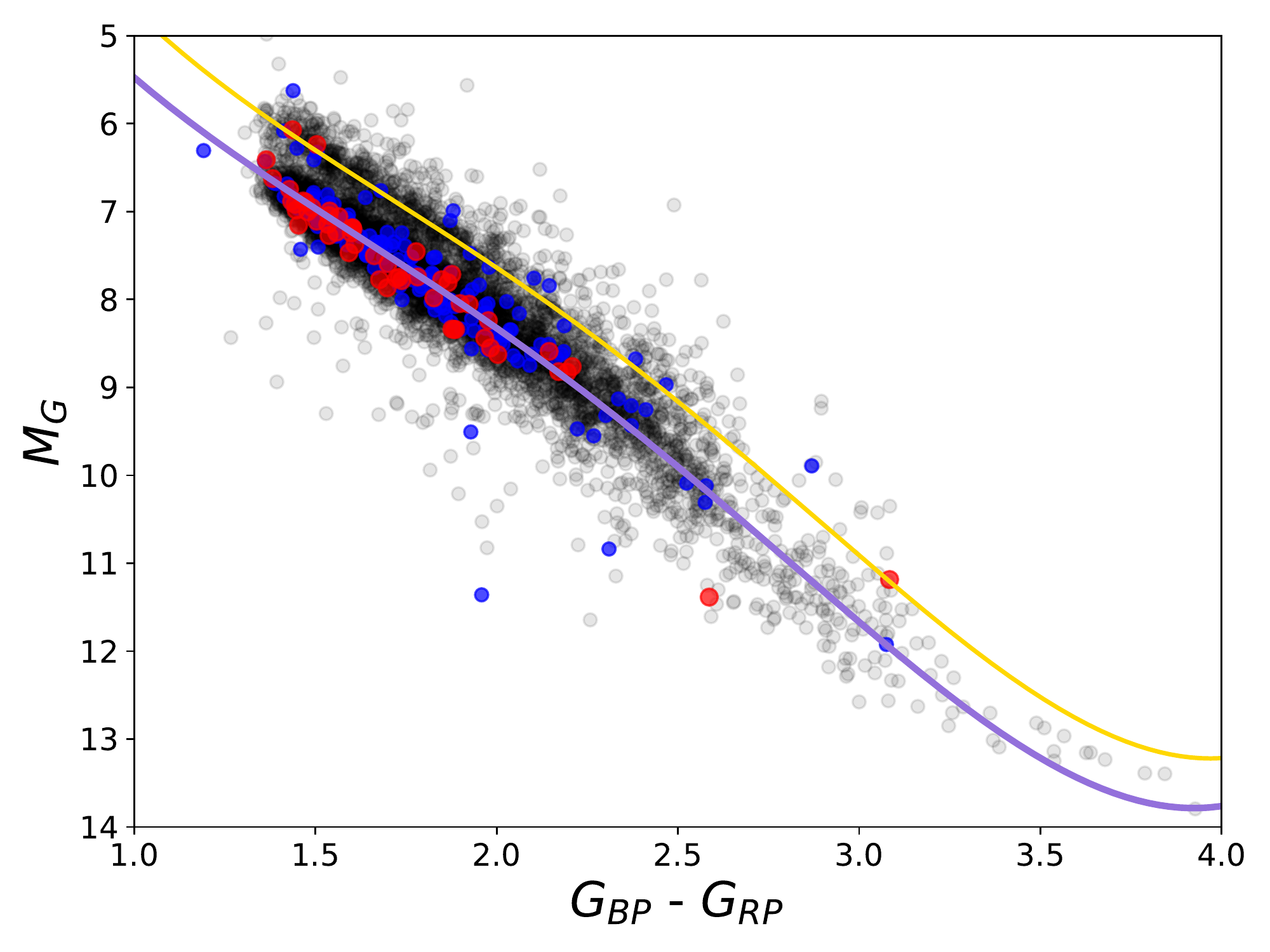}
\includegraphics[scale=0.43]{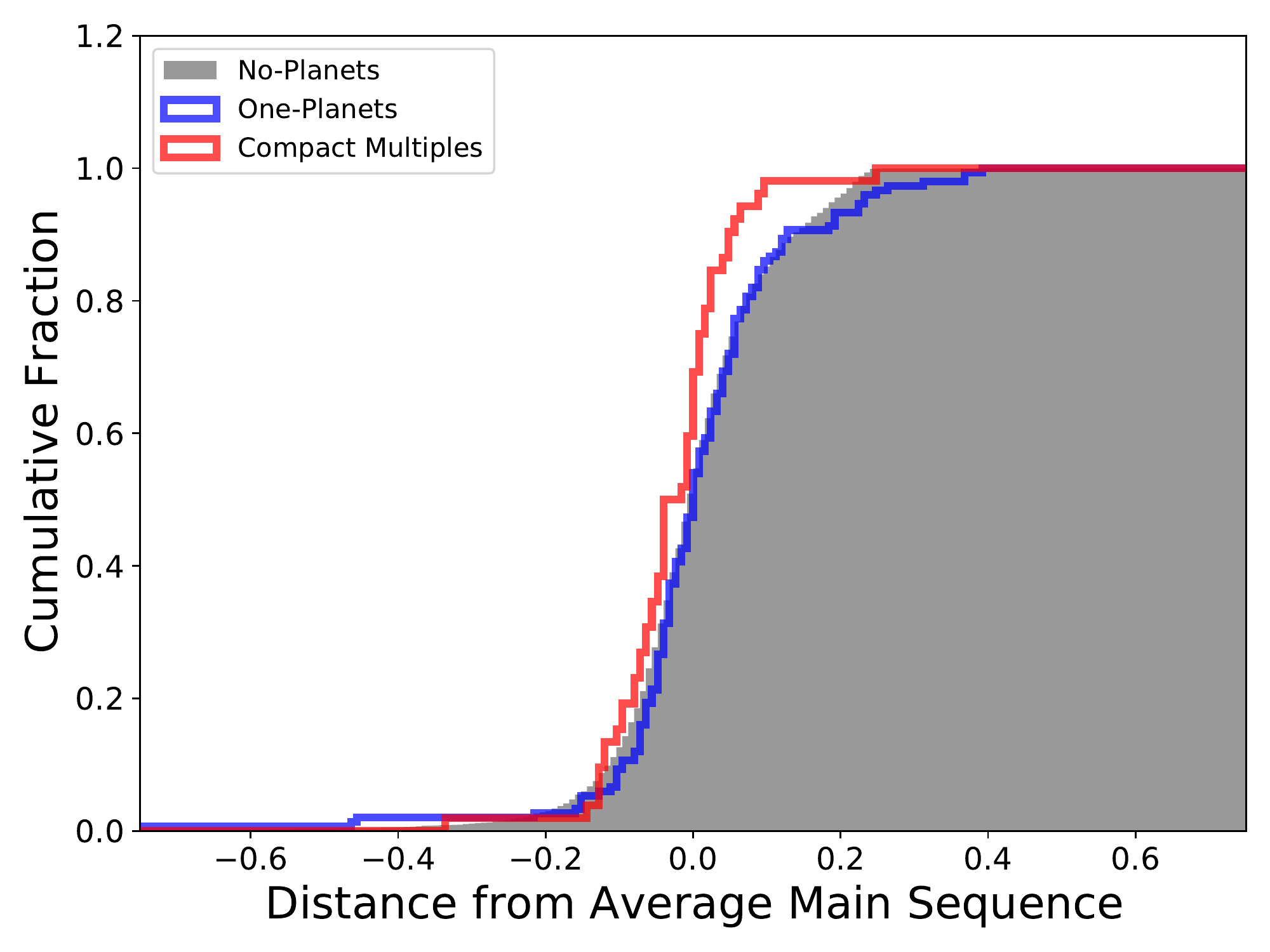}
\caption{The left plot shows the color-magnitude diagram for the $M_G$ vs. $G_B - G_R$ bands, with the interpolated B-spline fit shown in purple, and the cutoff between the main sequence and upper sequence shown in yellow. Red points represent compact multiples, blue points represent single-planet systems, and black points represent compact multiples. In the cumulative histogram on the right, the same colors correspond to the distribution of distance from the main sequence line for compact multiples, single-planet systems, and no-planet systems. Compact multiples prefer metal poor stars at a significant level as determined by the KS, MW, and AD test (values listed in Tables \ref{KS_S}, \ref{MW_S}, \ref{AD_stat_S}, and \ref{AD_sig_S}). The KS test p-value for the compact multiple systems as compared to the no-planet systems is 0.038, and the equivalent MW test value is 0.016.}
\label{GBR_Spline}
\end{figure}

In Figures \ref{igJ_Spline}, \ref{KBR_Spline}, and \ref{GBR_Spline}, we show the color-magnitude diagrams and histograms for these stars (similar to Figures \ref{linear_fit_color-mag_plots} and \ref{linear_fit_histograms}), including the interpolated main sequence used for this analysis shown in purple on the color-magnitude diagrams. The upper yellow line is 0.6 magnitudes brighter than (above) the purple line, and represents the cutoff between the main sequence and separate upper main sequence stars that were not used in the analysis. The cumulative histograms show horizontal distance between each star and the average-metallicity main sequence.

The KS test and MW test p-values, and AD test statistic and significance levels characterizing the differences between the compact multiple population, single-planet population, and no-planet population using the B-spline main sequence are listed below in Tables \ref{KS_S}, \ref{MW_S}, \ref{AD_stat_S} and \ref{AD_sig_S}. We see the same trends as with our previous analysis. Compact multiples are likely drawn from a different, more metal-poor metallicity distribution than the single-planet systems or broader stellar sample. This can be seen in the low KS and MW test p-values for the compact multiple population as compared to the no-planet population, in all three color bands. The AD test statistic exceeds the AD test significance level in all three bands for the compact multiple/no-planet case. The AD test result suggests that compact multiple population is statistically distinct in metallicity from the 1-planet and no-planet populations, but the latter two are not distinct from each other.

Therefore, we conclude that the compact multiple systems prefer to form (or are more likely to survive to the field age) around low metallicity stars, while the single transiting planet systems are indistinguishable from the field population of stars.

\begin{table}%[H]
\begin{center}
\caption{KS Test p-values - Spline}
\label{KS_S} 
\begin{tabular}{crrr}
\tableline\tableline
Comparison & $M_i$ vs. $g - J$ & $M_K$ vs. $G_B - G_R$ & $M_G$ vs. $G_B - G_R$ \\
\tableline
0 and multi-planet & 0.015 & 0.016 & 0.038 \\
single and multi-planet & 0.041 & 0.034 & 0.019 \\
0 and single-planet & 0.416 & 0.378 & 0.443 \\
\tableline 
\tableline
\end{tabular}
\end{center}
\end{table}

\begin{table}%[H]
\begin{center}
\caption{Mann-Whitney U Test p-values - Spline}
\label{MW_S} 
\begin{tabular}{crrr}
\tableline\tableline
Comparison & $M_i$ vs. $g - J$ & $M_K$ vs. $G_B - G_R$ & $M_G$ vs. $G_B - G_R$ \\
\tableline
0 and multi-planet & 0.023 & 0.005 & 0.016 \\
single and multi-planet & 0.013 & 0.002 & 0.006 \\
0 and single-planet & 0.190 & 0.161 & 0.132 \\
\tableline 
\tableline
\end{tabular}
\end{center}
\end{table}

\begin{table}%[H]
\begin{center}
\caption{K-sample Anderson-Darling Test Statistic - Spline}
\label{AD_stat_S} 
\begin{tabular}{crrr}
\tableline\tableline
Comparison & $M_i$ vs. $g - J$ & $M_K$ vs. $G_B - G_R$ & $M_G$ vs. $G_B - G_R$ \\
\tableline
0 and multi-planet & 2.500 & 3.716 & 2.574 \\
single and multi-planet & 2.652 & 4.545 & 3.455 \\
0 and single-planet & 0.025 & 0.530 & 1.229 \\
\tableline 
\tableline
\end{tabular}
\end{center}
\end{table}

\begin{table}%[H]
\begin{center}
\caption{K-sample Anderson-Darling Test Significance Level - Spline}
\label{AD_sig_S} 
\begin{tabular}{crrr}
\tableline\tableline
Comparison & $M_i$ vs. $g - J$ & $M_K$ vs. $G_B - G_R$ & $M_G$ vs. $G_B - G_R$ \\
\tableline
0 and multi-planet & 0.031 & 0.010 & 0.029 \\
single and multi-planet & 0.027 & 0.005 & 0.013 \\
0 and single-planet & 0.250 & 0.200 & 0.101 \\
\tableline 
\tableline
\end{tabular}
\end{center}
\end{table}

In the following subsections, we address complicating factors that could affect our interpretation of this observed trend. We describe the analysis we have done to confirm that these effects are not likely to have significantly affected our results.

\subsection{Gaia Goodness of Fit}
\label{GOF}
Although our B-spline fitting procedure mitigates the effect of possible binary systems, we attempt to determine the extent to which our results are still possibly affected by stellar binaries using the distribution of the Gaia database ``Astrometric Goodness of Fit" metric \citep{Gaia2018}. This metric represents the degree to which each star's position and proper motion fits within global model of the entire Gaia dataset and has been shown to correlate with the probability that a star is an unresolved binary system \citep{Evans2018}. \citet{Evans2018} suggested a cutoff of GOF = 20 to indicate an unresolved binary system, and as all of our stars are well under that value, we do not expect unresolved binaries to be a significant issue. However, we proceed with our analysis recognizing that the Goodness of Fit metric and cutoff do not eliminate the possibility of binary systems in our stellar sample, especially for faraway stars.

\begin{figure}
\centering
\includegraphics[scale=0.7]{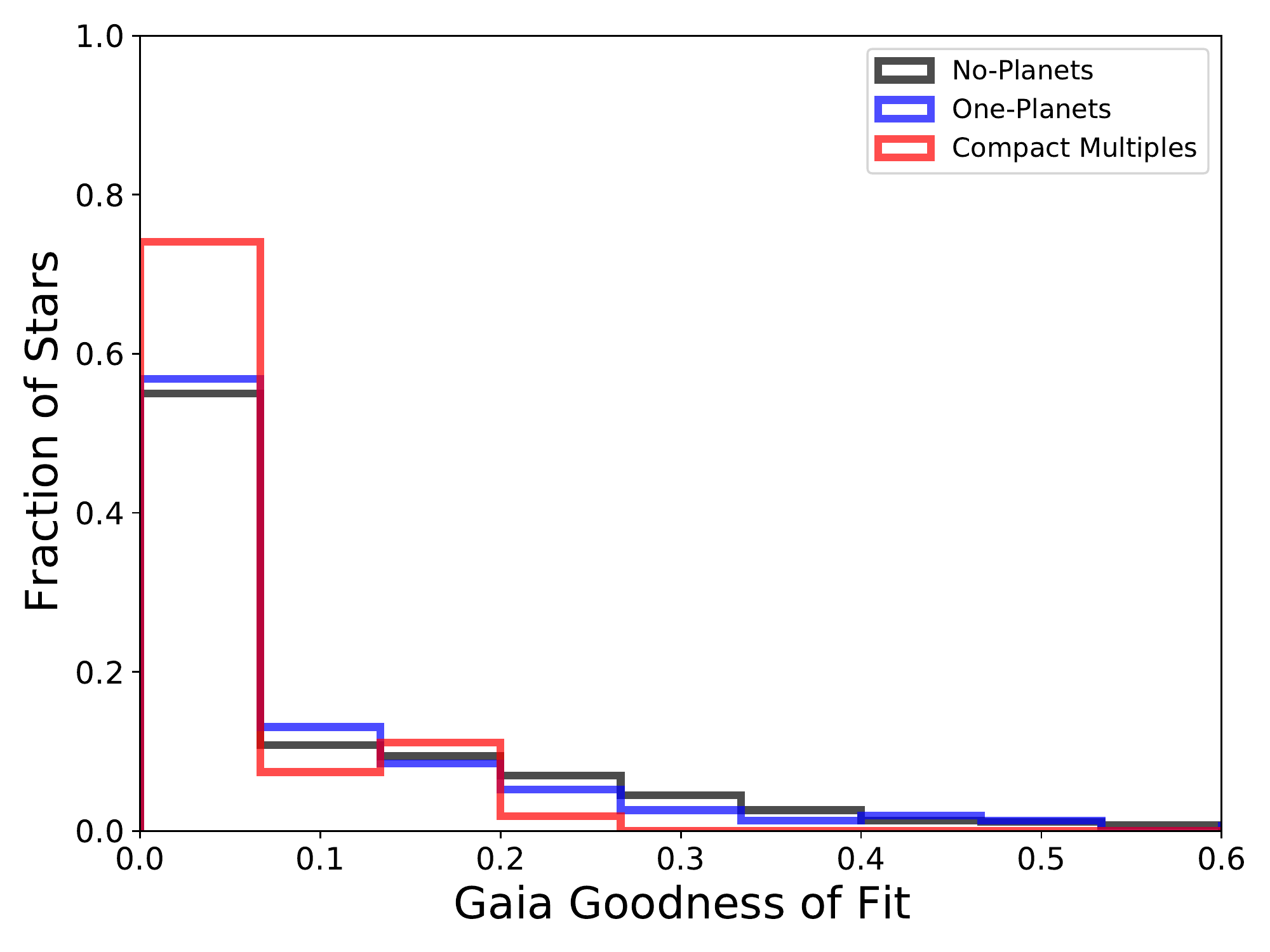} \\
\caption{Histogram using the Gaia Astrometric Goodness of Fit metric, which has been shown to correlate with unresolved binaries. Compact multiples are shown in red, single-planet stars in blue, and stars without detected transiting planets in black. All three populations show a similar distribution, and include very few stars with high Goodness of Fit values where we expect unresolved binary stars to appear. Therefore, we do not believe that these populations consist of stars with different binarity fractions.}
\label{goodness_of_fit}
\end{figure}

We show the distribution of the Gaia Goodness of Fit for each of our star populations in Figure \ref{goodness_of_fit}. We find that the shapes of these distributions are similar, particularly the tail of the distribution more likely to represent possible binary star systems. The Gaia data indicate that these stellar populations are not preferentially drawn from astrometrically-distinct populations or significantly contaminated by unresolved binaries.

\subsection{Possibility of Multiple Cross Matches}
\label{multiple_crossmatches_section}
We use the publicly available 1" Gaia-Kepler cross match table created by coauthor MB. Multiple cross matches for the same star could cause ambiguity in our data and error in our results. We ran through all of the M and late K-dwarf stars in our sample and checked each for multiple matches in the 1" table. We found only 49 stars with more than one match, less than 1.5\% of our sample. The Gaia data is precise enough that we did not consider using a larger radius table for our analysis. In summary, the fraction of stars with multiple cross matches in the 1" table is small enough that it any ambiguity does not significantly affect our results. Additionally, mismatches between Gaia DR2, 2MASS, and the KIC would likely place stars far from the main sequence. The fact that we robustly identify the low-mass stellar main sequence likely means that there are not a significant number of mismatches between these catalogs.

\subsection{Orbital Period Analysis}
We have both compact multiples and single-planet stars in our sample of planetary systems. If those groups had drastically different orbital periods, it could suggest a difference in formation mechanism. In order to confirm that we are comparing two samples derived from the same population with respect to orbital period (and possible formation mechanism), we eliminated all single planet stars with a longer period than that of the longest-period compact multiple and repeated our analysis. We eliminated 17 stars in total, and their removal yielded no change in the significance of our results. The largest change occurred in the statistical values comparing single-planet systems to zero-planet systems, which increased by an average of 0.16 in p-value for the KS test, indicating that the single-planet and no-planet stars are drawn from the same population. The effect of orbital period is a parameter that might be worth exploring in depth in future studies.

\subsection{The Effect of Differential Dust Extinction within the Kepler Stellar Sample}
\begin{figure}
\centering
\includegraphics[scale=0.7]{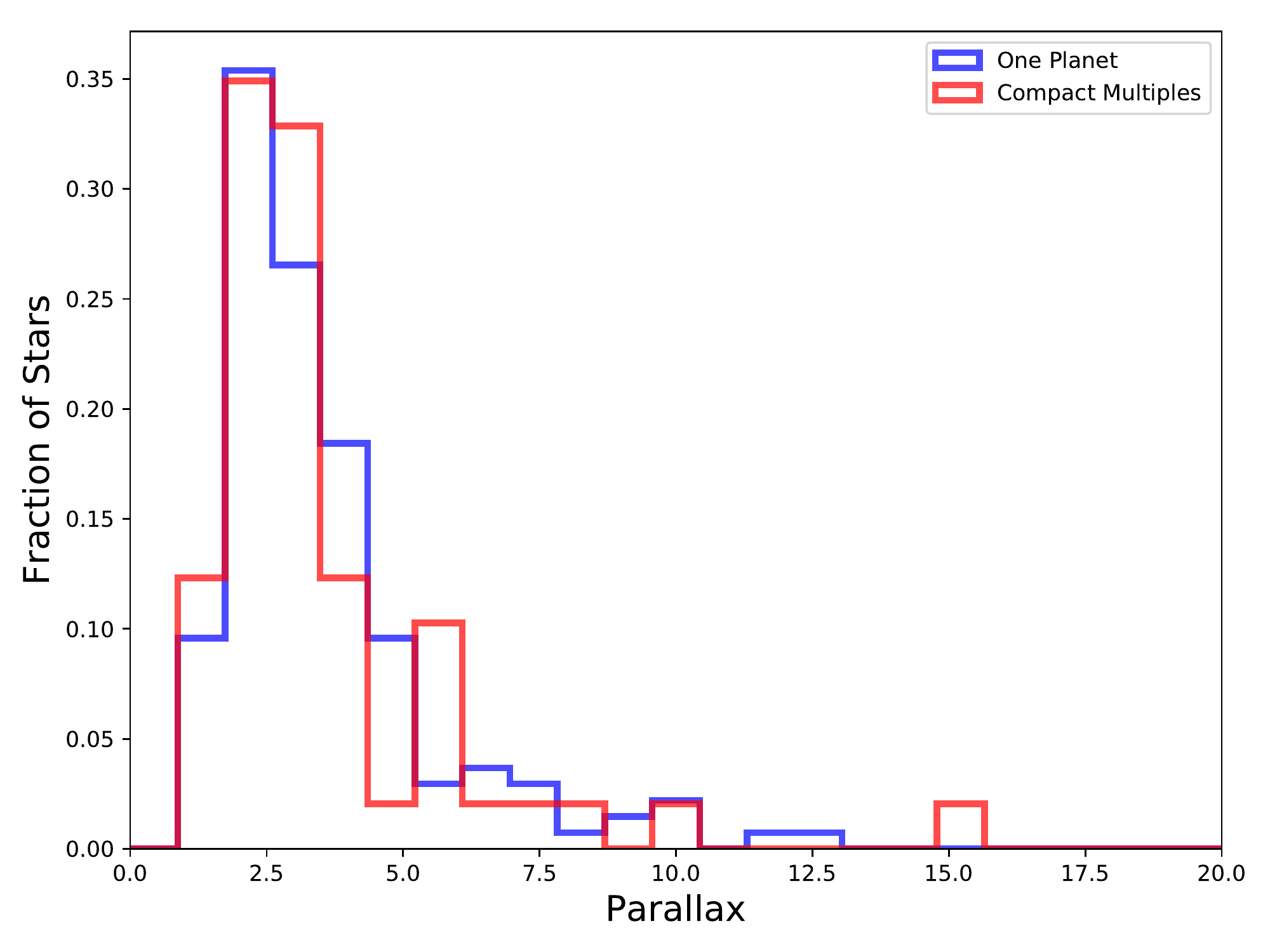}\\
\caption{Histogram showing the parallax distributions of the single and multi-planet populations, as a proxy for distance distribution. Since the two populations are appear similar in distribution (KS test p-value of 0.86), it is unlikely that differential reddening effects between stars in our compact multiple and single planet systems are responsible for this effect.}
\label{parallax}
\end{figure}

We have yet to consider interstellar extinction as a possible source of bias. Blue light is preferentially extincted by dust, and as a result stars that are further away could appear redder, independent of metallicity. Since we use color as a metallicity proxy in this analysis, dust interference could bias our results. We investigated this possible effect by comparing the distance distributions of our single and multi-planet stellar samples. If the single transiting planet and compact multiple systems were drawn from different distance distribution, they would be offset from each other in a main sequence diagram due to interstellar dust. In Figure \ref{parallax}, we show a histogram of the parallax distributions for the two populations. We performed a standard KS test, and the p-value was 0.86 indicating that the two populations are drawn from the same underlying sample. It is therefore unlikely that the radial dust distribution along the line of sight is responsible for our results. 

\begin{figure}
\centering
\includegraphics[scale=0.43]{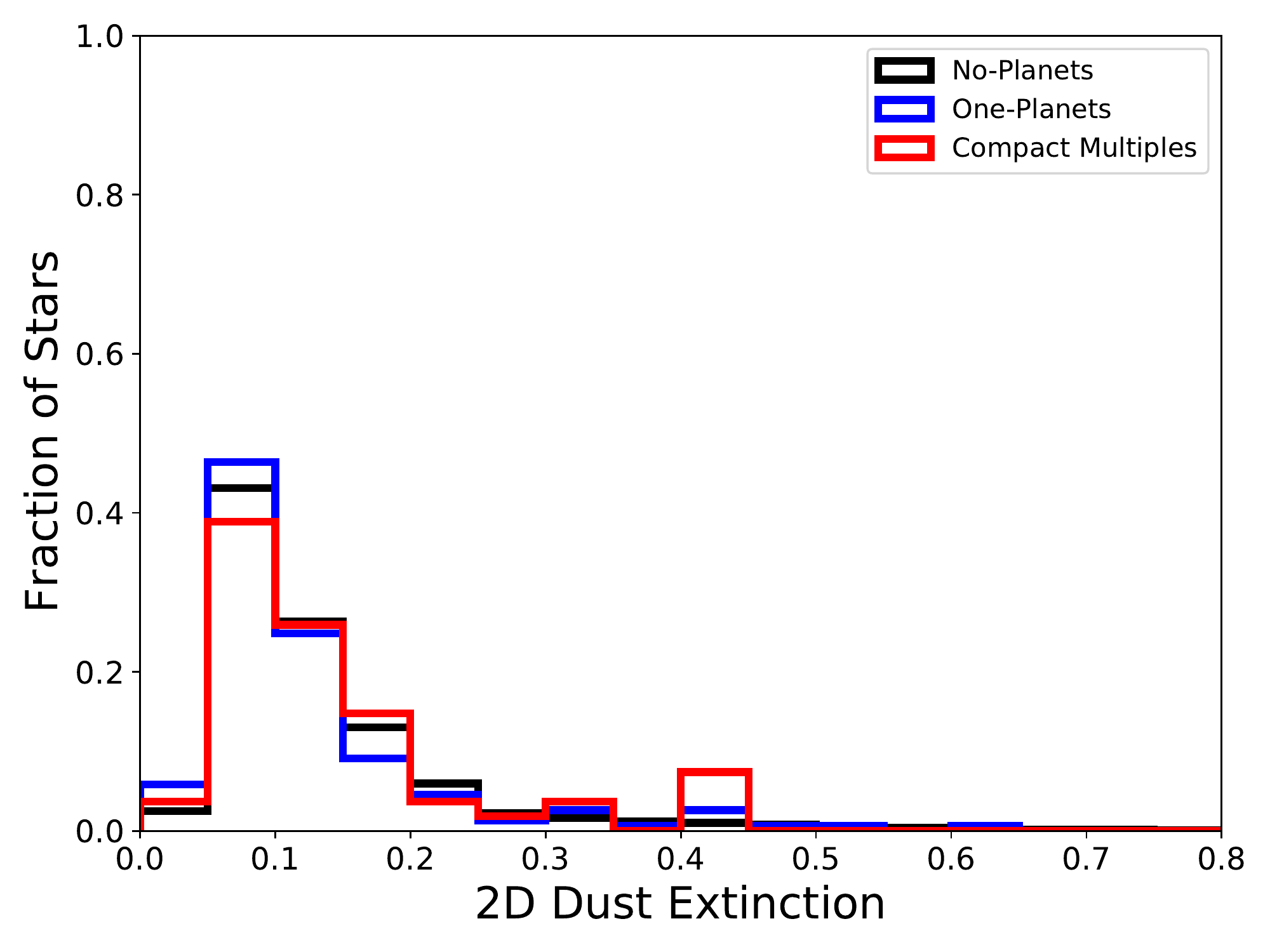}
\includegraphics[scale=0.43]{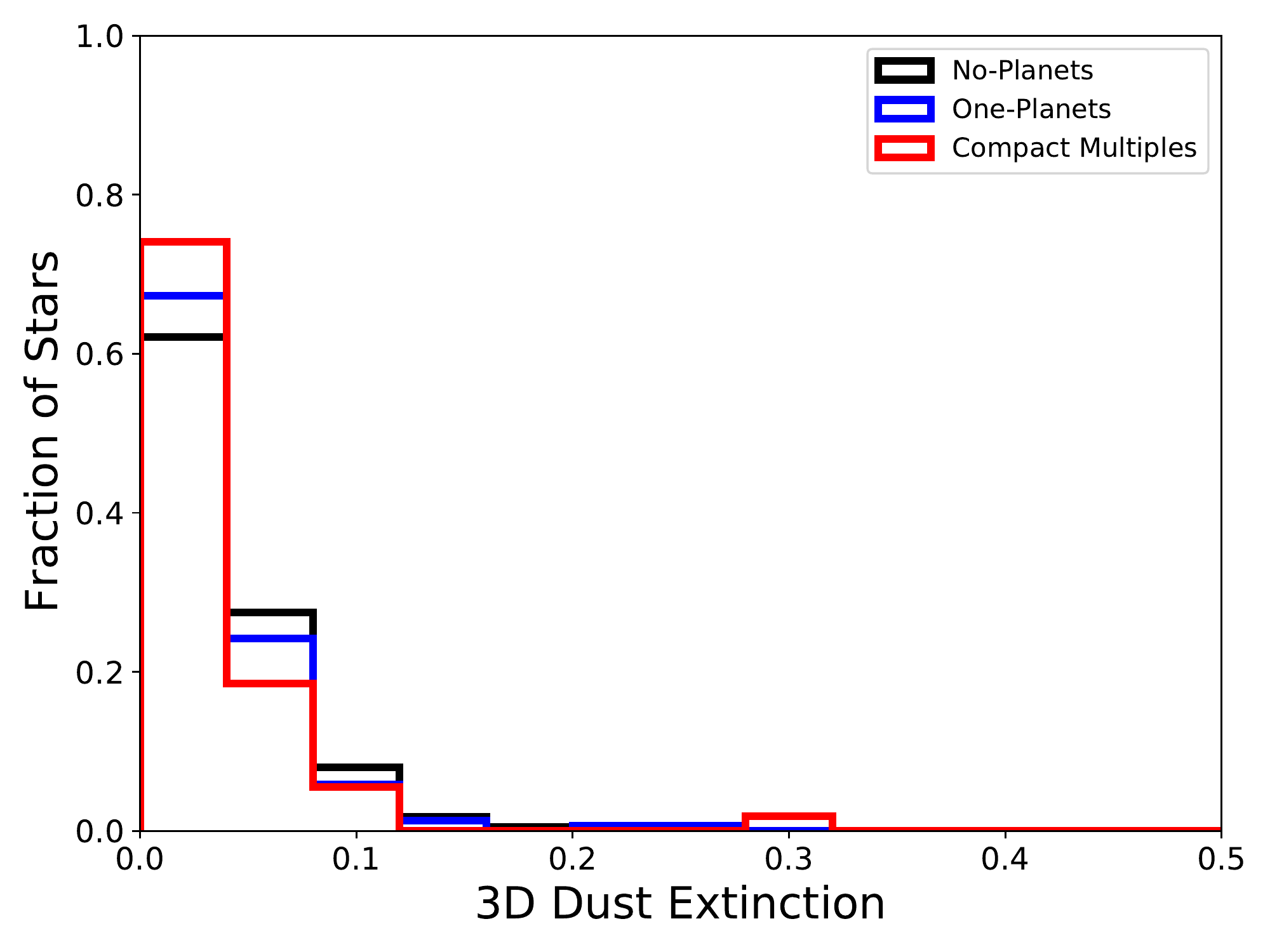}
\caption{Left, cumulative histogram of the 2D dust extinction from compact multiples (red), single planet systems (blue), and the field population (black), using data from the \citet{Schlegel1998} ``SFD" dust reddening map. We find that these populations of stars are impacted by dust in similar manners and therefore it is unlikely that differential dust extinction is responsible for the differences between compact multiple and single-planet systems. Right, identical but using the 3D dust distribution from the ``Bayestar" dust map \citep{Green2019}.}
\label{dust}
\end{figure}

The distribution of dust in the galaxy is non-uniform, so a comparison of the distance distributions is not sufficient to eliminate the possibility of extinction affecting our results. We plotted histograms of the 2D and 3D dust extinction distributions for each population using the \textit{dustmaps} Python package \citep{Green2018} and data from \citet{Schlegel1998} and \citet{Green2019}. These plots are shown in Figure \ref{dust}, and there is no statistically significant difference between the populations. The KS test p-value for the compact multiple and single planet systems is 0.336 in the 2D case and 0.839 in the 3D case.

The 2D dust map estimates the total amount of dust present along the line of sight, and we show the histogram of the total dust extinction from the 2D dust map for our stellar sample in Figure \ref{dust}, left panel. However, while the Kepler mission looked towards the bulge of the galaxy, most of the Kepler stellar sample lie in the foreground (a 14th magnitude Kepler star is relatively dim). Therefore, this 2D dust map overestimates the total extinction. Utilizing 3-dimensional dust maps can give us a better measurement of the actual extinction to our target stars. In Figure \ref{dust}, right, we show the total extinction to our sample stars using the 3D extinction calculated using the Gaia distances. In both the 2D and 3D cases we find no significant differences in our stellar samples, and conclude that differential dust extinction across the Kepler field of view is unlikely to be the origin of our results.

\subsection{The Effect of Detection Efficiency}
We have yet to consider detection efficiency, and it is especially relevant considering recent results from \cite{Zink19} that introduce detection efficiency as a possible cause for the Kepler dichotomy. Detection efficiency decreases for higher detection order planets as transit detection likelihood drops after the first detection. We account for any possible difference in detection efficiency in our sample in the following section and plots.
The NASA Exoplanet Archive includes two statistics that are relevant to planet detection efficiency: the maximum multiple event statistic and signal-to-noise. We created simple histograms showing the distribution of each over our sample of single planet and compact multiple systems separately. A similar distribution for the two populations would indicate that detection efficiency is not playing a large role in our results. We find the shapes of the distributions to be similar, with KS test p-values of 0.272 for the MES and 0.477 for the SNR. Both of these plots are included below in Figures \ref{MES} and \ref{SNR}.
\begin{figure}
\centering
\includegraphics[scale=0.7]{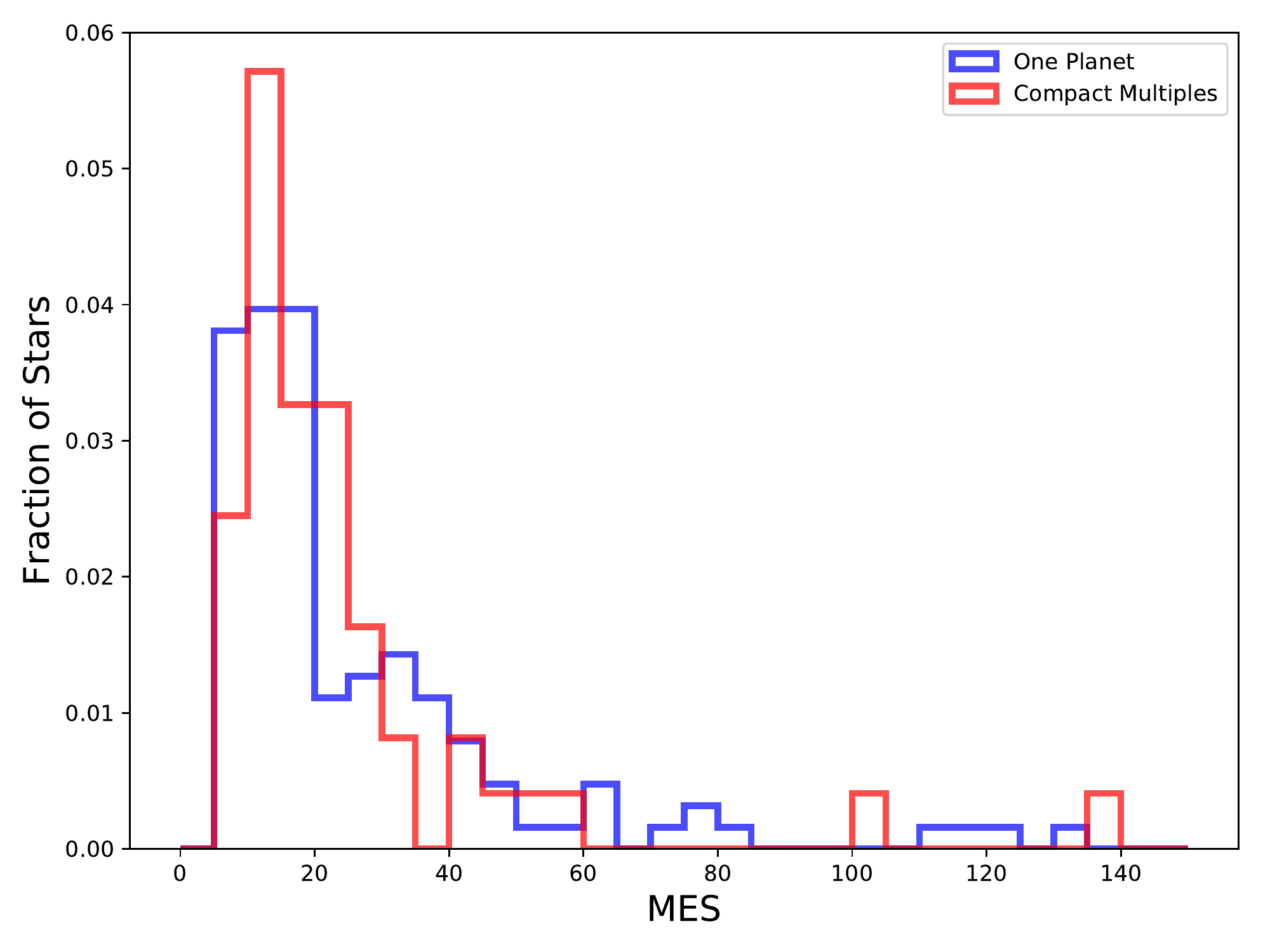} \\
\caption{Histogram showing the MES distributions for the single planet systems and compact multiple systems. The two are similar, with a KS test p-value of 0.272, indicating that MES, or detection efficiency, does not play a significant role in the results that we observe.}
\label{MES}
\end{figure}

\begin{figure}
\centering
\includegraphics[scale=0.7]{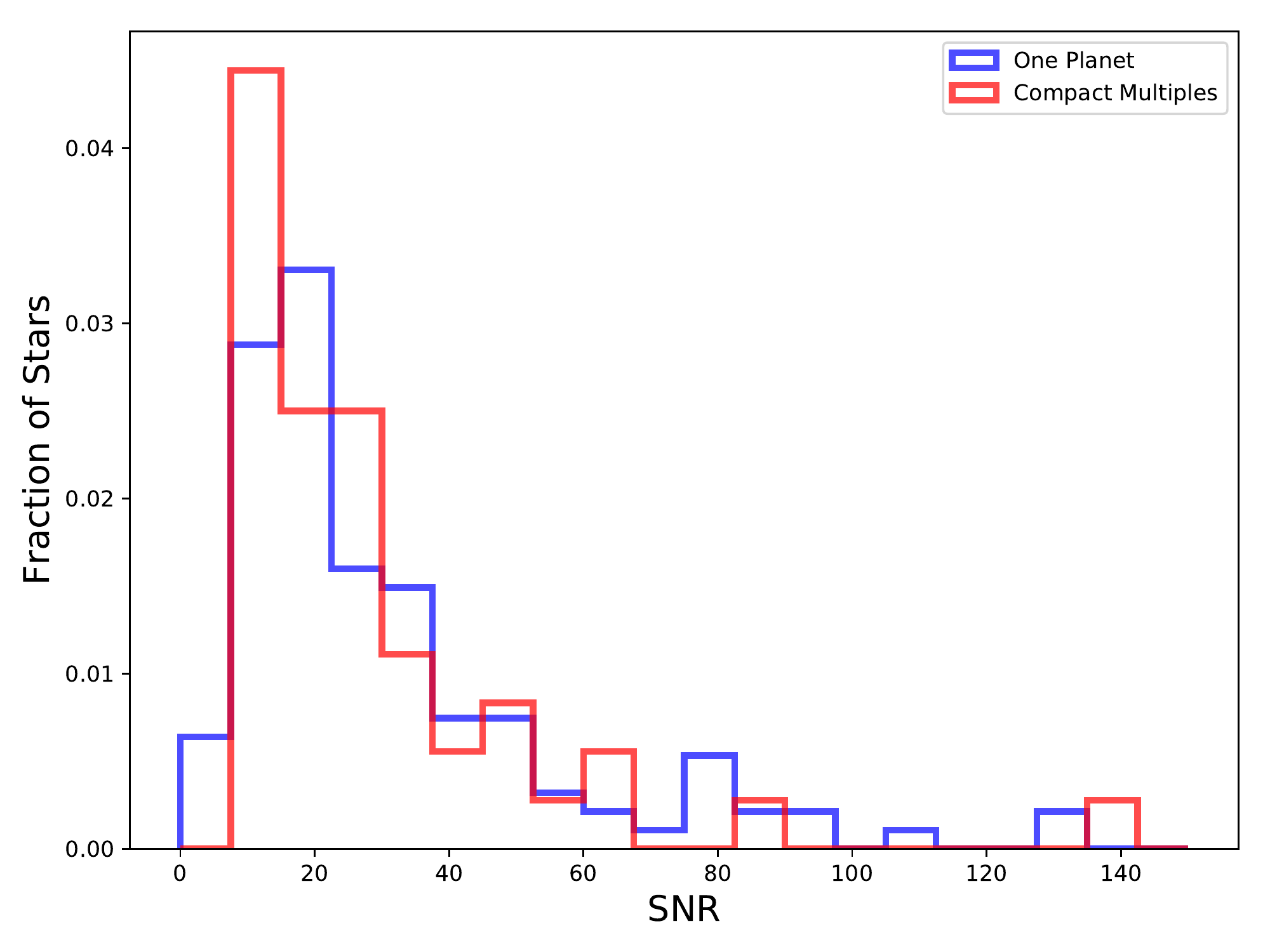} \\
\caption{Histogram showing SNR distributions for single and multi-planet systems. In this case, the KS test p-value is 0.477, again indicating no significant discrepancy in detection efficiency among these two populations within our sample}
\label{SNR}
\end{figure}

\subsection{Discussion of Measurement Uncertainty}
Each photometry and parallax measurement for our stars has an associated error. To ensure that the errors were not large enough to affect our result, we performed a Monte Carlo analysis of these data using the errors. We set the g, r, and i errors equal to their typical values from the KIC: 0.025, 0.02, and 0.02 respectively \citep{KIC}. The J, H, and K magnitude errors come from the 2MASS catalog and are unique for each photometric band and star \citep{Jarrett_2000}. The Gaia magnitude and parallax errors are from Gaia DR2 \citep{Gaia2018}. All of these error values are included in our data table (Table \ref{Data Table}). 

\begin{figure}
\centering
\includegraphics[scale=0.44]{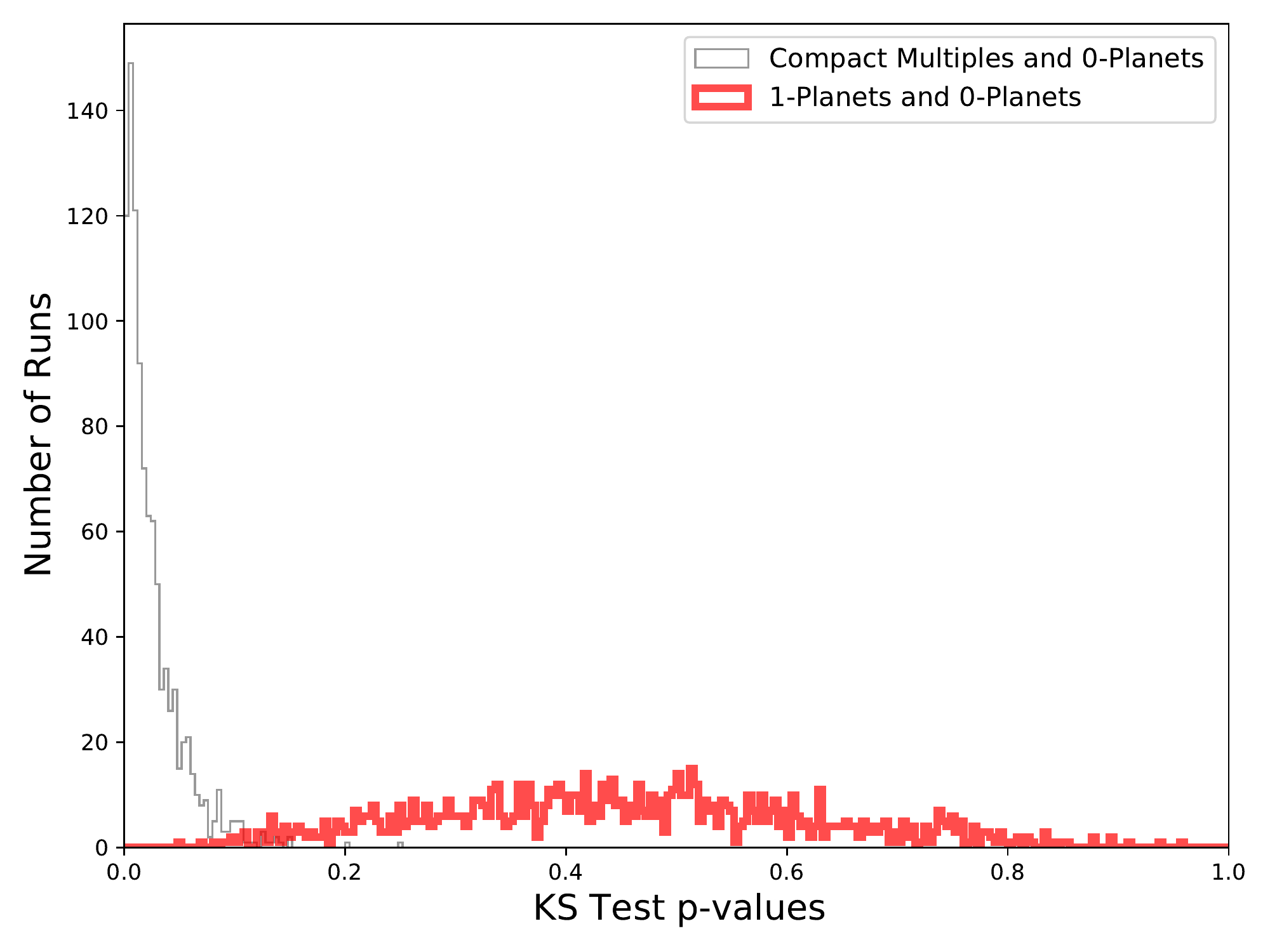}
\includegraphics[scale=0.44]{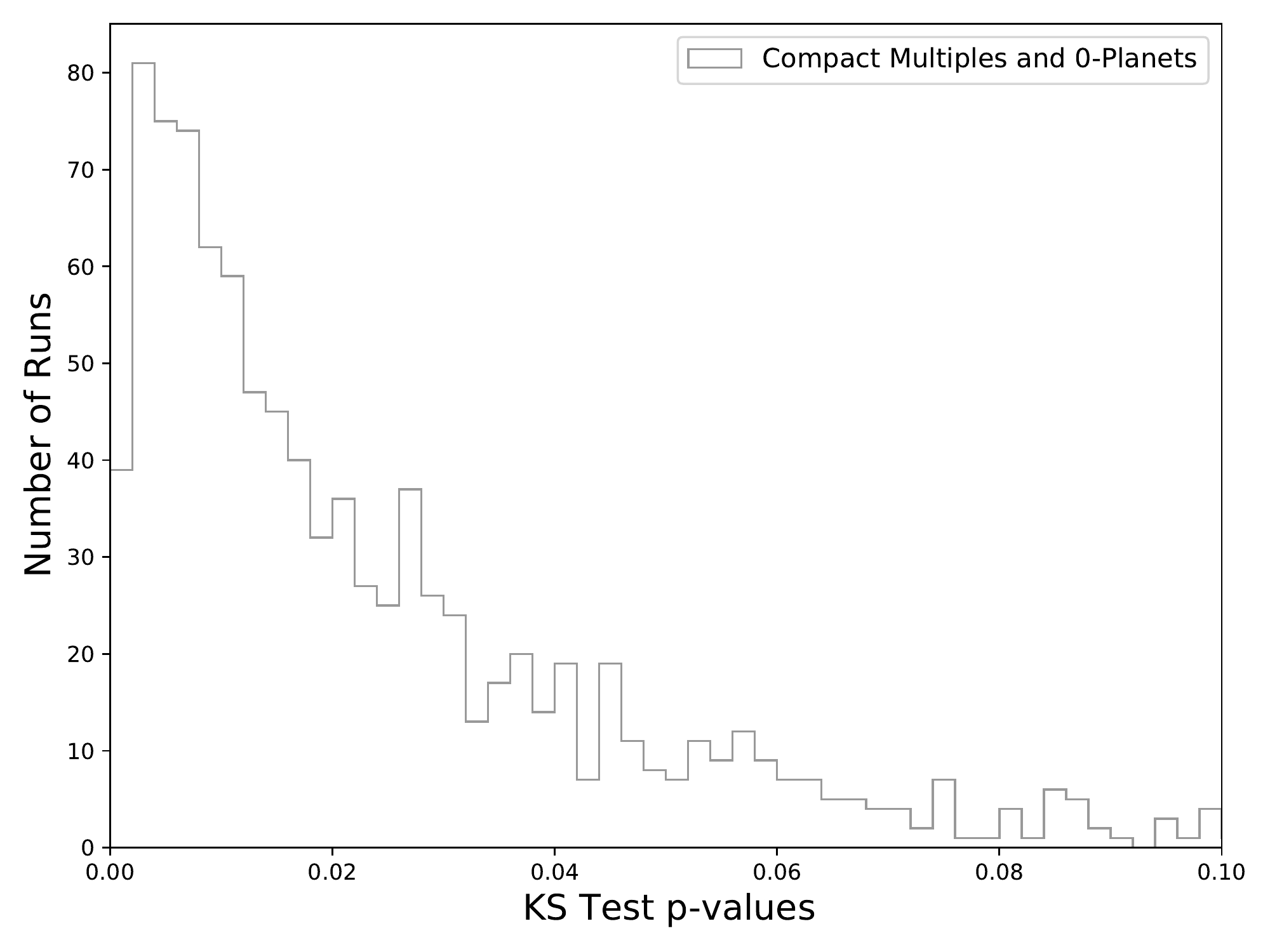}\\
\includegraphics[scale=0.44]{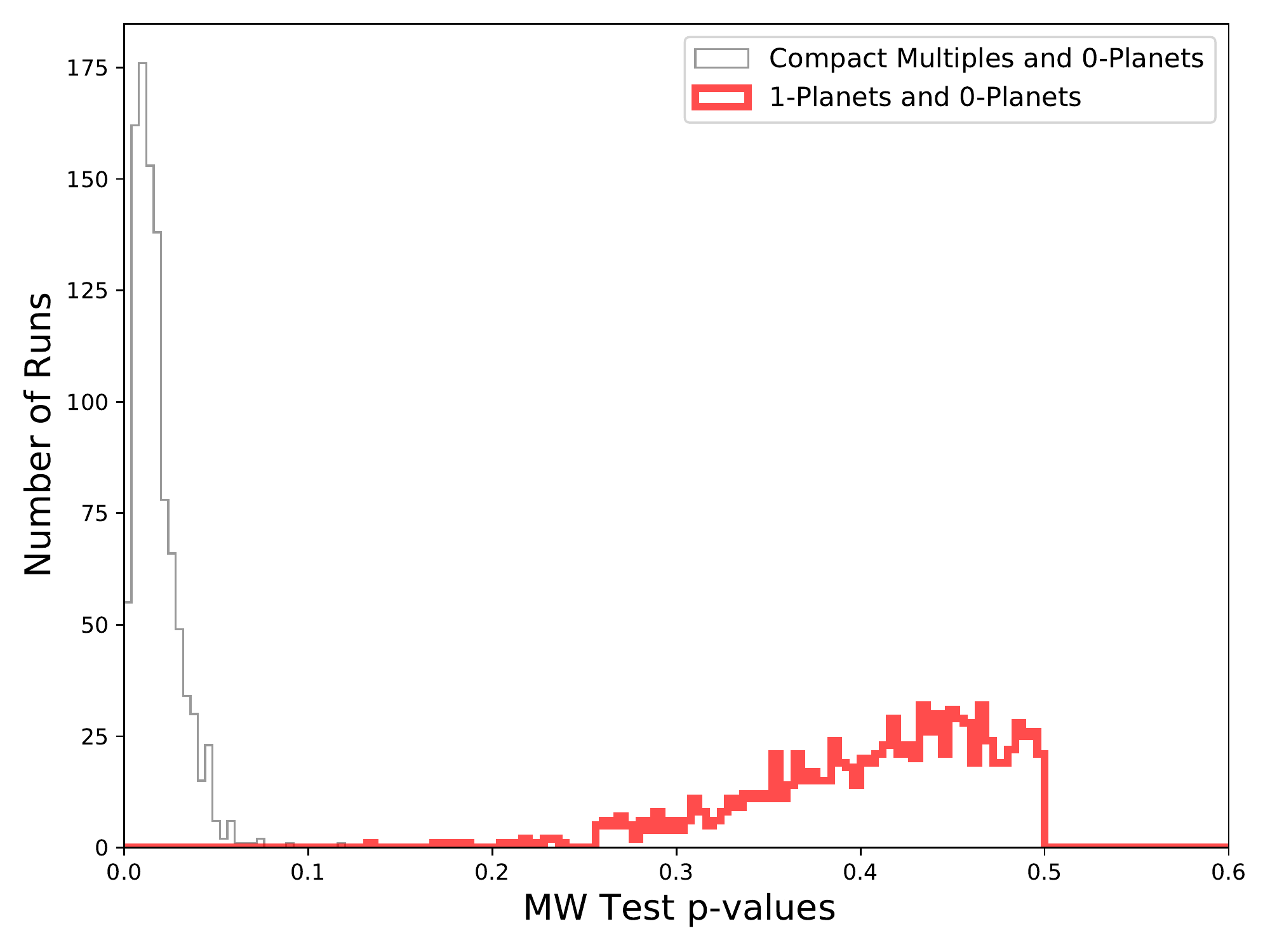}
\includegraphics[scale=0.44]{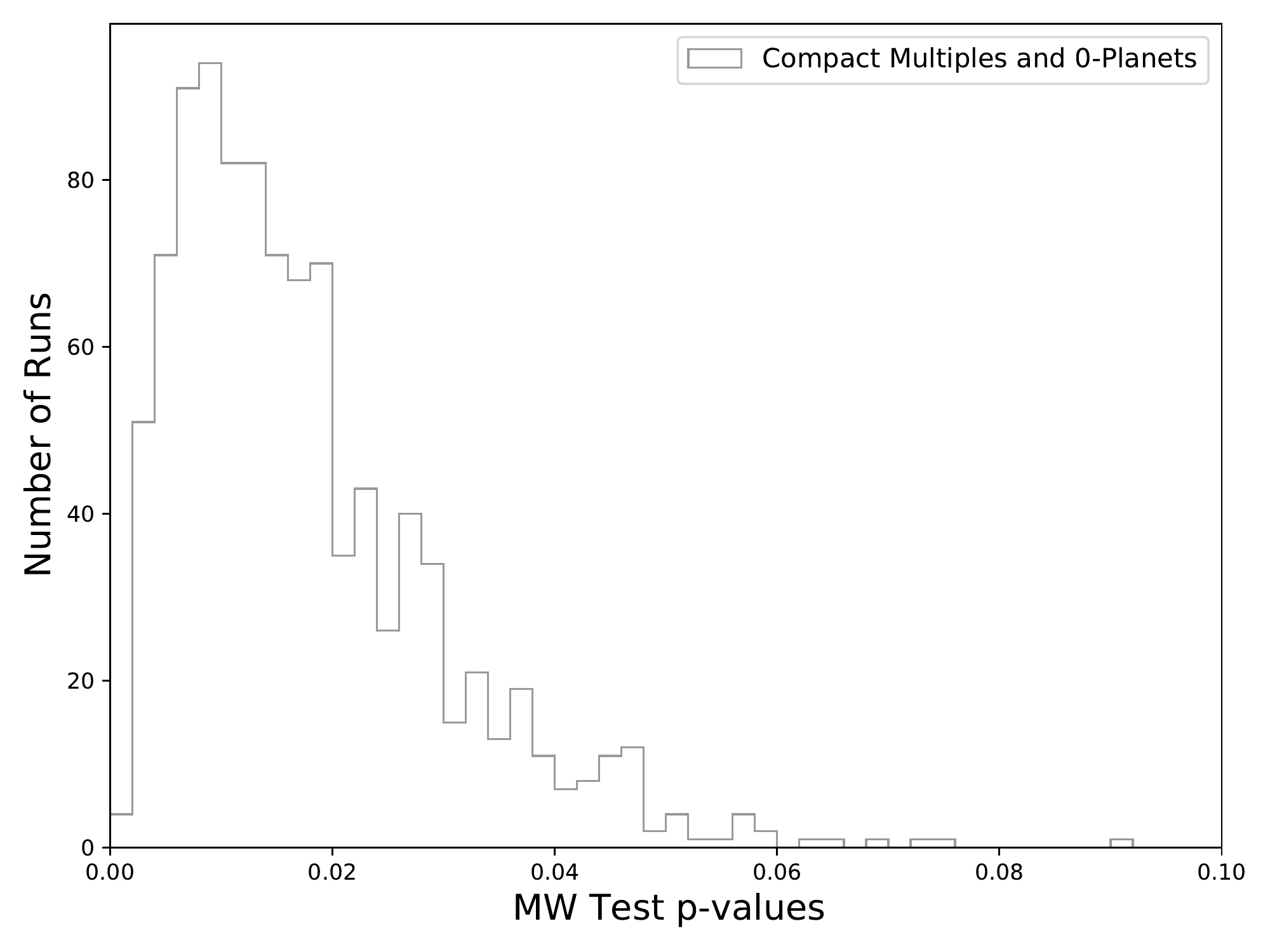}\\
\includegraphics[scale=0.44]{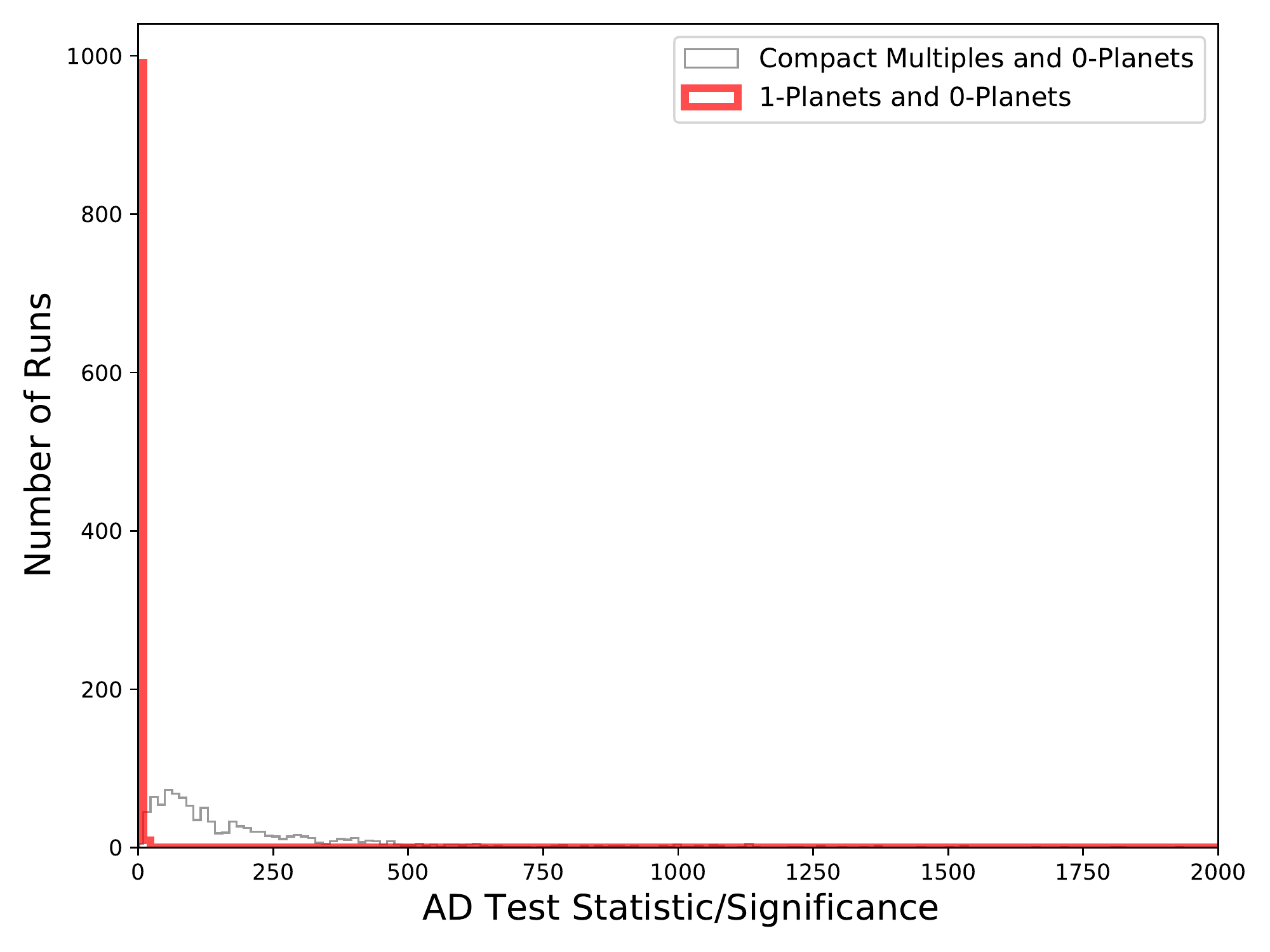}
\includegraphics[scale=0.44]{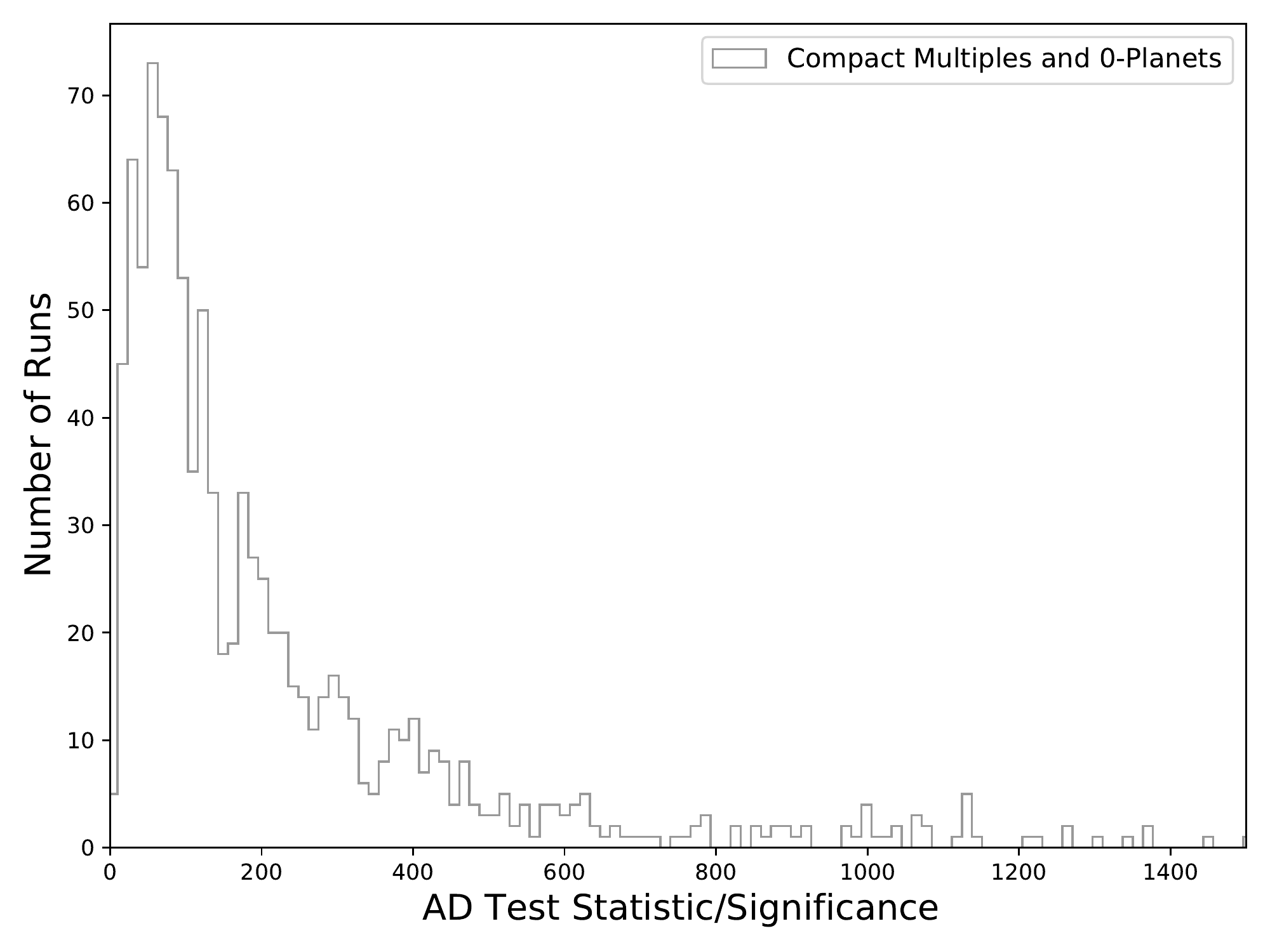}\\
\caption{Histograms showing KS and Mann Whitney U Test p-values, and Anderson-Darling test statistic/significance from the Monte Carlo simulation with randomly drawn errors for the $M_i$ vs. $M_g - M_J$ color magnitude case. The original p-values for the compact-multiple/0-planet populations and the 0-planet/1-planet populations respectively, without error added are: 0.015 and 0.416 for the KS test, 0.023 and 0.190 for the Mann Whitney test, and 80.6 and 0.1 for the AD statistic/significance. We again conclude that even with error our p-values for the compact multiples as compared to the no-planet population are statistically significant.}
\label{Error_igJ}
\end{figure}

\begin{figure}
\centering
\includegraphics[scale=0.44]{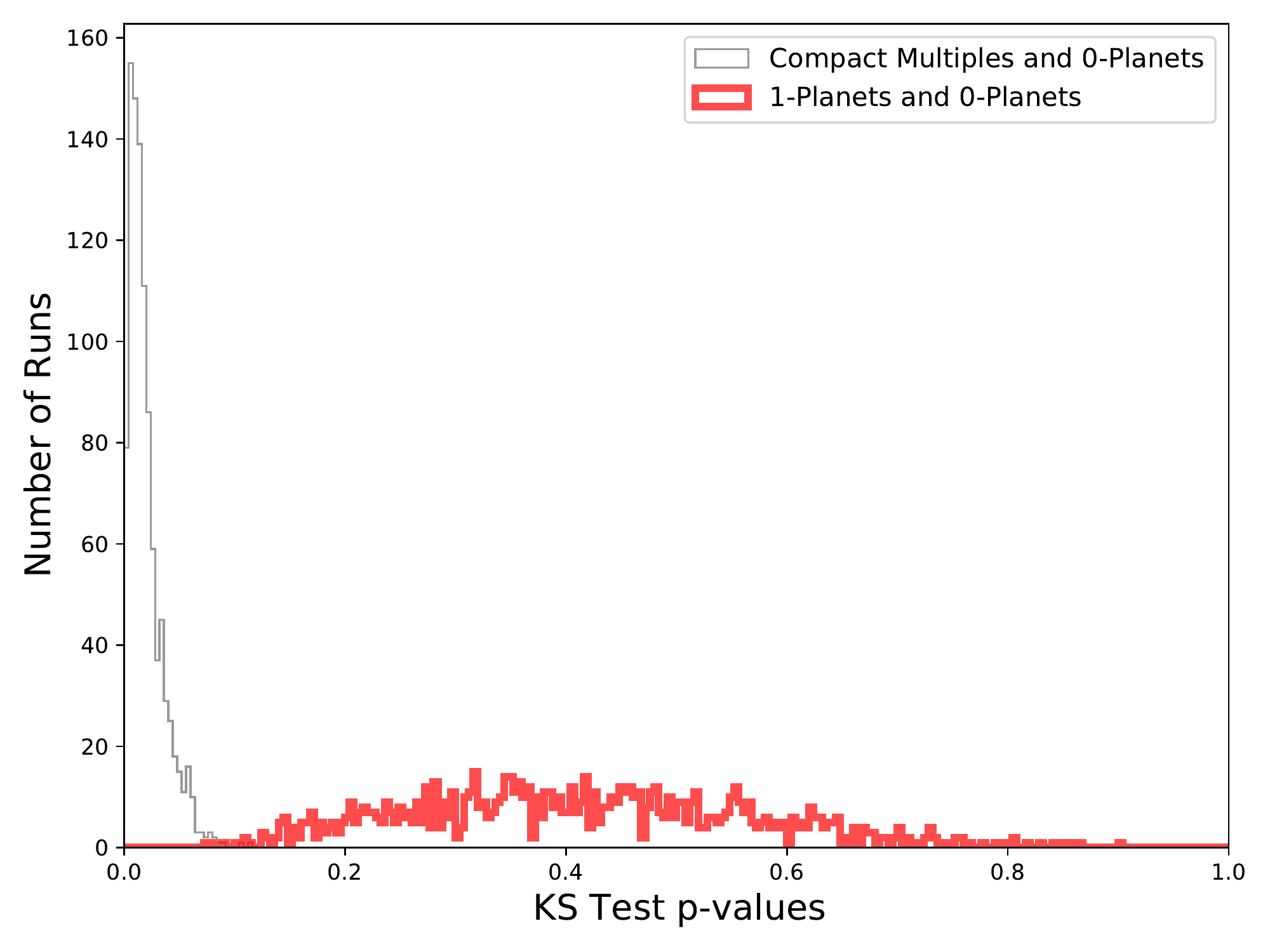}
\includegraphics[scale=0.44]{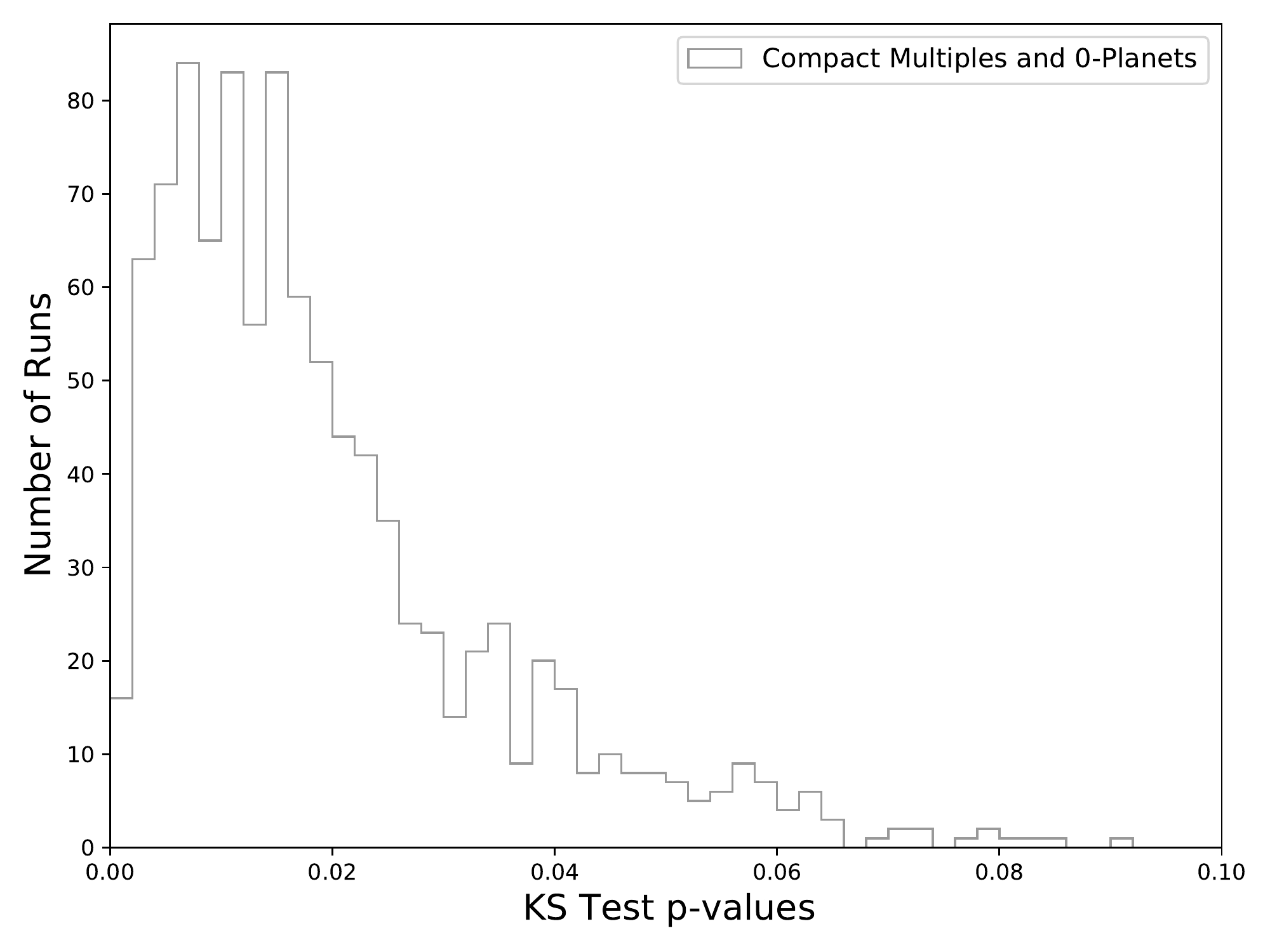}\\
\includegraphics[scale=0.44]{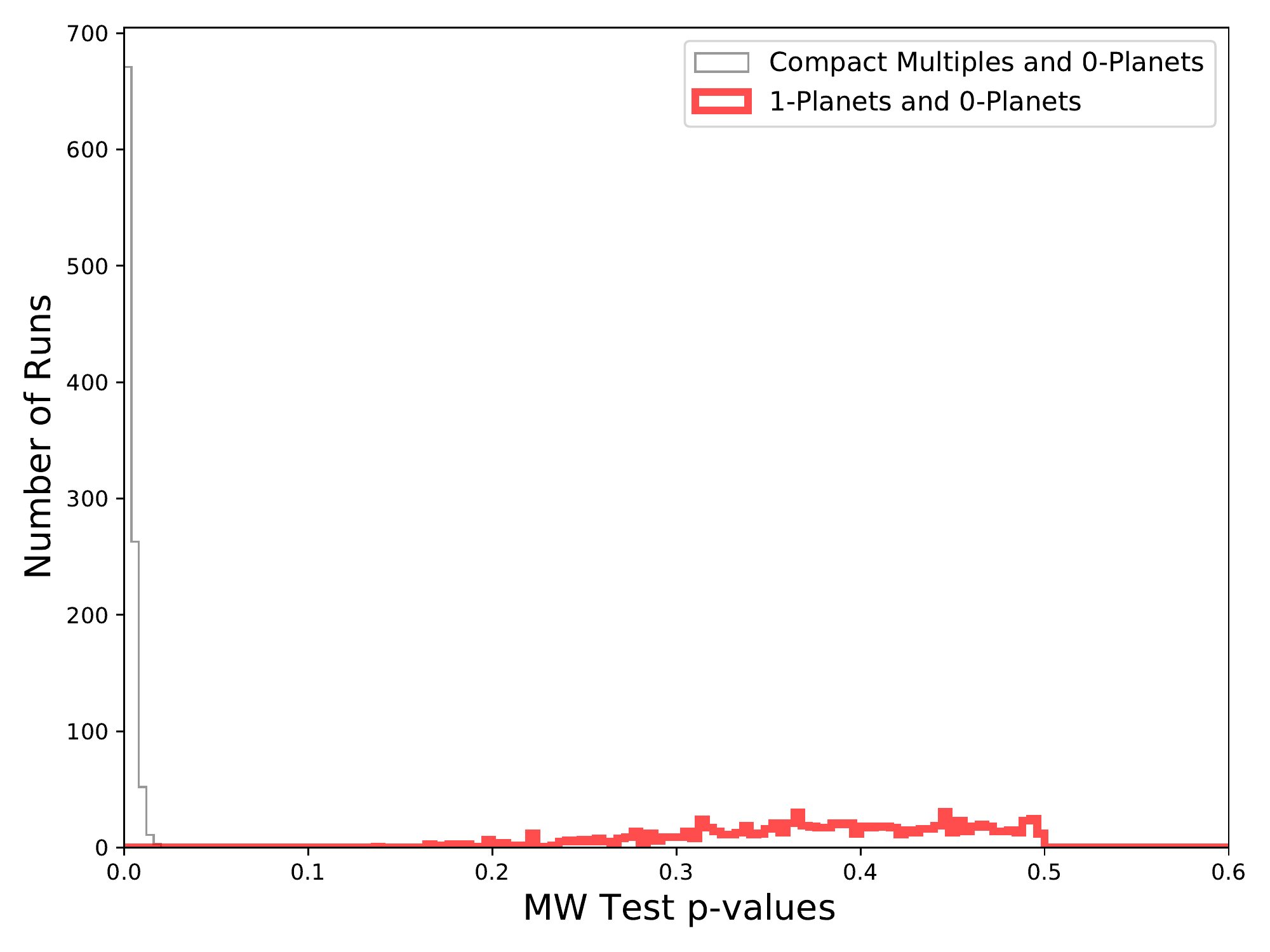}
\includegraphics[scale=0.44]{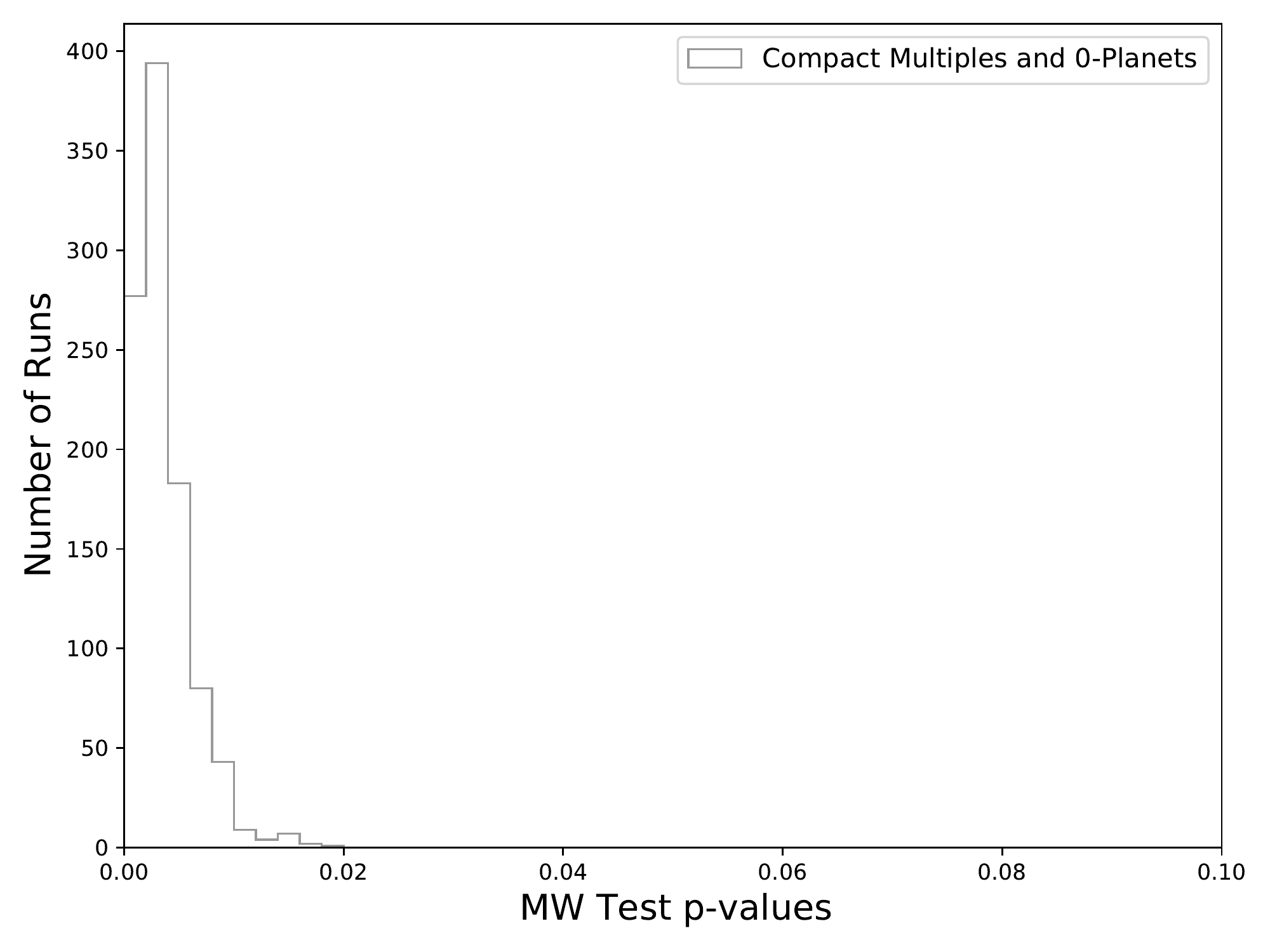}\\
\includegraphics[scale=0.44]{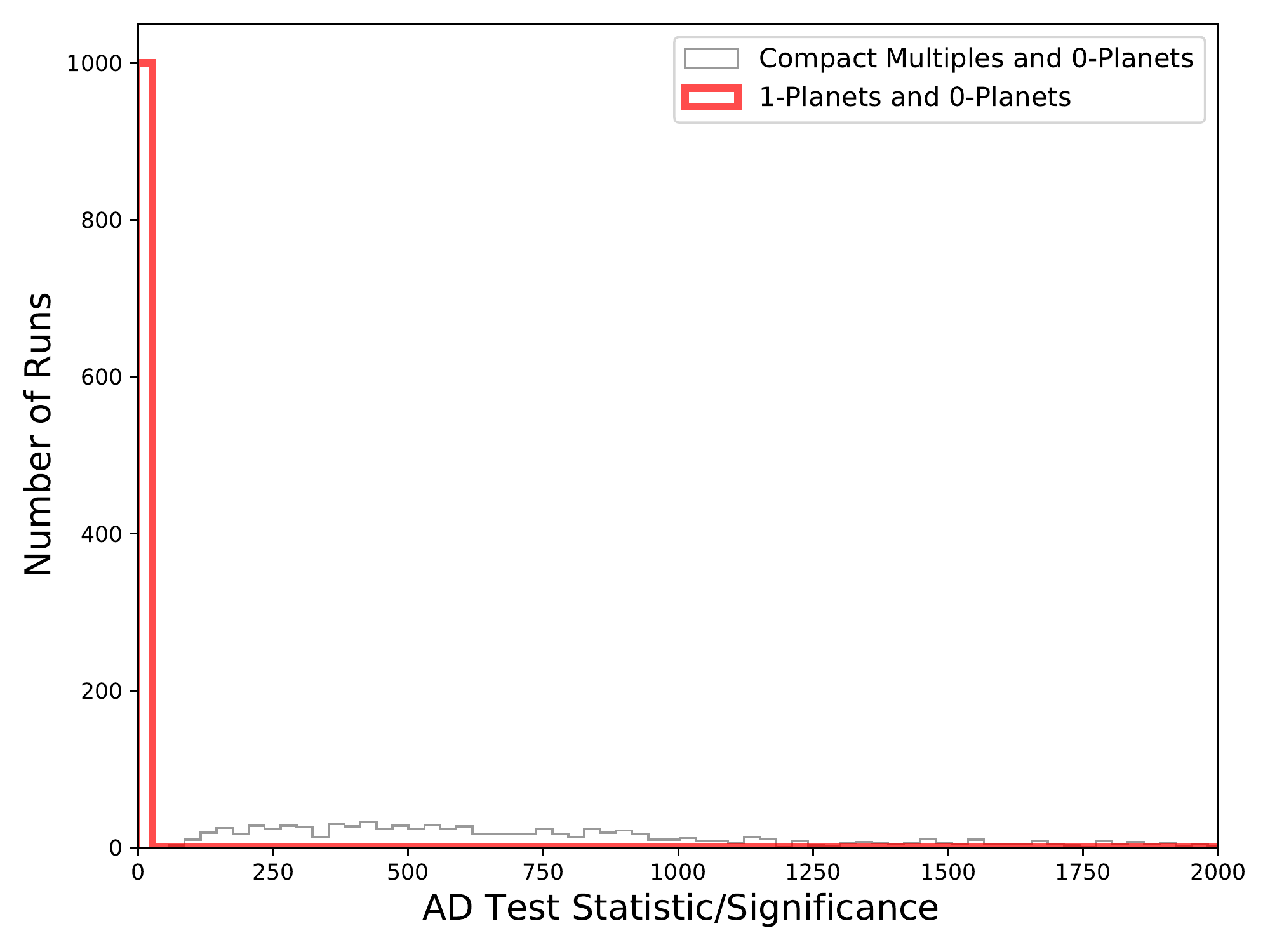}
\includegraphics[scale=0.44]{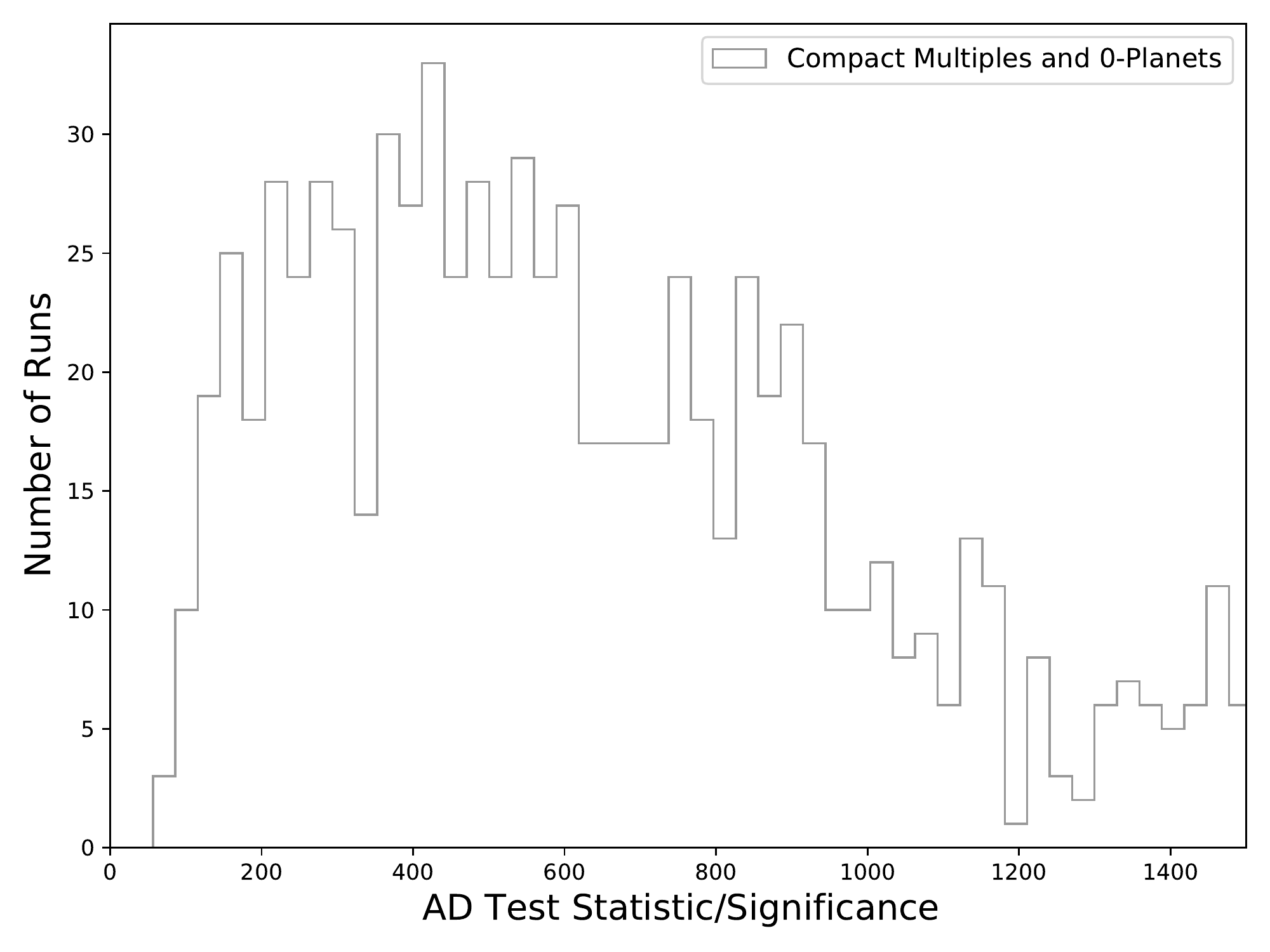}\\
\caption{Histograms showing KS and MW U Test p-values, and AD test statistic/significance from the Monte Carlo simulation with randomly drawn errors for the $M_K$ vs. $G_{BP} - G_{RP}$ color magnitude case. The KS test is shown in the two plots on the top, the Mann Whitney test in the middle, and the AD test on the bottom. In this case, our original KS test p-values were 0.016 and 0.378, the MW test p-values were 0.005 and 0.161, and the AD test statistic/significance values were 371.6 and 2.7 for the compact-multiple/0-planet populations and the 0-planet/1-planet populations respectively, without error added.}
\label{Error_KBR}
\end{figure}

\begin{figure}
\centering
\includegraphics[scale=0.44]{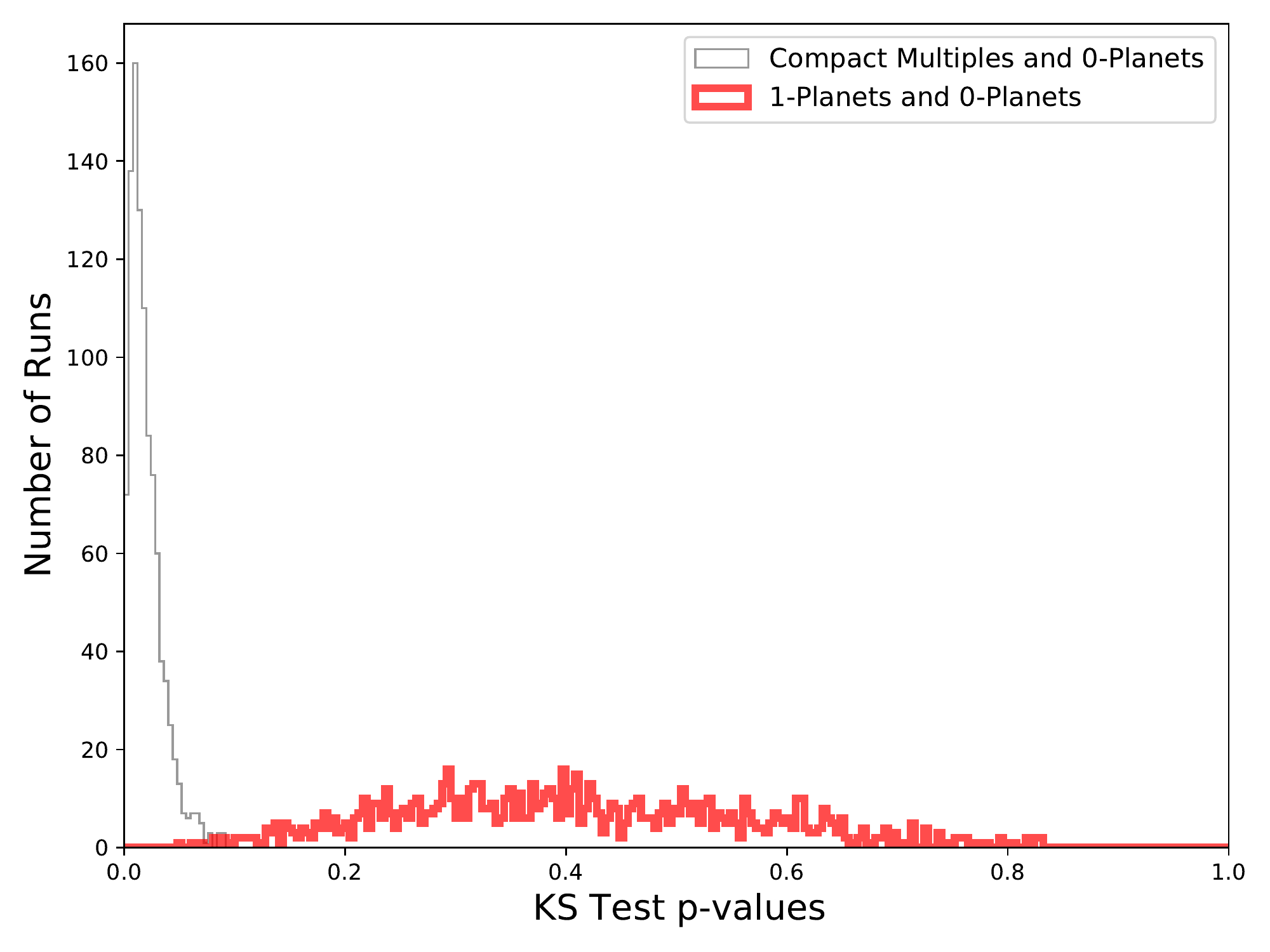}
\includegraphics[scale=0.44]{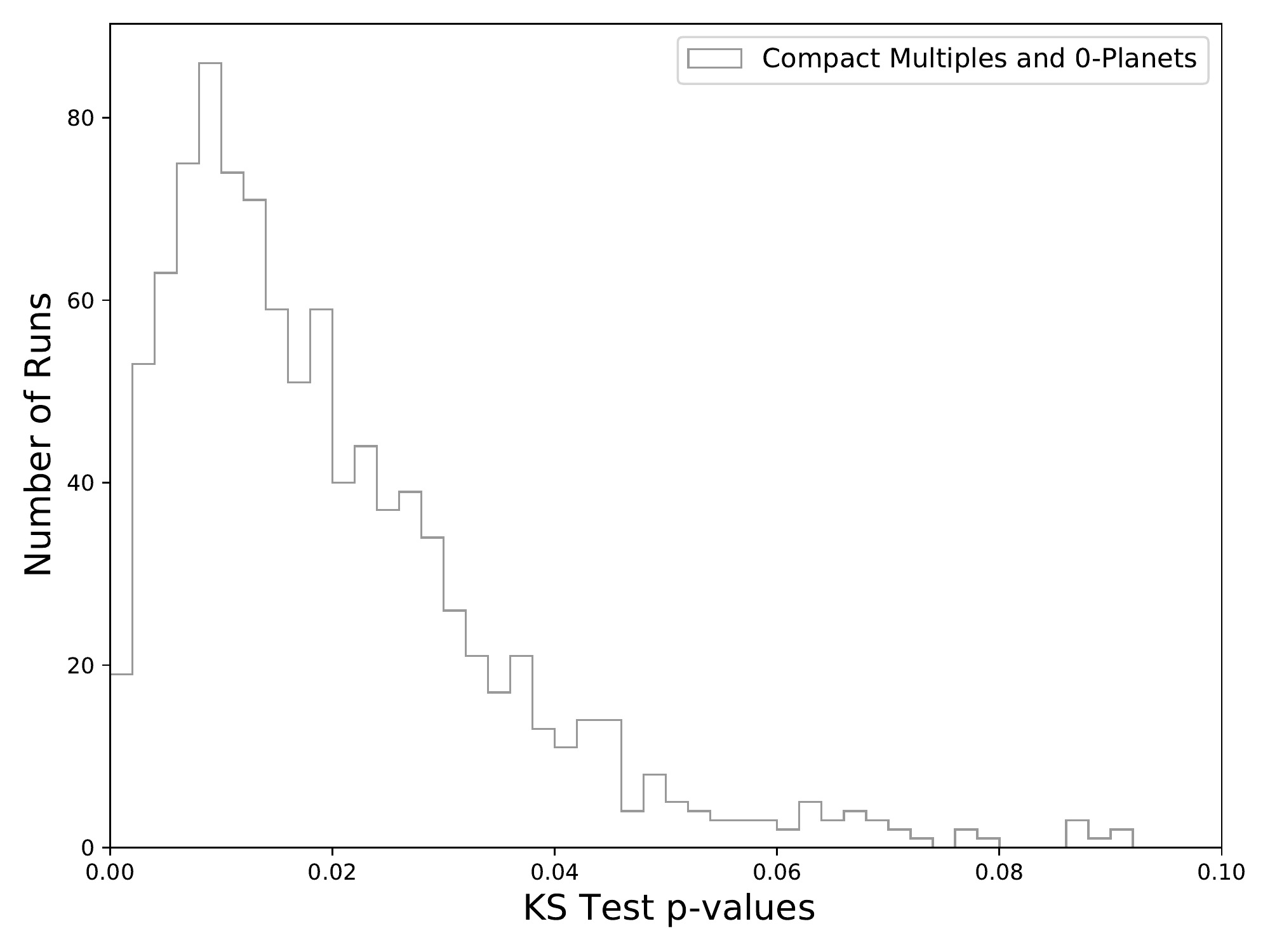}\\
\includegraphics[scale=0.44]{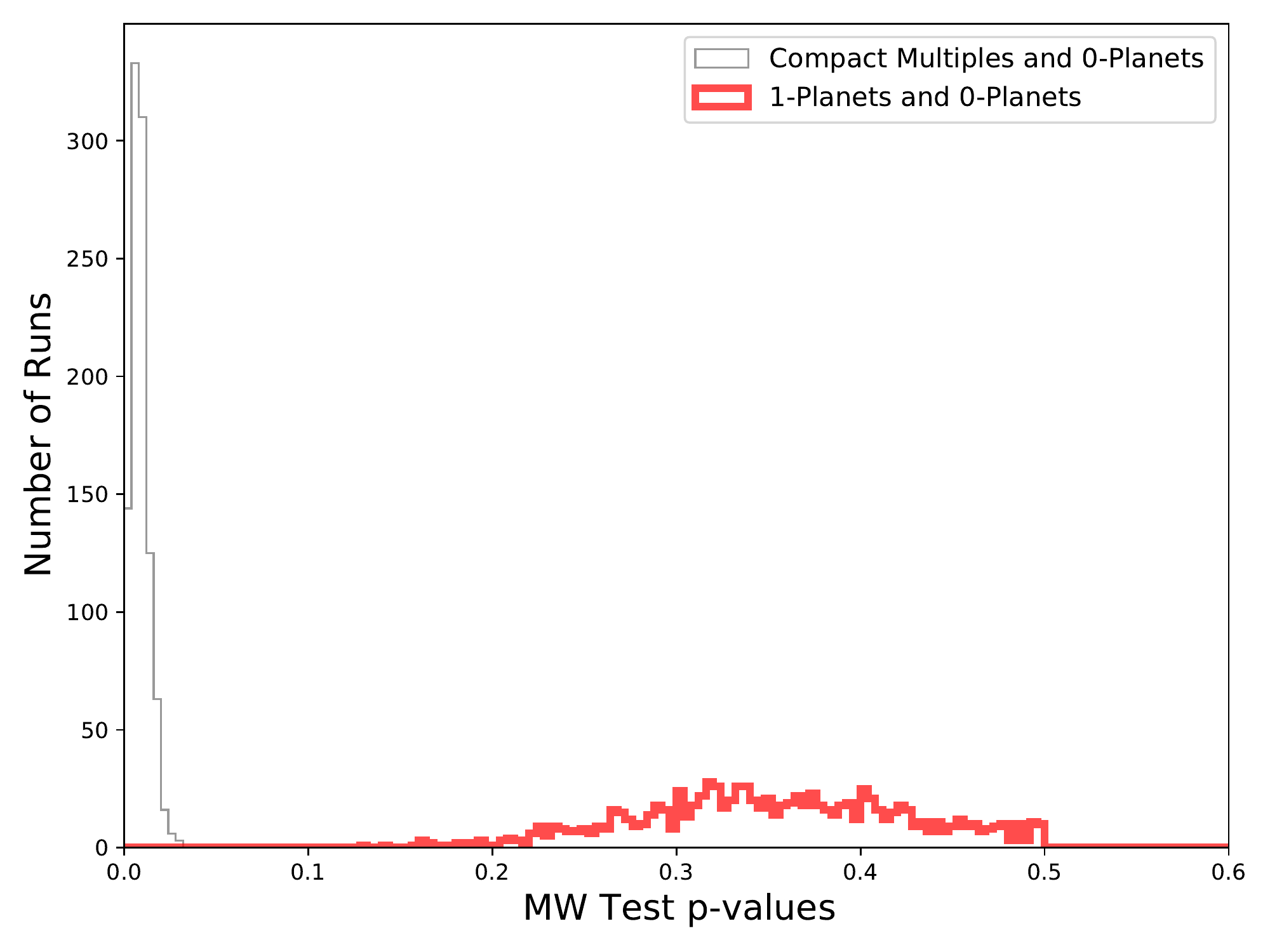}
\includegraphics[scale=0.44]{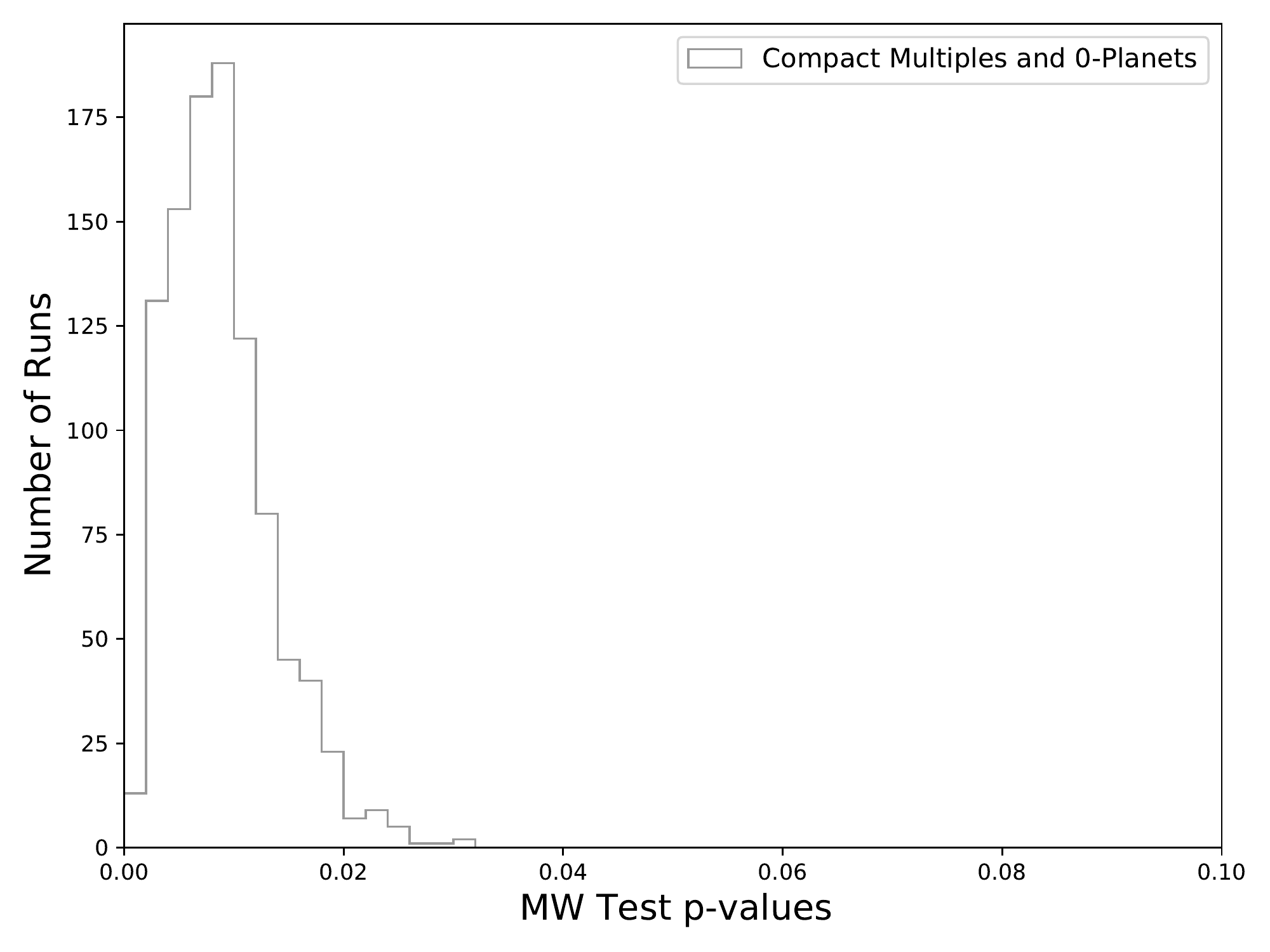}\\
\includegraphics[scale=0.44]{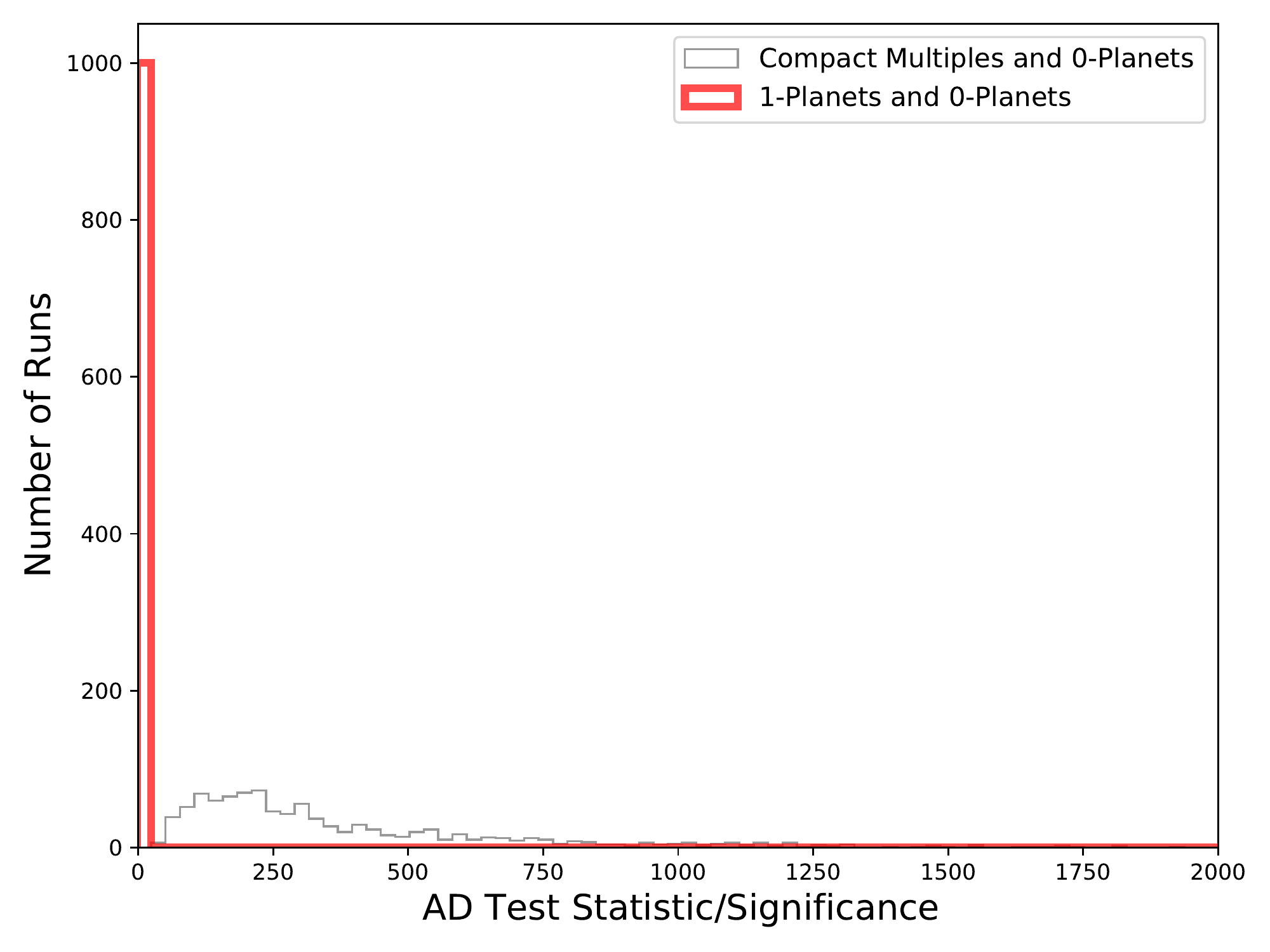}
\includegraphics[scale=0.44]{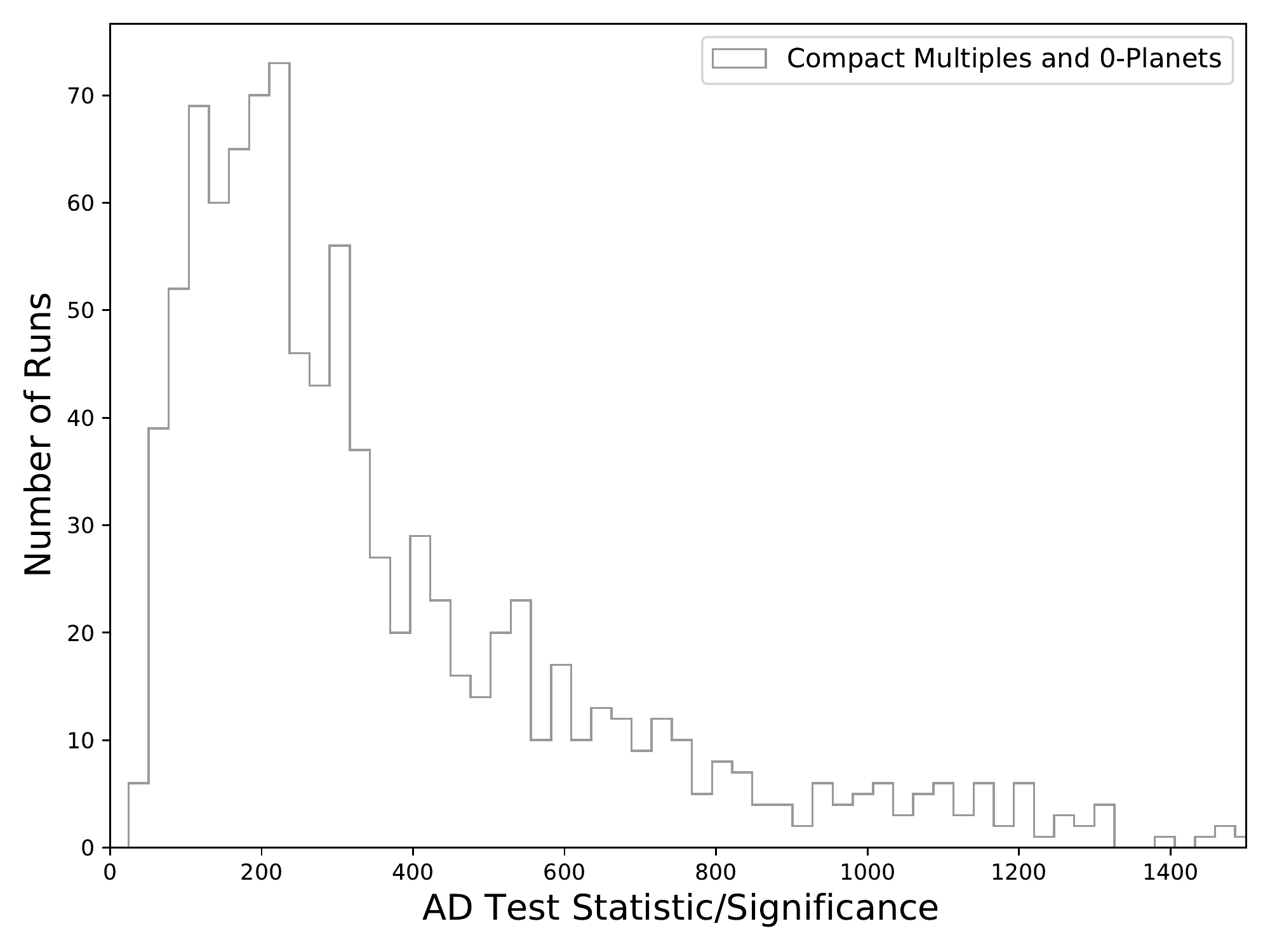}\\
\caption{Histograms showing KS and Mann Whitney U Test p-values and AD test statistic/significance from the Monte Carlo simulation with randomly drawn errors for the $M_G$ vs. $G_{BP} - G_{RP}$ color magnitude case. The original p-values without error added for the compact-multiple/0-planet populations and the 0-planet/1-planet populations are: 0.038 and 0.443 for the KS test, 0.016 and 0.132 for the Mann Whitney test, and 88.8 and 12.2 for the AD statistic/significance. We again conclude from the distributions shown in the plots that, even with error, our p-values for the compact multiples as compared to the no-planet population are statistically significant.}
\label{Error_GBR}
\end{figure}

For each magnitude and parallax for each star, we created a new value consisting of the error multiplied by a random number drawn from a Gaussian distribution, added to the error-less value. We reproduced our color-magnitude diagrams and calculated the KS test, MW test, and AD test statistics using the error-adjusted values for magnitudes and parallax. We repeated this over 1000 runs in a Monte Carlo simulation, and we show in the histograms below that measurement uncertainty has no significant effect on the results. In all of the left-column histograms in Figures \ref{Error_igJ}, \ref{Error_KBR} and \ref{Error_GBR}, there is a clear distinction between the distributions of the compact multiple-no planets statistic (grey) and the one planet-no planet statistic (red), with little overlap in the tails of the distributions. For each color-magnitude combination, we created an additional ``zoomed in" histogram (right column plots) showing the p-values for the compact multiple population as compared to the no-planet population. For the AD test, the histograms show the ratio of AD statistic to significance level. A high ratio indicates statistical significance (as seen in the distribution comparing compact multiples to zero-planet systems) and a low ratio indicates no statistical difference between the populations being compared. In the caption of each set of plots, we list the statistical values obtained previously, without error added, and see that in each case the value we found was within the range of the distributions with error added. We present these histograms below in Figures \ref{Error_igJ}, \ref{Error_KBR} and \ref{Error_GBR}, demonstrating statistical significance in all three filter choices.

\subsection{Pure M-Dwarf Sub-Sample}

We performed our analysis using a temperature upper bound of 4500 K. We chose this value so that our results would better connect with those of \citet{Brewer18}, which used 4500 K as a temperature lower bound and found a similar preference for compact multiples around metal poor stars. This choice includes both M-dwarf stars and low-mass late K-dwarf stars in our sample, and we recognize that the Kepler dichotomy may be principally constrained to the M-dwarf planetary systems \citep{Ballard2016}. We can use temperature to divide our sample into M-dwarfs, with a temperature upper bound of 4000 K, and a hotter late K-dwarf sub-sample with temperature between 4000 and 4500 K. We prefer temperature over color as a metric to determine stellar type because we use color as a metallicity proxy and because we used temperature to derive our initial sample. We find that our stellar sample includes 3809 M-dwarfs, 3691 with no planets, 87 with a single planet, and 31 compact multiples. Due to the already limited number of low-mass stars surveyed by Kepler and even smaller number of M-dwarfs, we leave investigation of the the relation between metallicity and compact multiple occurrence with a pure M-dwarf sample for a future work. The TESS Survey would provide a larger sample of M-dwarfs on which to perform this analysis \citep{Ricker2015}.

\subsection{Comparison with Model Population Predictions}
We have shown evidence that the colors of compact multiples, singles, and non-planets hosts are drawn from different underlying distributions. However, there must be some inherent overlap between these populations: a bona-fide compact multiple may present a single transiting planet or none at all, depending upon its transit geometry. Here we compare the observational result with model predictions from previous work on underlying M-dwarf planetary system occurrence and architecture. 

\subsubsection{Generation of Model Sample}
We generate a sample of synthetic host stars, drawing from the distribution of effective temperatures of the observed sample. Our synthetic sample contains $N$=7156 stars (equal to our original sample size of stars with temperatures cooler than 4500 Kelvin from \citet{Berger2018}). We assign absolute $K$ magnitude to our synthetic sample, interpolating from from the real sample of effective temperature and absolute K magnitude. We assign random noise in absolute $K$ with $\sigma$-0.4 magnitudes from the standard deviation in the real relation between $T_{\mbox{eff}}$ and absolute $K$. We make the simplifying assumption that the main sequence in this temperature range is well-described by the best-fit line shown in Figure 1, and assign a base $G_{B}-G_{R}$ color from this line at each absolute $K$ value. We assign planets to each host star using a model that combines the work of \cite{Dressing2015} with \cite{Ballard2016} and \cite{Muirhead2015}. The latter two studies assume a mixture model for planet occurrence. The first ``compact multiple" component is characterized by $N\ge5$ planets per stars (and mutual inclinations drawn from a Rayleigh distribution with peak at 1-2$^{\circ}$): these comprise a fraction $f$ of the total host star sample. The second component of the mixture model is characterized by $N=1$ or 2 planets per star, where two planet systems have higher mutual inclinations than the first model population (5$^{\circ}$ on average). We draw sets of $\{N,\sigma\}$ (where $N$ is the number of planets per stars and $\sigma$ is the peak of the Rayleigh distribution for mutual inclination) from the joint posterior distributions for these parameters from \cite{Ballard2016}. 

We assign periods and radii for each planetary system from the occurrence rates shown in \cite{Dressing2015}. The fraction $f$ of systems in the first configuration of ``compact multiples" varies with spectral type even around the M subtype: from ~15\% for M0V dwarfs to between 20 and 40\% for mid-M dwarfs \citep{Muirhead2015, Hardegree19}. This number is lower for FGK dwarfs, for which the fraction of analogous systems (i.e. $\ge2$ planets with orbital periods less than 100 days) is 3--5\% \citep{Lissauer2011,Moriarty16}. For the spectral range of stars in our synthetic sample (from late K to mid M), this fraction may therefore change by a factor of 10. We adopt a compact multiple rate of ~20\% uniformly, corresponding to the mean effective temperature of 3900 K of our sample.  We assume an isotropic distribution for the angle of the planetary midplane, and determine which planets transit the host star. We then apply detection sensitivity criteria to determine which transiting planets are ``detected." We employ the sensitivity maps of \cite{Christiansen16,Thompson18} for K-dwarfs (effective temperatures $\ge4100$) and from \cite{Dressing2015} for M-dwarfs. 

\subsubsection{Results of Comparison}

With our synthesized population of stars and planets, with associated planet detections, we then assign a color offset from the main sequence line. We employ only two free parameters: first, the offset in the blue direction assigned to ``compact multiples" and second, the inherent noise associated with that offset. To keep the mean color offset at zero (so that the main sequence is still centered on the best-fit line from Figure 1), we assign a red offset to the second population of planets with the same noise parameter. The red offset in our model corresponding to more sparsely-populated planetary systems is therefore directly determined from the blue ``compact multiple" offset. Said differently, if the blue offset is given by $B$ and the red offset by $R$, then $B \cdot f$ + $R \cdot (1-f)=0$. For example, for $f=0.2$ (a compact multiple fraction of 20\% of planetary systems) and an offset of -0.2 in $G_{B}-G_{R}$ (that is, compact multiples are bluer on average by 0.2 magnitudes), to keep the main sequence centered at zero we must assign a red offset of 0.2$\cdot$0.2/(1-0.2)=+0.05. 

\begin{figure}
\centering
\includegraphics[scale=0.8]{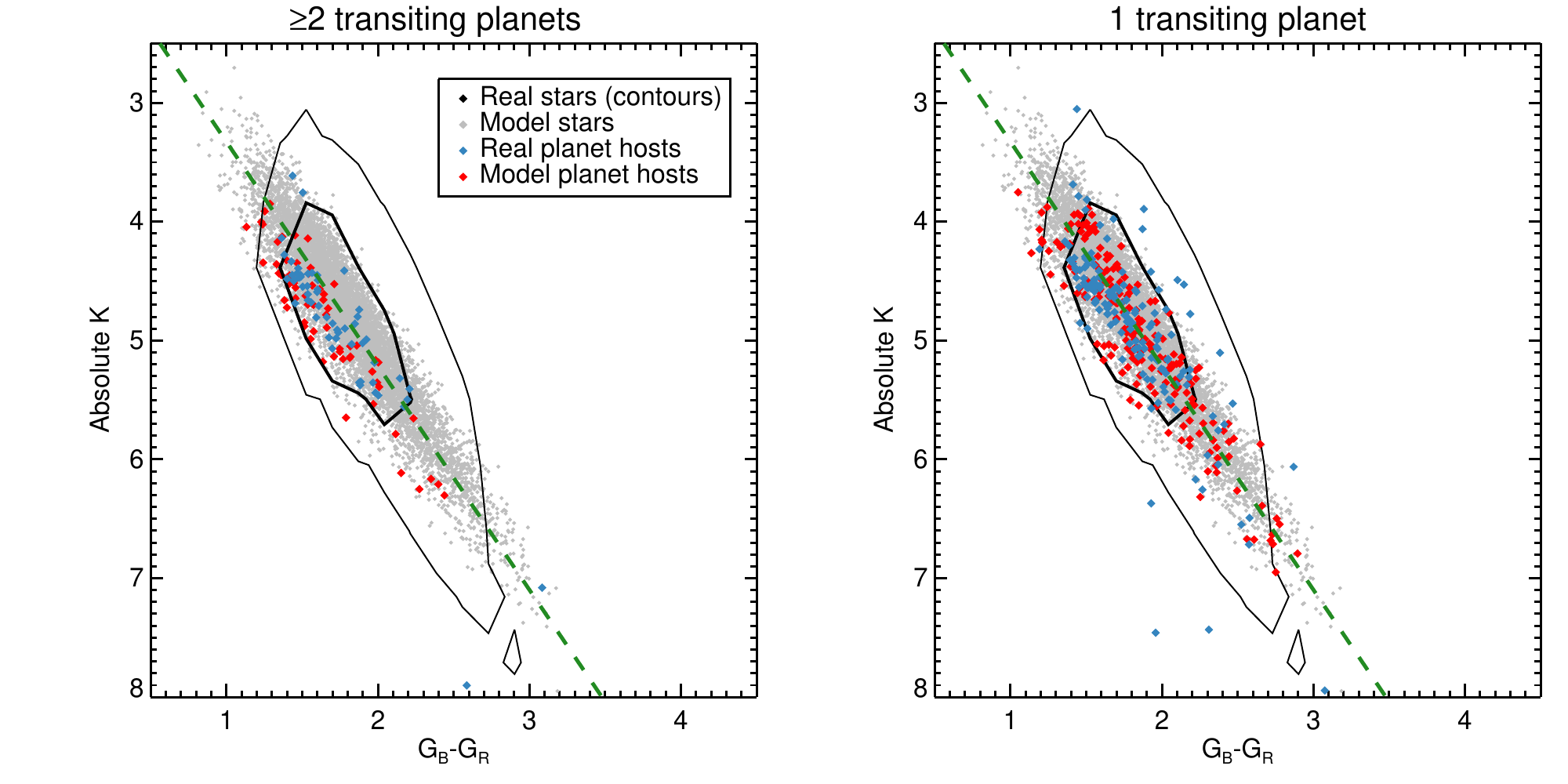} \\
\caption{Synthetic color-magnitude diagrams for compact multiple planetary systems (left) and single-transiting planet systems (right). The contours are the observed stellar population from Kepler and reflect 1 and 2 $\sigma$ contour lines. Model stars, Kepler planet hosts, and model planet hosts are shown as the colored points. Model planets are drawn from a mixture model presented by \cite{Ballard2016}, and compact multiple systems are offset blueward from the main sequence in order to reproduce the observed results.}
\label{CMD1}
\end{figure}

In Figure \ref{CMD1} we show our synthesized population of stars with associated planet detections. We first verify that the ``detected" number of synthetic planets is similar to the true number. Among the 7156 surveyed {\it Kepler} host stars with $T_{\mbox{eff}}<4500$ K in our original sample, 6944 host 0 detected transiting planets, 156 host 1, 33 host 2, and 23 host $\ge3$. In comparison, in our synthetic sample, 6916$\pm20$ host 0 detected transiting planets, 179$\pm$19 host 1, 37$\pm$10 host 2, and 22$\pm$6 host $\ge3$. This consistency is sensible, given the fact that value of $f$ from \cite{Ballard2016} was determined by fitting to the distribution of the number of transiting planets per star.  Using a blue offset of -0.15 magnitudes, coupled with an inherent noise in color of $\sigma$=0.2 magnitudes, we qualitatively reproduce the colors in the multi and single transiting systems in the observed sample (shown in the left and right panels of Figure \ref{CMD1}, respectively). 

\begin{figure}
\centering
\includegraphics[scale=1.2]{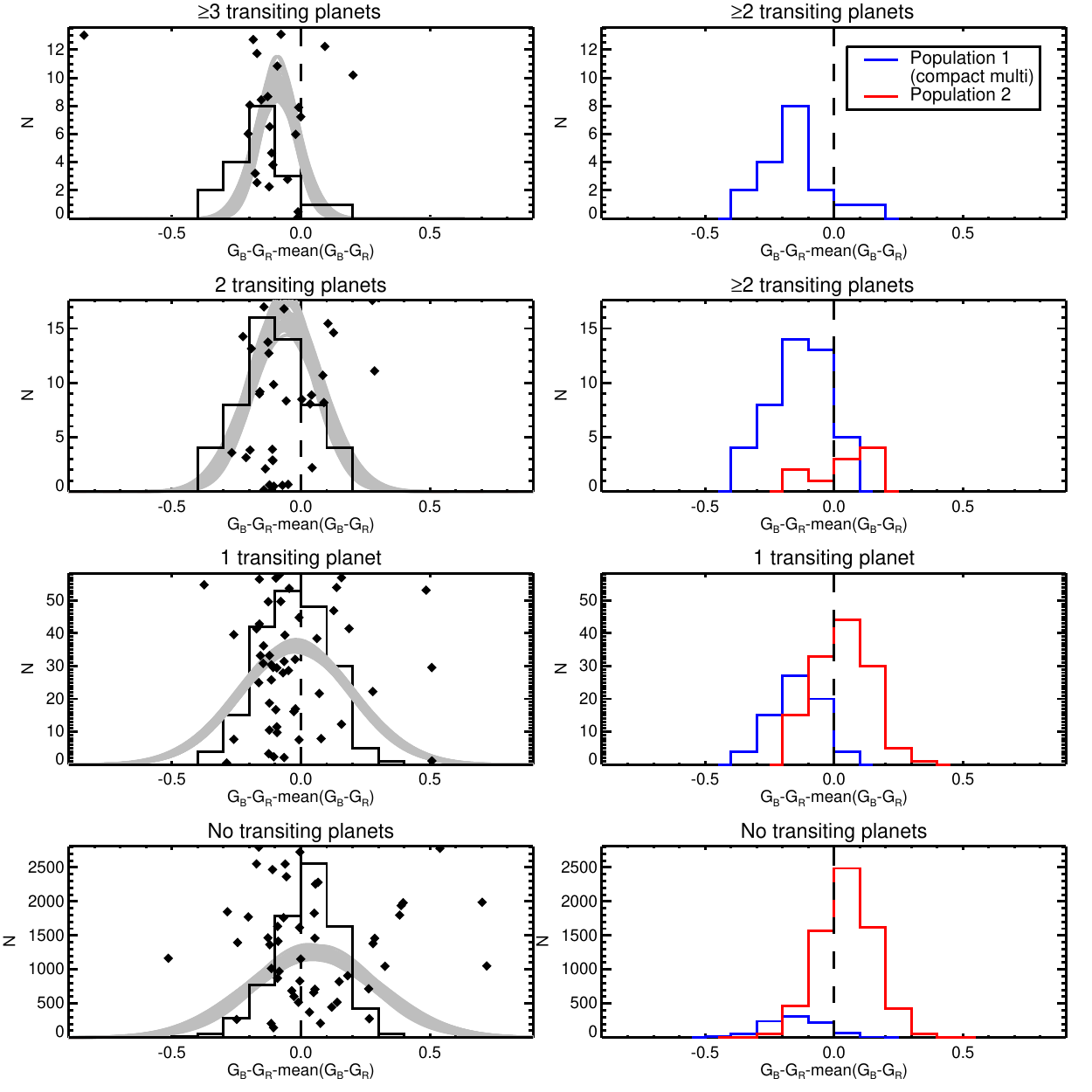} \\
\caption{Synthetic color-magnitude diagram cross sections for various planetary populations in the synthetic sample. The left plot shows the actual distribution Kepler planetary systems as diamonds with the synthetic population as the underlying histogram. The grey Gaussian represents the best-fit Gaussian distribution that represents the underlying data for each planetary system sub-sample. We find that bluer Gaussians are preferred for the compact multiple transiting systems, suggesting that they are more likely to orbit more metal poor stars. The right hand side shows the synthetic population, with each model component shown as a separate histogram.}
\label{CMD2}
\end{figure}

In Figure \ref{CMD2} we show how the underlying population of synthetic planet hosts maps to the observed detections. For systems hosting 0, 1, 2, and $\ge3$ detected transiting planets, we show how the predicted color offset (that is, the distance in $G_{BP}-G_{RP}$ from the best-fit main sequence line) compares with the observed values. We note here that we neglect possible effects of binarity and the potential binary sequence for this model. At left, we show the a sample synthetic color distribution (histogram) with real detections (individual diamonds). For each subset ($N$=0, 1, 2, and $\ge3$ transiting planets) we evaluate the Bayesian likelihood (assuming a Gaussian likelihood function) of the mean color and its standard deviation, conditioned on the observations. We overplot the best-fit Gaussian model and associated $1\sigma$ uncertainty, to enable a direct comparison between the predicted and observed colors.  

At right in Figure \ref{CMD2}, we show how each of the two underlying planet occurrence models contribute to the synthetic color distribution. We can now see how the mixture model in planet occurrence maps to the number of detected transiting planets: while 100\% of systems with $\ge3$ detected transits belong to the ``compact multiple" population, only 60\% of single transiting systems and 20\% of non-transiting systems belong to the same population. This provides a natural explanation for why host star color shifts progressively more blue as the number of detected transiting planets increases: the number of transiting planets is related to the likelihood of observing a ``bona fide" compact multiple system. 
We defer a more sophisticated forward modeling analysis to future work. Here we have demonstrated qualitative consistency between the data and a model with blue offset of -0.15 in $G_{BP}-G_{RP}$ among compact multiples. However, we have not evaluated the uncertainty in this color offset between different populations within a Bayesian framework. 

\section{Discussion}
%=============== What is the discussion? ===============
%The discussion section is an extension of the analysis section and you should relate how the quantities you have measured in the analysis question relate to or answer the scientific questions posed in the paper. 

%IE This section should contain all of the "big picture" things. How does your result compare to other similar studies? What have you done that is new? Do your observations support a particular physical picture or rule out another one? 

%All of the interpretation goes in this section
%============================================================

We have shown that small rocky planets in compact multiple systems orbiting M and late K-dwarfs preferentially form and survive to the present epoch around stars that are more blue (more metal-poor) than the typical Kepler star of the same spectral types. This result is similar to \cite{Brewer18}, with a few notable differences in our sample and methods. Both our study and \citet{Brewer18} filtered by surface gravity in an effort to remove contamination from giant stars. However, the \cite{Brewer18} sample was drawn from solar-like stars with temperatures above 4500 K, while our sample consists solely of M and K-dwarfs with effective temperatures $<$ 4500 K, effectively making our samples mutually exclusive. We also differed in our methods for determining metallicity. \cite{Brewer18} directly analyzed optical spectra from the Keck HIRES spectrograph to obtain 15 elemental abundances. The work we present here utilizes a photometric method developed in \cite{Dittmann2016} and extended in \citet{Dittmann2020} in order to determine relative metallicities for statistical comparison among our sample, as it is difficult to directly determine the metallicity of an M or late K-dwarf solely from optical spectra due to the excessive molecular features present. Finally, \cite{Brewer18} includes hot Jupiters, cool Jupiters, and compact multiples, whereas we differentiate our stellar samples only by number of planets and not by type. A direct comparison among Jovian-sized planets between our samples is not possible as the number of transiting Jupiter-mass companions orbiting M-dwarfs is too small for a meaningful statistical statement. Our result suggests that the planet metallicity correlation that \cite{Brewer18} identified for larger stars likely continues down the main sequence to the smallest stellar masses. 

Interestingly, \cite{Weiss18} found no statistical difference between the metallicity of compact multiple and single-planet systems. Their stellar sample was similar to that of \citet{Brewer18}, consisting of Kepler stars with masses ranging from 0.5 to 1.6 \msun and temperatures between 4500 and 6300 K. They determined metallicity for these larger stars spectroscopically. Since the \cite{Brewer18} and \cite{Weiss18} samples overlap significantly, the discrepancy in their results is likely due to differing interpretations of the spectra and how they measured spectral metallicities. Our results suggest that the \citet{Brewer18} interpretation of these spectra is likely correct. %\cite{Weiss18}'s lack of differences in properties between single-planet and multi-planet host stars implies that the two types of systems have similar formation processes. 

Any preferential formation channel of single-planet systems vs. compact multiple systems that can be correlated with bulk properties present at formation, such as metallicity, can possibly point to differences in the planet formation process. For M-dwarfs, this could be a potential explanation for the Kepler dichotomy \citep{Ballard2016}, although detailed planet formation simulations will need to be performed to show how this mechanism may work. It is also possible that the planet formation process proceeds differently or is more sensitive to external variables around M-dwarfs and low mass K-dwarfs, as the protoplanetary disk is much less massive and the relative proportion of metals from which to make planets may play a larger role in the formation outcome. \citet{Dawson15} studied the relation between surface density of the parent disk and tendency to form rocky or gaseous planets, using metallicity as a proxy for surface density. The planets we address in this study are rocky, and the results of \citet{Dawson15} indicate that lower surface densities (lower metallicities) yield purely rocky planets, because the gas has dissipated by the time the smaller planetary embryos have formed full planets. In the future, it would be interesting to investigate and compare the radius distributions for single and multi-planet low metallicity systems.

\section{Conclusion}
%=============== What is the Conclusion ===============
%This section is mostly brief and directly summarizes the paper. What you did, and what the main takeaway message is. It's sort of a longer version of the abstract and more detailed. If you imagine yourself giving a conference presentation on this paper, imagine your conclusion slide / takeaway message in paragraph form and that is your conclusion section. 
%============================================================

It has long been established that giant planets are more likely to form around metal rich stars (\cite{Fisher05}, \cite{Everett2013}, \cite{Buchhave2014}, \cite{Wang2014}). However, there is no consensus as to how metallicity affects the formation and orbital architecture of small rocky planets. We focused here on small planets around small stars. 

We utilized photometric data and high precision trigonometric distances to investigate the planet-metallicity correlation at small planet masses around M-dwarfs and late K-dwarfs. We used a technique developed by \cite{Dittmann2016} to relate these stars' location in the color-magnitude diagram to relative metallicity.

%Plotting average trend lines through our $M_K$ vs. $G_{BP} - G_{RP}$ and $M_G$ vs. $G_{BP} - G_{RP}$  plots provided our reference point for average metallicity. We then considered the distance from each star to the average and compared the outcomes for three populations of stars: no-planet, single-planet, and compact multiples. 
We find that compact multiples preferentially form around the more metal-poor M and late K-dwarfs in the Kepler sample, using the KS, MW, and AD statistical tests to verify our result. %with KS test p-values of 0.0087 for the $M_K$ vs. $G_B - G_R$ plot and 0.0067 for the $M_G$ vs. $G_B - G_R$ plot, comparing compact multiples to no-planet stars. When we include only M-dwarfs and eliminate the late K-dwarfs in our sample with temperature between 4000 and 4500K, \textbf{this result is still significant}. 
This result is consistent with \cite{Brewer18} and provides a new angle in the larger debate about a planet-metallicity relation across the main sequence. 

\section*{Acknowledgments}
JAD and SGA would acknowledge support from the Heising-Simons Foundation, without which this work would not have been possible. SGA would also like to acknowledge Professors Ian Crossfield and Julien de Wit, as well as the MIT UROP program which helped to support this work. Finally, we would like to thank the referee for their comments.

\bibliography{bibliography.bib}

\begin{thebibliography}{}
\expandafter\ifx\csname natexlab\endcsname\relax\def\natexlab#1{#1}\fi
\providecommand{\url}[1]{\href{#1}{#1}}
\providecommand{\dodoi}[1]{doi:~\href{http://doi.org/#1}{\nolinkurl{#1}}}
\providecommand{\doeprint}[1]{\href{http://ascl.net/#1}{\nolinkurl{http://ascl.net/#1}}}
\providecommand{\doarXiv}[1]{\href{https://arxiv.org/abs/#1}{\nolinkurl{https://arxiv.org/abs/#1}}}

\bibitem[{Akeson {et~al.}(2013)Akeson, Chen, Ciardi, Crane, Good, Harbut,
  Jackson, Kane, Laity, Leifer, Lynn, McElroy, Papin, Plavchan, Ram{\'{\i}}rez,
  Rey, von Braun, Wittman, Abajian, Ali, Beichman, Beekley, Berriman, Berukoff,
  Bryden, Chan, Groom, Lau, Payne, Regelson, Saucedo, Schmitz, Stauffer, Wyatt,
  \& Zhang}]{NASAexo}
Akeson, R.~L., Chen, X., Ciardi, D., {et~al.} 2013, Publications of the
  Astronomical Society of the Pacific, 125, 989, \dodoi{10.1086/672273}

\bibitem[{{Ballard} \& {Johnson}(2016)}]{Ballard2016}
{Ballard}, S., \& {Johnson}, J.~A. 2016, \apj, 816, 66,
  \dodoi{10.3847/0004-637X/816/2/66}

\bibitem[{{Benedict} {et~al.}(2016){Benedict}, {Henry}, {Franz}, {McArthur},
  {Wasserman}, {Jao}, {Cargile}, {Dieterich}, {Bradley}, {Nelan}, \&
  {Whipple}}]{Benedict2016}
{Benedict}, G.~F., {Henry}, T.~J., {Franz}, O.~G., {et~al.} 2016, \aj, 152,
  141, \dodoi{10.3847/0004-6256/152/5/141}

\bibitem[{Berger {et~al.}(2018)Berger, Huber, Gaidos, \& van
  Saders}]{Berger2018}
Berger, T.~A., Huber, D., Gaidos, E., \& van Saders, J.~L. 2018, The
  Astrophysical Journal, 866, 99, \dodoi{10.3847/1538-4357/aada83}

\bibitem[{{Borucki} {et~al.}(2010){Borucki}, {Koch}, {Basri}, {Batalha},
  {Brown}, {Caldwell}, {Caldwell}, {Christensen-Dalsgaard}, {Cochran}, \&
  {DeVore}}]{Borucki_Kepler}
{Borucki}, W.~J., {Koch}, D., {Basri}, G., {et~al.} 2010, Science, 327, 977,
  \dodoi{10.1126/science.1185402}

\bibitem[{{Brewer} {et~al.}(2018){Brewer}, {Wang}, {Fischer}, \&
  {Foreman-Mackey}}]{Brewer18}
{Brewer}, J.~M., {Wang}, S., {Fischer}, D.~A., \& {Foreman-Mackey}, D. 2018,
  \apjl, 867, L3, \dodoi{10.3847/2041-8213/aae710}

\bibitem[{Brown {et~al.}(2011)Brown, Latham, Everett, \& Esquerdo}]{KIC}
Brown, T.~M., Latham, D.~W., Everett, M.~E., \& Esquerdo, G.~A. 2011, The
  Astronomical Journal, 142, 112, \dodoi{10.1088/0004-6256/142/4/112}

\bibitem[{{Buchhave} {et~al.}(2012){Buchhave}, {Latham}, {Johansen},
  {Bizzarro}, {Torres}, {Rowe}, {Batalha}, {Borucki}, {Brugamyer}, \&
  {Caldwell}}]{Buchhave2012}
{Buchhave}, L.~A., {Latham}, D.~W., {Johansen}, A., {et~al.} 2012, \nat, 486,
  375, \dodoi{10.1038/nature11121}

\bibitem[{{Buchhave} {et~al.}(2014){Buchhave}, {Bizzarro}, {Latham},
  {Sasselov}, {Cochran}, {Endl}, {Isaacson}, {Juncher}, \&
  {Marcy}}]{Buchhave2014}
{Buchhave}, L.~A., {Bizzarro}, M., {Latham}, D.~W., {et~al.} 2014, \nat, 509,
  593, \dodoi{10.1038/nature13254}

\bibitem[{{Christiansen} {et~al.}(2016){Christiansen}, {Clarke}, {Burke},
  {Jenkins}, {Bryson}, {Coughlin}, {Mullally}, {Thompson}, {Twicken},
  {Batalha}, {Haas}, {Catanzarite}, {Campbell}, {Kamal Uddin}, {Zamudio},
  {Smith}, \& {Henze}}]{Christiansen16}
{Christiansen}, J.~L., {Clarke}, B.~D., {Burke}, C.~J., {et~al.} 2016, \apj,
  828, 99, \dodoi{10.3847/0004-637X/828/2/99}

\bibitem[{Dawson {et~al.}(2015)Dawson, Chiang, \& Lee}]{Dawson15}
Dawson, R.~I., Chiang, E., \& Lee, E.~J. 2015, Monthly Notices of the Royal
  Astronomical Society, 453, 1471, \dodoi{10.1093/mnras/stv1639}

\bibitem[{{Delfosse} {et~al.}(2000){Delfosse}, {Forveille}, {S{\'e}gransan},
  {Beuzit}, {Udry}, {Perrier}, \& {Mayor}}]{Delfosse2000}
{Delfosse}, X., {Forveille}, T., {S{\'e}gransan}, D., {et~al.} 2000, \aap, 364,
  217.
\newblock \doarXiv{astro-ph/0010586}

\bibitem[{Dittmann {et~al.}(2020)Dittmann, Irwin, Charbonneau, \&
  Bieryla}]{Dittmann2020}
Dittmann, J.~A., Irwin, J.~M., Charbonneau, D., \& Bieryla, A. 2020, The
  Astrophysical Journal, submitted

\bibitem[{Dittmann {et~al.}(2016)Dittmann, Irwin, Charbonneau, \&
  Newton}]{Dittmann2016}
Dittmann, J.~A., Irwin, J.~M., Charbonneau, D., \& Newton, E.~R. 2016, The
  Astrophysical Journal, 818, 153, \dodoi{10.3847/0004-637x/818/2/153}

\bibitem[{Dotter {et~al.}(2017)Dotter, Conroy, Cargile, \&
  Asplund}]{Dotter2017}
Dotter, A., Conroy, C., Cargile, P., \& Asplund, M. 2017, The Astrophysical
  Journal, 840, 99, \dodoi{10.3847/1538-4357/aa6d10}

\bibitem[{{Dressing} \& {Charbonneau}(2013)}]{Dressing2013}
{Dressing}, C.~D., \& {Charbonneau}, D. 2013, \apj, 767, 95,
  \dodoi{10.1088/0004-637X/767/1/95}

\bibitem[{Dressing \& Charbonneau(2015)}]{Dressing2015}
Dressing, C.~D., \& Charbonneau, D. 2015, 807, 45,
  \dodoi{10.1088/0004-637x/807/1/45}

\bibitem[{Duchêne \& Kraus(2013)}]{Duchene2013}
Duchêne, G., \& Kraus, A. 2013, Annual Review of Astronomy and Astrophysics,
  51, 269–310, \dodoi{10.1146/annurev-astro-081710-102602}

\bibitem[{Evans(2018)}]{Evans2018}
Evans, D.~F. 2018, Research Notes of the {AAS}, 2, 20,
  \dodoi{10.3847/2515-5172/aac173}

\bibitem[{Everett {et~al.}(2013)Everett, Howell, Silva, \&
  Szkody}]{Everett2013}
Everett, M.~E., Howell, S.~B., Silva, D.~R., \& Szkody, P. 2013, The
  Astrophysical Journal, 771, 107, \dodoi{10.1088/0004-637x/771/2/107}

\bibitem[{{Fischer} \& {Valenti}(2005)}]{Fisher05}
{Fischer}, D.~A., \& {Valenti}, J. 2005, \apj, 622, 1102,
  \dodoi{10.1086/428383}

\bibitem[{Furlan {et~al.}(2017)Furlan, Ciardi, Everett, Saylors, Teske, Horch,
  Howell, van Belle, Hirsch, Gautier, Adams, Barrado, Cartier, Dressing,
  Dupree, Gilliland, Lillo-Box, Lucas, \& Wang}]{Furlan2017}
Furlan, E., Ciardi, D.~R., Everett, M.~E., {et~al.} 2017, The Astronomical
  Journal, 153, 71, \dodoi{10.3847/1538-3881/153/2/71}

\bibitem[{{Gaia Collaboration} {et~al.}(2018){Gaia Collaboration}, {Brown, A.
  G. A.}, {Vallenari, A.}, {Prusti, T.}, {de Bruijne, J. H. J.}, {Babusiaux,
  C.}, {Bailer-Jones, C. A. L.}, {Biermann, M.}, {Evans, D. W.}, {Eyer, L.},
  {Jansen, F.}, {Jordi, C.}, {Klioner, S. A.}, {Lammers, U.}, {Lindegren, L.},
  {Luri, X.}, {Mignard, F.}, {Panem, C.}, {Pourbaix, D.}, {Randich, S.},
  {Sartoretti, P.}, {Siddiqui, H. I.}, {Soubiran, C.}, {van Leeuwen, F.},
  {Walton, N. A.}, {Arenou, F.}, {Bastian, U.}, {Cropper, M.}, {Drimmel, R.},
  {Katz, D.}, {Lattanzi, M. G.}, {Bakker, J.}, {Cacciari, C.}, {Casta\~neda,
  J.}, {Chaoul, L.}, {Cheek, N.}, {De Angeli, F.}, {Fabricius, C.}, {Guerra,
  R.}, {Holl, B.}, {Masana, E.}, {Messineo, R.}, {Mowlavi, N.}, {Nienartowicz,
  K.}, {Panuzzo, P.}, {Portell, J.}, {Riello, M.}, {Seabroke, G. M.}, {Tanga,
  P.}, {Th\'evenin, F.}, {Gracia-Abril, G.}, {Comoretto, G.},
  {Garcia-Reinaldos, M.}, {Teyssier, D.}, {Altmann, M.}, {Andrae, R.}, {Audard,
  M.}, {Bellas-Velidis, I.}, {Benson, K.}, {Berthier, J.}, {Blomme, R.},
  {Burgess, P.}, {Busso, G.}, {Carry, B.}, {Cellino, A.}, {Clementini, G.},
  {Clotet, M.}, {Creevey, O.}, {Davidson, M.}, {De Ridder, J.}, {Delchambre,
  L.}, {Dell\'{}Oro, A.}, {Ducourant, C.}, {Fern\'andez-Hern\'andez, J.},
  {Fouesneau, M.}, {Fr\'emat, Y.}, {Galluccio, L.}, {Garc\'{\i}a-Torres, M.},
  {Gonz\'alez-N\'u\~nez, J.}, {Gonz\'alez-Vidal, J. J.}, {Gosset, E.}, {Guy, L.
  P.}, {Halbwachs, J.-L.}, {Hambly, N. C.}, {Harrison, D. L.}, {Hern\'andez,
  J.}, {Hestroffer, D.}, {Hodgkin, S. T.}, {Hutton, A.}, {Jasniewicz, G.},
  {Jean-Antoine-Piccolo, A.}, {Jordan, S.}, {Korn, A. J.}, {Krone-Martins, A.},
  {Lanzafame, A. C.}, {Lebzelter, T.}, {L\"offler, W.}, {Manteiga, M.},
  {Marrese, P. M.}, {Mart\'{\i}n-Fleitas, J. M.}, {Moitinho, A.}, {Mora, A.},
  {Muinonen, K.}, {Osinde, J.}, {Pancino, E.}, {Pauwels, T.}, {Petit, J.-M.},
  {Recio-Blanco, A.}, {Richards, P. J.}, {Rimoldini, L.}, {Robin, A. C.},
  {Sarro, L. M.}, {Siopis, C.}, {Smith, M.}, {Sozzetti, A.}, {S\"uveges, M.},
  {Torra, J.}, {van Reeven, W.}, {Abbas, U.}, {Abreu Aramburu, A.}, {Accart,
  S.}, {Aerts, C.}, {Altavilla, G.}, {\'Alvarez, M. A.}, {Alvarez, R.}, {Alves,
  J.}, {Anderson, R. I.}, {Andrei, A. H.}, {Anglada Varela, E.}, {Antiche, E.},
  {Antoja, T.}, {Arcay, B.}, {Astraatmadja, T. L.}, {Bach, N.}, {Baker, S. G.},
  {Balaguer-N\'u\~nez, L.}, {Balm, P.}, {Barache, C.}, {Barata, C.}, {Barbato,
  D.}, {Barblan, F.}, {Barklem, P. S.}, {Barrado, D.}, {Barros, M.}, {Barstow,
  M. A.}, {Bartholom\'e Mu\~noz, S.}, {Bassilana, J.-L.}, {Becciani, U.},
  {Bellazzini, M.}, {Berihuete, A.}, {Bertone, S.}, {Bianchi, L.}, {Bienaym\'e,
  O.}, {Blanco-Cuaresma, S.}, {Boch, T.}, {Boeche, C.}, {Bombrun, A.},
  {Borrachero, R.}, {Bossini, D.}, {Bouquillon, S.}, {Bourda, G.}, {Bragaglia,
  A.}, {Bramante, L.}, {Breddels, M. A.}, {Bressan, A.}, {Brouillet, N.},
  {Br\"usemeister, T.}, {Brugaletta, E.}, {Bucciarelli, B.}, {Burlacu, A.},
  {Busonero, D.}, {Butkevich, A. G.}, {Buzzi, R.}, {Caffau, E.}, {Cancelliere,
  R.}, {Cannizzaro, G.}, {Cantat-Gaudin, T.}, {Carballo, R.}, {Carlucci, T.},
  {Carrasco, J. M.}, {Casamiquela, L.}, {Castellani, M.}, {Castro-Ginard, A.},
  {Charlot, P.}, {Chemin, L.}, {Chiavassa, A.}, {Cocozza, G.}, {Costigan, G.},
  {Cowell, S.}, {Crifo, F.}, {Crosta, M.}, {Crowley, C.}, {Cuypers+, J.},
  {Dafonte, C.}, {Damerdji, Y.}, {Dapergolas, A.}, {David, P.}, {David, M.},
  {de Laverny, P.}, {De Luise, F.}, {De March, R.}, {de Martino, D.}, {de
  Souza, R.}, {de Torres, A.}, {Debosscher, J.}, {del Pozo, E.}, {Delbo, M.},
  {Delgado, A.}, {Delgado, H. E.}, {Di Matteo, P.}, {Diakite, S.}, {Diener,
  C.}, {Distefano, E.}, {Dolding, C.}, {Drazinos, P.}, {Dur\'an, J.},
  {Edvardsson, B.}, {Enke, H.}, {Eriksson, K.}, {Esquej, P.}, {Eynard Bontemps,
  G.}, {Fabre, C.}, {Fabrizio, M.}, {Faigler, S.}, {Falc\~ao, A. J.}, {Farr\`as
  Casas, M.}, {Federici, L.}, {Fedorets, G.}, {Fernique, P.}, {Figueras, F.},
  {Filippi, F.}, {Findeisen, K.}, {Fonti, A.}, {Fraile, E.}, {Fraser, M.},
  {Fr\'ezouls, B.}, {Gai, M.}, {Galleti, S.}, {Garabato, D.},
  {Garc\'{\i}a-Sedano, F.}, {Garofalo, A.}, {Garralda, N.}, {Gavel, A.},
  {Gavras, P.}, {Gerssen, J.}, {Geyer, R.}, {Giacobbe, P.}, {Gilmore, G.},
  {Girona, S.}, {Giuffrida, G.}, {Glass, F.}, {Gomes, M.}, {Granvik, M.},
  {Gueguen, A.}, {Guerrier, A.}, {Guiraud, J.}, {Guti\'errez-S\'anchez, R.},
  {Haigron, R.}, {Hatzidimitriou, D.}, {Hauser, M.}, {Haywood, M.}, {Heiter,
  U.}, {Helmi, A.}, {Heu, J.}, {Hilger, T.}, {Hobbs, D.}, {Hofmann, W.},
  {Holland, G.}, {Huckle, H. E.}, {Hypki, A.}, {Icardi, V.}, {Jan\ss{}en, K.},
  {Jevardat de Fombelle, G.}, {Jonker, P. G.}, {Juh\'asz, \'A. L.}, {Julbe,
  F.}, {Karampelas, A.}, {Kewley, A.}, {Klar, J.}, {Kochoska, A.}, {Kohley,
  R.}, {Kolenberg, K.}, {Kontizas, M.}, {Kontizas, E.}, {Koposov, S. E.},
  {Kordopatis, G.}, {Kostrzewa-Rutkowska, Z.}, {Koubsky, P.}, {Lambert, S.},
  {Lanza, A. F.}, {Lasne, Y.}, {Lavigne, J.-B.}, {Le Fustec, Y.}, {Le
  Poncin-Lafitte, C.}, {Lebreton, Y.}, {Leccia, S.}, {Leclerc, N.},
  {Lecoeur-Taibi, I.}, {Lenhardt, H.}, {Leroux, F.}, {Liao, S.}, {Licata, E.},
  {Lindstr\o{}m, H. E. P.}, {Lister, T. A.}, {Livanou, E.}, {Lobel, A.},
  {L\'opez, M.}, {Managau, S.}, {Mann, R. G.}, {Mantelet, G.}, {Marchal, O.},
  {Marchant, J. M.}, {Marconi, M.}, {Marinoni, S.}, {Marschalk\'o, G.},
  {Marshall, D. J.}, {Martino, M.}, {Marton, G.}, {Mary, N.}, {Massari, D.},
  {Matijevic, G.}, {Mazeh, T.}, {McMillan, P. J.}, {Messina, S.}, {Michalik,
  D.}, {Millar, N. R.}, {Molina, D.}, {Molinaro, R.}, {Moln\'ar, L.},
  {Montegriffo, P.}, {Mor, R.}, {Morbidelli, R.}, {Morel, T.}, {Morris, D.},
  {Mulone, A. F.}, {Muraveva, T.}, {Musella, I.}, {Nelemans, G.}, {Nicastro,
  L.}, {Noval, L.}, {O\'{}Mullane, W.}, {Ord\'enovic, C.}, {Ord\'o\~nez-Blanco,
  D.}, {Osborne, P.}, {Pagani, C.}, {Pagano, I.}, {Pailler, F.}, {Palacin, H.},
  {Palaversa, L.}, {Panahi, A.}, {Pawlak, M.}, {Piersimoni, A. M.}, {Pineau,
  F.-X.}, {Plachy, E.}, {Plum, G.}, {Poggio, E.}, {Poujoulet, E.}, {Prsa, A.},
  {Pulone, L.}, {Racero, E.}, {Ragaini, S.}, {Rambaux, N.}, {Ramos-Lerate, M.},
  {Regibo, S.}, {Reyl\'e, C.}, {Riclet, F.}, {Ripepi, V.}, {Riva, A.}, {Rivard,
  A.}, {Rixon, G.}, {Roegiers, T.}, {Roelens, M.}, {Romero-G\'omez, M.},
  {Rowell, N.}, {Royer, F.}, {Ruiz-Dern, L.}, {Sadowski, G.}, {Sagrist\`a
  Sell\'es, T.}, {Sahlmann, J.}, {Salgado, J.}, {Salguero, E.}, {Sanna, N.},
  {Santana-Ros, T.}, {Sarasso, M.}, {Savietto, H.}, {Schultheis, M.}, {Sciacca,
  E.}, {Segol, M.}, {Segovia, J. C.}, {S\'egransan, D.}, {Shih, I-C.},
  {Siltala, L.}, {Silva, A. F.}, {Smart, R. L.}, {Smith, K. W.}, {Solano, E.},
  {Solitro, F.}, {Sordo, R.}, {Soria Nieto, S.}, {Souchay, J.}, {Spagna, A.},
  {Spoto, F.}, {Stampa, U.}, {Steele, I. A.}, {Steidelm\"uller, H.},
  {Stephenson, C. A.}, {Stoev, H.}, {Suess, F. F.}, {Surdej, J.}, {Szabados,
  L.}, {Szegedi-Elek, E.}, {Tapiador, D.}, {Taris, F.}, {Tauran, G.}, {Taylor,
  M. B.}, {Teixeira, R.}, {Terrett, D.}, {Teyssandier, P.}, {Thuillot, W.},
  {Titarenko, A.}, {Torra Clotet, F.}, {Turon, C.}, {Ulla, A.}, {Utrilla, E.},
  {Uzzi, S.}, {Vaillant, M.}, {Valentini, G.}, {Valette, V.}, {van Elteren,
  A.}, {Van Hemelryck, E.}, {van Leeuwen, M.}, {Vaschetto, M.}, {Vecchiato,
  A.}, {Veljanoski, J.}, {Viala, Y.}, {Vicente, D.}, {Vogt, S.}, {von Essen,
  C.}, {Voss, H.}, {Votruba, V.}, {Voutsinas, S.}, {Walmsley, G.}, {Weiler,
  M.}, {Wertz, O.}, {Wevers, T.}, {Wyrzykowski, L.}, {Yoldas, A.}, {Zerjal,
  M.}, {Ziaeepour, H.}, {Zorec, J.}, {Zschocke, S.}, {Zucker, S.}, {Zurbach,
  C.}, \& {Zwitter, T.}}]{Gaia2018}
{Gaia Collaboration}, {Brown, A. G. A.}, {Vallenari, A.}, {et~al.} 2018, A\&A,
  616, A1, \dodoi{10.1051/0004-6361/201833051}

\bibitem[{Green {et~al.}(2019)Green, Schlafly, Zucker, Speagle, \&
  Finkbeiner}]{Green2019}
Green, G.~M., Schlafly, E., Zucker, C., Speagle, J.~S., \& Finkbeiner, D. 2019,
  The Astrophysical Journal, 887, 93, \dodoi{10.3847/1538-4357/ab5362}

\bibitem[{{Green} {et~al.}(2018){Green}, {Schlafly}, {Finkbeiner}, {Rix},
  {Martin}, {Burgett}, {Draper}, {Flewelling}, {Hodapp}, {Kaiser}, {Kudritzki},
  {Magnier}, {Metcalfe}, {Tonry}, {Wainscoat}, \& {Waters}}]{Green2018}
{Green}, G.~M., {Schlafly}, E.~F., {Finkbeiner}, D., {et~al.} 2018, \mnras,
  478, 651, \dodoi{10.1093/mnras/sty1008}

\bibitem[{Hardegree-Ullman {et~al.}(2019)Hardegree-Ullman, Cushing, Muirhead,
  \& Christiansen}]{Hardegree19}
Hardegree-Ullman, K.~K., Cushing, M.~C., Muirhead, P.~S., \& Christiansen,
  J.~L. 2019, The Astronomical Journal, 158, 75,
  \dodoi{10.3847/1538-3881/ab21d2}

\bibitem[{{Henry} {et~al.}(2006){Henry}, {Jao}, {Subasavage}, {Beaulieu},
  {Ianna}, {Costa}, \& {M{\'e}ndez}}]{Henry2006}
{Henry}, T.~J., {Jao}, W.-C., {Subasavage}, J.~P., {et~al.} 2006, \aj, 132,
  2360, \dodoi{10.1086/508233}

\bibitem[{{Howard} {et~al.}(2012){Howard}, {Marcy}, {Bryson}, {Jenkins},
  {Rowe}, {Batalha}, {Borucki}, {Koch}, {Dunham}, {Gautier}, {Van Cleve},
  {Cochran}, {Latham}, {Lissauer}, {Torres}, {Brown}, {Gilliland}, {Buchhave},
  {Caldwell}, {Christensen-Dalsgaard}, {Ciardi}, {Fressin}, {Haas}, {Howell},
  {Kjeldsen}, {Seager}, {Rogers}, {Sasselov}, {Steffen}, {Basri},
  {Charbonneau}, {Christiansen}, {Clarke}, {Dupree}, {Fabrycky}, {Fischer},
  {Ford}, {Fortney}, {Tarter}, {Girouard}, {Holman}, {Johnson}, {Klaus},
  {Machalek}, {Moorhead}, {Morehead}, {Ragozzine}, {Tenenbaum}, {Twicken},
  {Quinn}, {Isaacson}, {Shporer}, {Lucas}, {Walkowicz}, {Welsh}, {Boss},
  {Devore}, {Gould}, {Smith}, {Morris}, {Prsa}, {Morton}, {Still}, {Thompson},
  {Mullally}, {Endl}, \& {MacQueen}}]{Howard2012}
{Howard}, A.~W., {Marcy}, G.~W., {Bryson}, S.~T., {et~al.} 2012, The
  Astrophysical Journal Supplement Series, 201, 15,
  \dodoi{10.1088/0067-0049/201/2/15}

\bibitem[{Hsu {et~al.}(2019)Hsu, Ford, Ragozzine, \& Ashby}]{Hsu2019}
Hsu, D.~C., Ford, E.~B., Ragozzine, D., \& Ashby, K. 2019, The Astronomical
  Journal, 158, 109, \dodoi{10.3847/1538-3881/ab31ab}

\bibitem[{Huber {et~al.}(2014)Huber, Aguirre, Matthews, Pinsonneault, Gaidos,
  Garc{\'{\i}}a, Hekker, Mathur, Mosser, Torres, Bastien, Basu, Bedding,
  Chaplin, Demory, Fleming, Guo, Mann, Rowe, Serenelli, Smith, \&
  Stello}]{Huber2014}
Huber, D., Aguirre, V.~S., Matthews, J.~M., {et~al.} 2014, The Astrophysical
  Journal Supplement Series, 211, 2, \dodoi{10.1088/0067-0049/211/1/2}

\bibitem[{Jarrett {et~al.}(2000)Jarrett, Chester, Cutri, Schneider, Skrutskie,
  \& Huchra}]{Jarrett_2000}
Jarrett, T.~H., Chester, T., Cutri, R., {et~al.} 2000, The Astronomical
  Journal, 119, 2498–2531, \dodoi{10.1086/301330}

\bibitem[{{Kraus} {et~al.}(2016){Kraus}, {Ireland}, {Huber}, {Mann}, \&
  {Dupuy}}]{Krauss_binaries}
{Kraus}, A.~L., {Ireland}, M.~J., {Huber}, D., {Mann}, A.~W., \& {Dupuy}, T.~J.
  2016, \aj, 152, 8, \dodoi{10.3847/0004-6256/152/1/8}

\bibitem[{{L{\'e}pine} {et~al.}(2007){L{\'e}pine}, {Rich}, \&
  {Shara}}]{Lepine2007}
{L{\'e}pine}, S., {Rich}, R.~M., \& {Shara}, M.~M. 2007, \apj, 669, 1235,
  \dodoi{10.1086/521614}

\bibitem[{Lindgren {et~al.}(2016)Lindgren, Heiter, \& Seifahrt}]{Lindgren2016}
Lindgren, S., Heiter, U., \& Seifahrt, A. 2016, Astronomy \& Astrophysics, 586,
  A100, \dodoi{10.1051/0004-6361/201526602}

\bibitem[{{Lissauer} {et~al.}(2011){Lissauer}, {Ragozzine}, {Fabrycky},
  {Steffen}, {Ford}, {Jenkins}, {Shporer}, {Holman}, {Rowe}, {Quintana},
  {Batalha}, {Borucki}, {Bryson}, {Caldwell}, {Carter}, {Ciardi}, {Dunham},
  {Fortney}, {Gautier}, {Howell}, {Koch}, {Latham}, {Marcy}, {Morehead}, \&
  {Sasselov}}]{Lissauer2011}
{Lissauer}, J.~J., {Ragozzine}, D., {Fabrycky}, D.~C., {et~al.} 2011, The
  Astrophysical Journal Supplement Series, 197, 8,
  \dodoi{10.1088/0067-0049/197/1/8}

\bibitem[{Mann {et~al.}(2014)Mann, Deacon, Gaidos, Ansdell, Brewer, Liu,
  Magnier, \& Aller}]{Mann2014}
Mann, A.~W., Deacon, N.~R., Gaidos, E., {et~al.} 2014, The Astronomical
  Journal, 147, 160, \dodoi{10.1088/0004-6256/147/6/160}

\bibitem[{Mann {et~al.}(2012)Mann, Gaidos, L{\'{e}}pine, \& Hilton}]{Mann2012}
Mann, A.~W., Gaidos, E., L{\'{e}}pine, S., \& Hilton, E.~J. 2012, The
  Astrophysical Journal, 753, 90, \dodoi{10.1088/0004-637x/753/1/90}

\bibitem[{Mann {et~al.}(2019)Mann, Dupuy, Kraus, Gaidos, Ansdell, Ireland,
  Rizzuto, Hung, Dittmann, Factor, Feiden, Martinez,
  Ru{\'{\i}}z-Rodr{\'{\i}}guez, \& Thao}]{Mann2019}
Mann, A.~W., Dupuy, T., Kraus, A.~L., {et~al.} 2019, The Astrophysical Journal,
  871, 63, \dodoi{10.3847/1538-4357/aaf3bc}

\bibitem[{{Moriarty} \& {Ballard}(2016)}]{Moriarty16}
{Moriarty}, J., \& {Ballard}, S. 2016, \apj, 832, 34,
  \dodoi{10.3847/0004-637X/832/1/34}

\bibitem[{Muirhead {et~al.}(2012)Muirhead, Hamren, Schlawin, Rojas-Ayala,
  Covey, \& Lloyd}]{Muirhead2012}
Muirhead, P.~S., Hamren, K., Schlawin, E., {et~al.} 2012, The Astrophysical
  Journal, 750, L37, \dodoi{10.1088/2041-8205/750/2/l37}

\bibitem[{Muirhead {et~al.}(2014)Muirhead, Becker, Feiden, Rojas-Ayala,
  Vanderburg, Price, Thorp, Law, Riddle, Baranec, Hamren, Schlawin, Covey,
  Johnson, \& Lloyd}]{Muirhead2014}
Muirhead, P.~S., Becker, J., Feiden, G.~A., {et~al.} 2014, The Astrophysical
  Journal Supplement Series, 213, 5, \dodoi{10.1088/0067-0049/213/1/5}

\bibitem[{{Muirhead} {et~al.}(2015){Muirhead}, {Mann}, {Vanderburg}, {Morton},
  {Kraus}, {Ireland}, {Swift}, {Feiden}, {Gaidos}, \& {Gazak}}]{Muirhead2015}
{Muirhead}, P.~S., {Mann}, A.~W., {Vanderburg}, A., {et~al.} 2015, \apj, 801,
  18, \dodoi{10.1088/0004-637X/801/1/18}

\bibitem[{{Mulders} {et~al.}(2015){Mulders}, {Pascucci}, \&
  {Apai}}]{Mulders2015}
{Mulders}, G.~D., {Pascucci}, I., \& {Apai}, D. 2015, \apj, 814, 130,
  \dodoi{10.1088/0004-637X/814/2/130}

\bibitem[{{Mulders} {et~al.}(2016){Mulders}, {Pascucci}, {Apai}, {Frasca}, \&
  {Molenda-{\.Z}akowicz}}]{Mulders16}
{Mulders}, G.~D., {Pascucci}, I., {Apai}, D., {Frasca}, A., \&
  {Molenda-{\.Z}akowicz}, J. 2016, \aj, 152, 187,
  \dodoi{10.3847/0004-6256/152/6/187}

\bibitem[{Newton {et~al.}(2013)Newton, Charbonneau, Irwin, Berta-Thompson,
  Rojas-Ayala, Covey, \& Lloyd}]{Newton2014}
Newton, E.~R., Charbonneau, D., Irwin, J., {et~al.} 2013, The Astronomical
  Journal, 147, 20, \dodoi{10.1088/0004-6256/147/1/20}

\bibitem[{{Ricker} {et~al.}(2014){Ricker}, {Winn}, {Vanderspek}, {Latham},
  {Bakos}, {Bean}, {Berta-Thompson}, {Brown}, {Buchhave}, \& {Butler}}]{TESS}
{Ricker}, G.~R., {Winn}, J.~N., {Vanderspek}, R., {et~al.} 2014, in Society of
  Photo-Optical Instrumentation Engineers (SPIE) Conference Series, Vol. 9143,
  \procspie, 914320

\bibitem[{{Ricker} {et~al.}(2015){Ricker}, {Winn}, {Vanderspek}, {Latham},
  {Bakos}, {Bean}, {Berta-Thompson}, {Brown}, {Buchhave}, {Butler}, {Butler},
  {Chaplin}, {Charbonneau}, {Christensen-Dalsgaard}, {Clampin}, {Deming},
  {Doty}, {De Lee}, {Dressing}, {Dunham}, {Endl}, {Fressin}, {Ge}, {Henning},
  {Holman}, {Howard}, {Ida}, {Jenkins}, {Jernigan}, {Johnson}, {Kaltenegger},
  {Kawai}, {Kjeldsen}, {Laughlin}, {Levine}, {Lin}, {Lissauer}, {MacQueen},
  {Marcy}, {McCullough}, {Morton}, {Narita}, {Paegert}, {Palle}, {Pepe},
  {Pepper}, {Quirrenbach}, {Rinehart}, {Sasselov}, {Sato}, {Seager},
  {Sozzetti}, {Stassun}, {Sullivan}, {Szentgyorgyi}, {Torres}, {Udry}, \&
  {Villasenor}}]{Ricker2015}
{Ricker}, G.~R., {Winn}, J.~N., {Vanderspek}, R., {et~al.} 2015, Journal of
  Astronomical Telescopes, Instruments, and Systems, 1, 014003,
  \dodoi{10.1117/1.JATIS.1.1.014003}

\bibitem[{{Rojas-Ayala} {et~al.}(2012){Rojas-Ayala}, {Covey}, {Muirhead}, \&
  {Lloyd}}]{Ayala2012}
{Rojas-Ayala}, B., {Covey}, K.~R., {Muirhead}, P.~S., \& {Lloyd}, J.~P. 2012,
  \apj, 748, 93, \dodoi{10.1088/0004-637X/748/2/93}

\bibitem[{Romero \& Kempton(2018)}]{Romero2018}
Romero, C. E.~M., \& Kempton, E. M.-R. 2018, The Astronomical Journal, 155,
  134, \dodoi{10.3847/1538-3881/aaab5e}

\bibitem[{Schlaufman(2015)}]{Schlaufman2015}
Schlaufman, K.~C. 2015, The Astrophysical Journal, 799, L26,
  \dodoi{10.1088/2041-8205/799/2/l26}

\bibitem[{Schlaufman \& Laughlin(2011)}]{Schlaufman2011}
Schlaufman, K.~C., \& Laughlin, G. 2011, The Astrophysical Journal, 738, 177,
  \dodoi{10.1088/0004-637x/738/2/177}

\bibitem[{{Schlegel} {et~al.}(1998){Schlegel}, {Finkbeiner}, \&
  {Davis}}]{Schlegel1998}
{Schlegel}, D.~J., {Finkbeiner}, D.~P., \& {Davis}, M. 1998, \apj, 500, 525,
  \dodoi{10.1086/305772}

\bibitem[{Souto {et~al.}(2018)Souto, Cunha, Smith, Prieto,
  Garc{\'{\i}}a-Hern{\'{a}}ndez, Pinsonneault, Holzer, Frinchaboy, Holtzman,
  Johnson, Jönsson, Majewski, Shetrone, Sobeck, Stringfellow, Teske, Zamora,
  Zasowski, Carrera, Stassun, Fernandez-Trincado, Villanova, Minniti, \&
  Santana}]{Souto2018}
Souto, D., Cunha, K., Smith, V.~V., {et~al.} 2018, The Astrophysical Journal,
  857, 14, \dodoi{10.3847/1538-4357/aab612}

\bibitem[{{Terrien} {et~al.}(2012){Terrien}, {Mahadevan}, {Bender},
  {Deshpande}, {Ramsey}, \& {Bochanski}}]{Terrien1}
{Terrien}, R.~C., {Mahadevan}, S., {Bender}, C.~F., {et~al.} 2012, \apjl, 747,
  L38, \dodoi{10.1088/2041-8205/747/2/L38}

\bibitem[{{Terrien} {et~al.}(2015){Terrien}, {Mahadevan}, {Deshpande}, \&
  {Bender}}]{Terrien2}
{Terrien}, R.~C., {Mahadevan}, S., {Deshpande}, R., \& {Bender}, C.~F. 2015,
  \apjs, 220, 16, \dodoi{10.1088/0067-0049/220/1/16}

\bibitem[{{Thompson} {et~al.}(2018){Thompson}, {Coughlin}, {Hoffman},
  {Mullally}, {Christiansen}, {Burke}, {Bryson}, {Batalha}, {Haas},
  {Catanzarite}, {Rowe}, {Barentsen}, {Caldwell}, {Clarke}, {Jenkins}, {Li},
  {Latham}, {Lissauer}, {Mathur}, {Morris}, {Seader}, {Smith}, {Klaus},
  {Twicken}, {Van Cleve}, {Wohler}, {Akeson}, {Ciardi}, {Cochran}, {Henze},
  {Howell}, {Huber}, {Pr{\v s}a}, {Ram{\'{\i}}rez}, {Morton}, {Barclay},
  {Campbell}, {Chaplin}, {Charbonneau}, {Christensen-Dalsgaard}, {Dotson},
  {Doyle}, {Dunham}, {Dupree}, {Ford}, {Geary}, {Girouard}, {Isaacson},
  {Kjeldsen}, {Quintana}, {Ragozzine}, {Shabram}, {Shporer}, {Silva Aguirre},
  {Steffen}, {Still}, {Tenenbaum}, {Welsh}, {Wolfgang}, {Zamudio}, {Koch}, \&
  {Borucki}}]{Thompson18}
{Thompson}, S.~E., {Coughlin}, J.~L., {Hoffman}, K., {et~al.} 2018, \apjs, 235,
  38, \dodoi{10.3847/1538-4365/aab4f9}

\bibitem[{Wang \& Fischer(2014)}]{Wang2014}
Wang, J., \& Fischer, D.~A. 2014, The Astronomical Journal, 149, 14,
  \dodoi{10.1088/0004-6256/149/1/14}

\bibitem[{{Weiss} {et~al.}(2018){Weiss}, {Isaacson}, {Marcy}, {Howard},
  {Petigura}, {Fulton}, {Winn}, {Hirsch}, {Sinukoff}, {Rowe}, \& {California
  Kepler Survey}}]{Weiss18}
{Weiss}, L.~M., {Isaacson}, H.~T., {Marcy}, G.~W., {et~al.} 2018, \aj, 156,
  254, \dodoi{10.3847/1538-3881/aae70a}

\bibitem[{Winters {et~al.}(2019)Winters, Henry, Jao, Subasavage, Chatelain,
  Slatten, Riedel, Silverstein, \& Payne}]{Winters2019}
Winters, J.~G., Henry, T.~J., Jao, W.-C., {et~al.} 2019, The Astronomical
  Journal, 157, 216, \dodoi{10.3847/1538-3881/ab05dc}

\bibitem[{Woolf {et~al.}(2009)Woolf, L{\'{e}}pine, \& Wallerstein}]{Woolf2009}
Woolf, V.~M., L{\'{e}}pine, S., \& Wallerstein, G. 2009, Publications of the
  Astronomical Society of the Pacific, 121, 117, \dodoi{10.1086/597433}

\bibitem[{Zhao {et~al.}(2012)Zhao, Zhao, Chu, Jing, \& Deng}]{LAMOST}
Zhao, G., Zhao, Y.-H., Chu, Y.-Q., Jing, Y.-P., \& Deng, L.-C. 2012, Research
  in Astronomy and Astrophysics, 12, 723, \dodoi{10.1088/1674-4527/12/7/002}

\bibitem[{{Zhu} {et~al.}(2018){Zhu}, {Petrovich}, {Wu}, {Dong}, \&
  {Xie}}]{Zhu18}
{Zhu}, W., {Petrovich}, C., {Wu}, Y., {Dong}, S., \& {Xie}, J. 2018, \apj, 860,
  101, \dodoi{10.3847/1538-4357/aac6d5}

\bibitem[{Ziegler {et~al.}(2018)Ziegler, Law, Baranec, Morton, Riddle, De~Lee,
  Huber, Mahadevan, \& Pepper}]{Ziegler2018}
Ziegler, C., Law, N.~M., Baranec, C., {et~al.} 2018, The Astronomical Journal,
  156, 259, \dodoi{10.3847/1538-3881/aad80a}

\bibitem[{{Zink} {et~al.}(2019){Zink}, {Christiansen}, \& {Hansen}}]{Zink19}
{Zink}, J.~K., {Christiansen}, J.~L., \& {Hansen}, B.~M.~S. 2019, \mnras, 483,
  4479, \dodoi{10.1093/mnras/sty3463}

\end{thebibliography}
%============== How do I make the bibliography? ==============
%Bibtex is great because it means you don't have to worry about bibliography formatting and that sort of thing. In LaTeX, you point the command \bibliography to a file (in this case, the creatively named bibliography.bib), and it will automatically compile the bibliography based on the sources that you cited in the main text via \cite{}, \citep{}, \citealt{}, etc. 

%In the bibliography.bib file is a list of sources in BibTeX format. In order to get these sources, you can go to NASA ADS (http://adsabs.harvard.edu/abstract_service.html) and find the paper that you want to cite. Each paper in the ADS service has a link to a BibTeX entry at the bottom of the. Click on this, and then copy and paste it into your bibliography file and you're done. If you want, you can alter the top line of each source to rename it to what you want to use to cite in in your \cite commands. 

\end{document}